\documentclass[a4paper,10pt]{elsarticle}
\usepackage{jcomp}
\usepackage{framed,multirow}
\usepackage[utf8]{inputenc}
\usepackage{bm}
\usepackage{amsmath}
\usepackage{mathrsfs}
\usepackage{graphicx}
\usepackage{epsfig, setspace}
\usepackage{url}
\usepackage{graphicx, transparent, color}
\usepackage{algorithm} 
\usepackage{algorithmicx}
\usepackage{etex}
\usepackage{dirtytalk}
\usepackage[mathscr]{euscript}
\usepackage[bookmarks=true]{hyperref}
\usepackage{amsmath}
\usepackage{xcolor, soul}
\usepackage{rotating} 
\usepackage{pdflscape} 
\usepackage{silence}
\usepackage{ulem,cancel}

\usepackage{graphicx}
\usepackage{float}
\usepackage{subfigure}
\usepackage{subfigmat}

\usepackage{tikz}
\usepackage{capt-of}
\usepackage{setspace,lipsum}
\usepackage{lineno}
\usepackage{gensymb}
\usepackage{amssymb}
\biboptions{sort&compress}

\usepackage{setspace,lipsum}
\usepackage{listings}
\usepackage{xcolor}
\definecolor{aquamarine}{rgb}{0.5, 1.0, 0.83}
\definecolor{OliveGreen}{rgb}{0,0.6,0}
\usepackage[bottom]{footmisc}

\sethlcolor{aquamarine}

\definecolor{codegreen}{rgb}{0,0.6,0}
\definecolor{codegray}{rgb}{0.5,0.5,0.5}
\definecolor{codepurple}{rgb}{0.58,0,0.82}
\definecolor{backcolour}{rgb}{0.95,0.95,0.92}

\lstdefinestyle{mystyle}{
    backgroundcolor=\color{backcolour},   
    commentstyle=\color{codegreen},
    keywordstyle=\color{magenta},
    numberstyle=\tiny\color{codegray},
    stringstyle=\color{codepurple},
    basicstyle=\ttfamily\footnotesize,
    breakatwhitespace=false,         
    breaklines=true,                 
    captionpos=b,                    
    keepspaces=true,                 
    numbers=left,                    
    numbersep=5pt,                  
    showspaces=false,                
    showstringspaces=false,
    showtabs=false,                  
    tabsize=2
}

\begin{document}

\begin{frontmatter}
\title{On the Application of Gradient Based Reconstruction for Flow Simulations on Generalized Curvilinear and Dynamic Mesh Domains}

\author[]{Hemanth Chandravamsi\footnote{Corresponding author, hemanthgrylls@gmail.com}}
\author[]{Amareshwara Sainadh Chamarthi}
\author[]{Natan Hoffmann}
\author[]{Steven H.\ Frankel}
\address{Faculty of Mechanical Engineering, Technion - Israel Institute of Technology, Haifa, Israel}


\begin{abstract}

Accurate high-speed flow simulations of practical interest require numerical methods with high-resolution properties. In this paper, we present an extension and demonstration of the high-accuracy Gradient-based reconstruction and $\alpha$-damping schemes introduced by Chamarthi (2022) \cite{chamx} for simulating high-speed flows in generalized curvilinear and dynamic mesh domains with the freestream preservation property. In the first part of this paper, the algorithms are detailed within the generalized curvilinear coordinate framework, with a focus on demonstration through stationary and dynamic mesh test cases. It has been shown both theoretically and through the use of test cases that the conservative metrics, including their interpolation to cell interfaces, must be numerically computed using a central scheme that is consistent with the inviscid flux algorithm to achieve the freestream preservation property. The second part of the paper illustrates the efficacy of the algorithm in simulating supersonic jet screech by displaying its capability to capture the screech tones and accurately characterize the unsteady lateral flapping mode of a Mach 1.35 under-expanded supersonic jet, in contrast to the WENO-Z scheme which fails to do so at the same grid resolution. In the final part of the paper, the parallelizability of the schemes on GPU architectures is demonstrated and performance metrics are evaluated. A significant speedup of over $200 \times$ (compared to a single core CPU) and a reduction in simulation completion time to 34.5 hours per simulation were achieved for the supersonic jet noise case at a grid resolution of 13 million cells.

\end{abstract}


\end{frontmatter}

\section{Introduction}\label{sec-1}

Accurate numerical simulations of high-speed flows demand methods that exhibit good shock-capturing and eddy-resolving capabilities. On this front, several high-resolution numerical methods and improvements are proposed in the literature every year \cite{chamarthi2021high,li2021low}. Many of the proposed methods are generally developed and tested on Cartesian grids and employing canonical test cases. However, extending these methods and testing their applicability on the non-uniform meshes is also important to enable complex flow simulations of practical interest. In the previous paper from our group by Chamarthi \cite{chamx}, a new gradient based reconstruction algorithm and $\alpha$-damping \cite{Nishikawa2010} viscous flux discretization scheme were proposed. In this paper, we extend the proposed numerical schemes of Chamarthi \cite{chamx} to a generalized curvilinear coordinate system and perform demonstrations on highly skewed and dynamically deforming meshes. Moreover, we also demonstrate the method's parallelizability and efficacy in resolving the physics of noise due to an under-expanded supersonic jet.

One of the first widely recognized high-order shock-capturing schemes in the literature was the Weighted Essentially Non-Oscillatory (WENO) class of schemes first introduced by Liu et al. \cite{liu1994weighted} and later improved by Jiang and Shu \cite{jiang1996efficient}. The performance of WENO schemes and the issues concerning its extension to simulate flows over non-uniform curvilinear grids were addressed in various studies \cite{cai2008performance,nonomura2010freestream,shadab2019fifth}. It was then followed by the wide usage of the WENO schemes to study the physics of numerous high-speed flow problems of practical interest \cite{cao2019gortler,cheng2005numerical}. Although the WENO class of schemes has gained significant popularity in the literature with several enhancements proposed over the years \cite{martin2006bandwidth,Henrick2005,Fu2016}, alternative high-order shock-capturing schemes with even better spectral properties were also studied and applied. A few examples include the limiter-based Monotonocity Preserving (MP) approach of Suresh and Huynh \cite{suresh1997accurate}, filtering/artificial dissipation based approaches \cite{kawai2008localized, kakumani2022use}, and hybrid approaches such as Boundary Variation Diminishing (BVD) algorithm \cite{sun2016boundary}. 
Efforts were also made by researchers to mitigate the dissipation coming from the shock capturing scheme. This was achieved through hybrid-central schemes, where a non-dissipative central scheme in smooth regions is combined with a shock capturing scheme near discontinuities. Examples of such attempts include the Hybrid Central-WENO schemes by Costa \& Don \cite{costa2007high} and Karami et al. \cite{karami2019high}. In addition, there have been attempts to design non-linearly stable numerical methods that retain the energy properties of the governing equations in a discrete sense. These methods, known as energy consistent schemes, have been found to enhance stability and eliminate numerical dissipation \cite{pirozzoli2011stabilized}. The inviscid scheme in the present study proposed by Chamarthi \cite{chamx} fundamentally operates on a novel reconstruction polynomial that employs explicit or implicit gradients of primitive variables to enhance the solution accuracy and an improved MP limiter for shock capturing. The two versions of the algorithm either use the standard central sixth-order explicit (E6) gradients or the fourth-order implicit gradients (IG4) (specific details of the gradient scheme will be presented in Sec. \ref{sec:disc}). The resultant schemes were named MEG6 and MIG4 where `M' stands for Monotonocity preserving, and `EG6' and `IG4' refer to sixth-order Explicit Gradients and fourth-order Implicit Gradients, respectively. \\

Historically, viscous flux discretization has received significantly less attention in the literature than the discretization of inviscid terms. Many of the widely used viscous flux discretization schemes suffer from the odd-even decoupling phenomenon and poor second derivative spectral properties \cite{chamarthi2022importance,sainadh2022spectral}. To counter this problem, Chamarthi \cite{chamx} has also proposed a new $\alpha$-damping discretization with superior spectral properties. One of the key benefits of the method is that the same cell-center primitive variable gradients used in inviscid scheme are used in the viscous flux discretization algorithm. This gradient-sharing strategy was demonstrated to balance the computational efficiency and solution resolution well. The new inviscid and viscous discretization schemes proposed in Ref. \cite{chamx} were primarily tested using two-dimensional test cases on Cartesian grids. However, the fine details of extending the MEG6/MIG4 and the new $\alpha$-damping schemes to curvilinear, stationary, and dynamic meshes have not been addressed thus far and require attention. This establishes the motivation for the first objective of the current work.

\noindent \textbf{Objective 1:} To adapt MEG6/MIG4 and $\alpha$-damping schemes to simulate flows on stationary and moving curvilinear grids. \\

Mesh non-uniformities and low-quality grid cells are unavoidable and difficult to control while designing grids for complex geometries. Often, such low-quality grids can affect the simulation results, mainly when the interest of the simulation is to capture sensitive flow features such as small amplitude acoustic waves, eddies, or instability waves. The freestream preservation property is vital while solving such flow problems on non-uniform grids (containing grid stretching and local skewness) since the errors associated with non-preserved freestream can grow to become comparable to the magnitude of essential flow features of interest. Visbal and Gaitonde \cite{Visbal2002} were the first to address this issue in the context of solving the compressible Navier-Stokes equations on curvilinear grids using high-order finite difference schemes. They employed the conservative metric term formulations of Thomas and Lombard \cite{thomas1979geometric} and maintained a consistent discretization between inviscid flux terms and metric terms to achieve freestream preservation on stationary and dynamically moving three-dimensional curvilinear meshes. Later, in 2010, Nonomura et al. \cite{nonomura2010freestream} pointed out that employing the same procedure for the WENO class of shock capturing schemes does not preserve freestream due to the nature of WENO reconstruction. They resolved this issue in the same work by proposing an alternative discretization approach. In the later years, other methods were proposed in the literature for freestream preservation in the context of the WENO class of schemes \cite{nonomura2015new,nonomura2010freestream,zhu2019free}. In addition to objective 1 stated above, the present work also explores the freestream preservation nature of the MEG6 and MIG4 schemes using theory and example numerical flow simulations. This motivates the second objective of this work.

\noindent \textbf{Objective 2:} To elucidate the required procedure to guarantee freestream preservation for the present methods on stationary and moving curvilinear grids. \\

The present study is part of the broader investigation initiated in the CFDLAB group at Technion to understand the fundamental and applied aspects of noise due to supersonic jets \cite{kakumani2023gpu}. Supersonic jet noise is one of the primary concerns in the aerospace industry. Although numerous experimental and computational studies have been conducted since the 1950s \cite{powell1953mechanism}, some of the fundamental aspects of supersonic jet noise remain unsolved. Of all the supersonic jet noise components, the noise due to aeroacoustic resonance (commonly referred to as screech) can be harmful due to its discrete tonal nature and requires a through physical understanding and control. The review article by Edgington \cite{edgington2019aeroacoustic} summarizes the current understanding of the fundamental aspects of aeroacoustic resonance loop in shock containing supersonic jets. It was believed that the high-accuracy methods (MEG6/MIG4 and $\alpha$-damping schemes) adapted in the current work could offer superior fidelity in resolving the three-dimensional physical aspects of supersonic jet noise that are otherwise challenging to capture even through state-of-the-art experimental diagnostics. This leads to the third objective of the current study.

\noindent \textbf{Objective 3:} To explore the efficacy of MEG6 and MIG4 schemes in resolving the screech tones and unsteady aspects of a supersonic jet. \\


Large Eddy Simulations (LES) and Direct Numerical Simulations (DNS) are the current state-of-the-art high-fidelity simulation strategies that enable us to understand the physics of fundamental and applied flow problems. However, the grid resolution requirements for a wall resolved LES roughly scales with the $\left(\frac{13}{7}\right)^{\text{th}}$ power of the flow Reynolds number \cite{choi2012grid}. As a result, the primary bottleneck in performing LES (particularly of high Reynolds number flows) is the computational time. Although the computing power of CPUs has been growing continuously over the years (owing to their increasing transistor concentration), significant speedup gains in recent times were achieved through dedicated accelerator cards, especially from Graphics Processing Units (GPUs). In recent years, there has been an increasing interest in the use of GPUs to accelerate Computational Fluid Dynamics (CFD) applications. A number of popular open-source GPU-based CFD solvers have been developed, such as PyFR \cite{witherden2014pyfr}, STREAmS \cite{bernardini2021streams}, ZEFR \cite{romero2020zefr}, and HTR \cite{di2021htr}. These solvers utilize some of the latest generations of high-order algorithms to simulate a range of flow regimes including incompressible \cite{witherden2014pyfr}, compressible \cite{bernardini2021streams,romero2020zefr}, and turbulent reacting flows \cite{di2021htr}. Additionally, several non-open-source GPU accelerated codes have also been developed, such as CharLES \cite{goc2021large} and COMP-SQUARE \cite{nampelly2022surface}. A study by Konrad et al. \cite{goc2021large} has shown that utilizing 96 NVIDIA V100 GPU cards can result in up to $26$ times faster performance than using 2000 CPU cores for the same task. This effectively means that each GPU can replace approximately 540 CPU cores. The exceptional acceleration capabilities of GPUs can be attributed to their unique hardware architecture, which is comprised of hundreds of weaker processing elements, referred to as threads, rather than a smaller number of more powerful processor cores as in CPUs. This allows GPUs to excel at processing data when a limited number of straightforward instructions are to be executed on large datasets, a requirement commonly seen in CFD algorithms. Research has also shown that GPUs are more energy efficient than CPUs \cite{huang2009energy,vspetko2021dgx}, which has led to the widespread adoption of GPU cards in the new exascale data center supercomputers and desktop-grade workstations. To take advantage of this increased computing power, it is important to adapt parallelization models in CFD applications to suit GPU accelerators. In this study, we focus on accelerating the MEG6/MIG4 + $\alpha$-damping algorithm using OpenACC. While most of the GPU accelerated CFD applications in the literature primarily use the CUDA based programming model \cite{bres2022gpu,goc2021large,terrana2020gpu,laufer2022gpu,cernetic2022high}, we investigate the performance gains that could be achieved by using OpenACC a directive-based programming language, which requires less development time due to its high-level nature. The fourth objective of the present study is stated below.

\noindent \textbf{Objective 4:} To accelerate the MEG6/MIG4 algorithms on GPUs and explore their computational efficiency and performance statistics on the latest generations of data center GPUs. \\


The rest of the article is organized as follows. Firstly, the governing equations are presented. This is followed by a brief description of the conservative finite difference discretization approach in section \ref{sec:coner-FD+algo}. The freestream preserving conservative metric formulations are presented in section \ref{conser-metrics}. The discretization approach employed for inviscid and viscous terms is presented in Section \ref{sec:disc}. The freestream preservation nature of MEG6 and MIG4 schemes is discussed in section \ref{sec:FP}. Next, the results and demonstrations are presented using a suite of standard and practical test cases (section \ref{sec:results}). In section \ref{sec:gpu-accel} the GPU acceleration models adapted for single and multi-GPU simulations to accelerate the MEG6 and MIG4 algorithms are presented with detailed analyses. Concluding remarks are laid out in the last section.



%
\section{Governing equations}   \label{sec:gov-eqns}
The unsteady three-dimensional compressible Navier-Stokes equations (dimensional) in a generalized curvilinear coordinate system with $\xi$, $\eta$, and $\zeta$ as the spatial coordinate directions and $t$ as the time can be written in the following vector form:
\begin{equation} \label{trans-eqn}
    \frac{\partial}{\partial t}\left(\frac{\mathbf{Q}}{J}\right)+\frac{\partial \mathbf{\hat{F}}}{\partial \xi}+\frac{\partial \mathbf{\hat{G}}}{\partial \eta}+\frac{\partial \mathbf{\hat{H}}}{\partial \zeta}=\frac{\partial \mathbf{\hat{F}_{v}}}{\partial \xi}+\frac{\partial \mathbf{\hat{G}_{v}}}{\partial \eta}+\frac{\partial \mathbf{\hat{H}_{v}}}{\partial \zeta},
\end{equation}\label{NS_TC}

\noindent where $\mathbf{Q}$ represents the vector of conservative variables, i.e. $\mathbf{Q}=[\rho, \rho u, \rho v, \rho w, E]^T$. The vectors of inviscid fluxes ($\mathbf{\hat{F}}$, $\mathbf{\hat{G}}$ and $\mathbf{\hat{H}}$) and viscous fluxes ($\mathbf{\hat{F}^{v}}$, $\mathbf{\hat{G}^{v}}$ and $\mathbf{\hat{H}^{v}}$) are,

\begin{subequations} \label{inv_fluxes}
    \begin{gather}
    \mathbf{\hat{F}}=\left[\begin{array}{c}
    \rho \hat{U} \\
    \rho u \hat{U}+\hat{\xi}_{x} p \\
    \rho v \hat{U}+\hat{\xi}_{y} p \\
    \rho w \hat{U}+\hat{\xi}_{z} p \\
    (E+p) \hat{U} -\hat{\xi}_{t} p
    \end{array}\right], \quad \mathbf{\hat{G}}=\left[\begin{array}{c}
    \rho \hat{V} \\
    \rho u \hat{V}+\hat{\eta}_{x} p \\
    \rho v \hat{V}+\hat{\eta}_{y} p \\
    \rho w \hat{V}+\hat{\eta}_{z} p \\
    (E+p) \hat{V} -\hat{\eta}_{t} p
    \end{array}\right], \quad \mathbf{\hat{H}}=\left[\begin{array}{c}
    \rho \hat{W} \\
    \rho u \hat{W}+\hat{\zeta}_{x} p \\
    \rho v \hat{W}+\hat{\zeta}_{y} p \\
    \rho w \hat{W}+\hat{\zeta}_{z} p \\
    (E+p) \hat{W} -\hat{\zeta}_{t} p
    \end{array}\right].
    \tag{\theequation a-\theequation c}
    \end{gather}
\end{subequations}

\begin{subequations}\label{visc-curvi}
    \begin{gather}
        \mathbf{\hat{F}^{v}}=\left[\begin{array}{c}
    0 \\
    \hat{\xi}_{x} \tau_{x x}+\hat{\xi}_{y} \tau_{x y}+\hat{\xi}_{z} \tau_{x z} \\
    \hat{\xi}_{x} \tau_{y x}+\hat{\xi}_{y} \tau_{y y}+\hat{\xi}_{z} \tau_{y z} \\
    \hat{\xi}_{x} \tau_{z x}+\hat{\xi}_{y} \tau_{z y}+\hat{\xi}_{z} \tau_{z z} \\
    \hat{\xi}_{x} \beta_{x}+\hat{\xi}_{y} \beta_{y}+\hat{\xi}_{z} \beta_{z}
    \end{array}\right], 
        \tag{\theequation a-\theequation b}
        \quad
         \mathbf{\hat{G}^{v}}=\left[\begin{array}{c}
        0 \\
        \hat{\eta}_{x} \tau_{x x}+\hat{\eta}_{y} \tau_{x y}+\hat{\eta}_{z} \tau_{x z} \\
        \hat{\eta}_{x} \tau_{y x}+\hat{\eta}_{y} \tau_{y y}+\hat{\eta}_{z} \tau_{y z} \\
        \hat{\eta}_{x} \tau_{z x}+\hat{\eta}_{y} \tau_{z y}+\hat{\eta}_{z} \tau_{z z} \\
        \hat{\eta}_{x} \beta_{x}+\hat{\eta}_{y} \beta_{y}+\hat{\eta}_{z} \beta_{z}
        \end{array}\right], \\
       \mathbf{\hat{H}^{v}}=\left[\begin{array}{c}
    0 \\
    \hat{\zeta}_{x} \tau_{x x}+\hat{\zeta}_{y} \tau_{x y}+\hat{\zeta}_{z} \tau_{x z} \\
    \hat{\zeta}_{x} \tau_{y x}+\hat{\zeta}_{y} \tau_{y y}+\hat{\zeta}_{z} \tau_{y z} \\
    \hat{\zeta}_{x} \tau_{z x}+\hat{\zeta}_{y} \tau_{z y}+\hat{\zeta}_{z} \tau_{z z} \\
    \hat{\zeta}_{x} \beta_{x} +\hat{\zeta}_{y} \beta_{y} +\hat{\zeta}_{z} \beta_{z}
    \end{array}\right]. \tag{\theequation c}
    \end{gather}
\end{subequations}

The quantity `$J$' here represents the Jacobian of grid transformation. The metric terms in the inviscid and viscous flux terms in Eqns.(\ref{inv_fluxes},\ref{visc-curvi}) with a hat $\hat{(\cdot)}$ denote Jacobian normalized quantities (e.g. $\hat{\xi}_x = \xi_x/J$).\\

\noindent The contravariant velocities $\hat{U}$, $\hat{V}$, and $\hat{W}$ are,

\begin{subequations}
\begin{gather}
    \hat{U}=\hat{\xi}_{t} + \hat{\xi}_{x} u+\hat{\xi}_{y} v+\hat{\xi}_{z} w, \quad
    \hat{V}=\hat{\eta}_{t} +\hat{\eta}_{x} u+\hat{\eta}_{y} v+\hat{\eta}_{z} w, \quad
    \hat{W}=\hat{\zeta}_{t} + \hat{\zeta}_{x} u+\hat{\zeta}_{y} v+\hat{\zeta}_{z} w \tag{\theequation a-\theequation c}
\end{gather}
\end{subequations}

\noindent The energy per unit volume $E$ is,
\begin{equation}
E = \frac{p}{\gamma-1} + \frac{1}{2} \rho (u^2 + v^2 + w^2)
\end{equation}

For stationary meshes, the computation of the terms $\hat{\xi}_{t}$, $\hat{\eta}_{t}$, and $\hat{\zeta}_{t}$, can be skipped as they are zero. The viscous and thermal stress related terms in the equations are defined as,

\begin{subequations} \label{eqn:shear-str}
\begin{gather}
\tau_{x x}= \lambda \nabla \cdot \bar{u}+2 \mu \frac{\partial u}{\partial x}, \quad
\tau_{y y}= \lambda \nabla \cdot \bar{u}+2 \mu \frac{\partial v}{\partial y}, \quad
\tau_{z z}= \lambda \nabla \cdot \bar{u}+2 \mu \frac{\partial w}{\partial z},  \tag{\theequation a-\theequation c}\\
\tau_{x y}=\tau_{y x}=\mu\left(\frac{\partial u}{\partial y}+\frac{\partial v}{\partial x}\right), \quad
\tau_{x z}=\tau_{z x}=\mu\left(\frac{\partial u}{\partial z}+\frac{\partial w}{\partial x}\right), \quad 
\tau_{y z}=\tau_{z y}=\mu\left(\frac{\partial v}{\partial z}+\frac{\partial w}{\partial y}\right) \tag{\theequation d-\theequation f}
\end{gather}
\end{subequations}

\begin{subequations} \label{eqn:thermal-work}
    \begin{gather}
    \beta_{x}=u \tau_{x x}+v \tau_{x y}+w \tau_{x z}+ \kappa_t \frac{\partial T}{\partial x}, \quad
    \beta_{y}=u \tau_{y x}+v \tau_{y y}+w \tau_{y z}+ \kappa_t \frac{\partial T}{\partial y}, \quad
    \beta_{z}=u \tau_{z x}+v \tau_{z y}+w \tau_{z z}+ \kappa_t \frac{\partial T}{\partial z} \tag{\theequation a-\theequation c}
    \end{gather}
\end{subequations}

\noindent Where $\bar{u}$ in the normal stress terms denotes the velocity vector. The thermodynamic equation of state $p=\rho RT$ is used to close the governing equations. $\mu$ and $\kappa_t$ denote the fluid's dynamic viscosity and thermal conductivity, respectively, which are temperature-dependent properties. The dynamic viscosity is computed using Sutherland's law \cite{Sutherland1893} while thermal conductivity is calculated using the Prandtl number ($Pr$) and dynamic viscosity. Additionally, bulk viscosity $\lambda=-\frac{2}{3}\mu$ is incorporated into the normal stress terms $\tau_{xx}$, $\tau_{yy}$, and $\tau_{zz}$ based on Stokes' hypothesis. The values of $\gamma$ and $Pr$ are set to 1.4 and 0.71, respectively, for air as the working fluid. Although the governing equations are presented in their dimensional form here, they are solved in their non-dimensional form to facilitate problem parameterization and minimize computational errors. The ambient speed of sound, density, temperature, viscosity, and nozzle exit diameter are used as reference quantities for scaling the flow variables.



\section{Review of conservative finite difference framework, and an overview of the algorithm} \label{sec:coner-FD+algo}

\begin{figure}[h!]
    \centering
    \includegraphics[width=140mm]{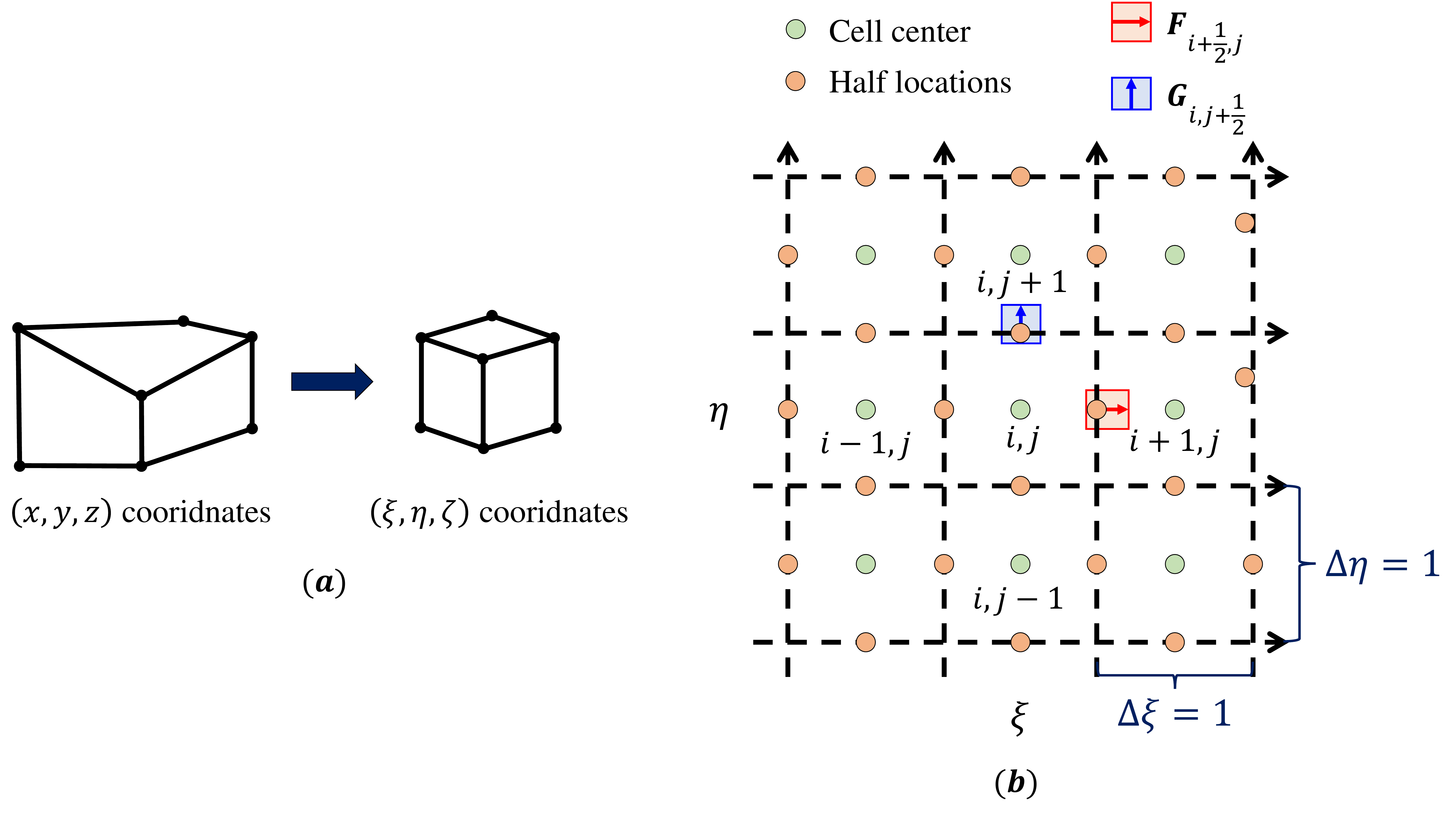}
    \caption{(a) A 3-dimensional grid cell before and after coordinate transformation, (b) A two-dimensional $\xi-\eta$ plane of the computational domain depicting, cell centers, interface centers (located at the centroid of cell interface), and interface fluxes.}
    \label{FD_grid}
\end{figure}

The current work employs MEG6, MIG4, and $\alpha$-damping schemes \cite{chamx}, which use a conservative finite difference approach to discretize the spatial derivative terms. Like most numerical schemes, these schemes are derived based on uniform stencils. Applying these schemes to non-uniform stencils (in curved/stretched grids) requires a recast of the Cartesian form of governing equations into generalized curvilinear coordinates form. The transformed equations are presented in Section \ref{sec:gov-eqns}. Such a coordinate transformation allows non-uniform skewed input cells of the mesh to be re-stretched into uniform cubical cells in the computational space, as shown in Fig. \ref{FD_grid}(a). Based on this idea, MEG6, MIG4, and $\alpha$-damping schemes are implemented in the solver to compute flow fields over non-uniform meshes.

The association between grid, solution, and conservative fluxes is as follows. The input mesh consists of hexahedral cells, identified by the coordinates of cell vertices provided in the grid file. A schematic of the two-dimensional solution domain in transformed coordinates is depicted in Fig. \ref{FD_grid}(b). The figure shows cell centers, interface centers (half locations), and fluxes. The solution $\textbf{Q}$ (conservative variables) is stored at the cell centers, which are located by the centroid of corresponding cell vertices. On the other hand, inviscid and diffusion fluxes ($\mathbf{F_{i+1/2}}, \mathbf{G_{i+1/2}}$ etc.) are computed at cell-interface centers to preserve conservation. The current paper also refers to these interface centers as half locations. After computing the inviscid and diffusion fluxes at all six half locations corresponding to the cell, the right-hand-side (RHS) residual corresponding to that cell denoted by $\mathbf{Res}_{i,j,k}$ is computed as follows:

\begin{equation}
 \frac{\partial}{\partial t}\left(\frac{\mathbf{Q}}{J}\right)=\mathbf{Res}_{i, j, k}=-\frac{\partial \left(\hat{\mathbf{F}} - \hat{\mathbf{F}}^{\mathrm{v}}\right)}{\partial \xi} -\frac{\partial \left(\hat{\mathbf{G}} - \hat{\mathbf{G}}^{\mathrm{v}}\right)}{\partial \eta} -\frac{\partial \left(\hat{\mathbf{H}} - \hat{\mathbf{H}}^{\mathrm{v}}\right)}{\partial \zeta}
\end{equation}

\begin{equation}
    \begin{aligned}
 \frac{\partial \left(\hat{\mathbf{F}} - \hat{\mathbf{F}}^{\mathrm{v}}\right)}{\partial \xi} = &\frac{1}{\Delta \xi}\left[\left(\hat{\mathbf{F}}_{i+\frac{1}{2}, j, k}-\hat{\mathbf{F}}_{i-\frac{1}{2}, j, k}\right)-\left(\hat{\mathbf{F}}_{i+\frac{1}{2}, j, k}^{\mathrm{v}}-\hat{\mathbf{F}}_{i-\frac{1}{2}, j, k}^{\mathrm{v}}\right)\right] \\
\frac{\partial \left(\hat{\mathbf{G}} - \hat{\mathbf{G}}^{\mathrm{v}}\right)}{\partial \eta} = &\frac{1}{\Delta \eta}\left[\left(\hat{\mathbf{G}}_{i, j+\frac{1}{2}, k}-\hat{\mathbf{G}}_{i, j-\frac{1}{2}, k}\right)-\left(\hat{\mathbf{G}}_{i, j+\frac{1}{2}, k}^{\mathrm{v}}-\hat{\mathbf{G}}_{i, j-\frac{1}{2}, k}^{\mathrm{v}}\right)\right] \\
\frac{\partial \left(\hat{\mathbf{H}} - \hat{\mathbf{H}}^{\mathrm{v}}\right)}{\partial \zeta} = &\frac{1}{\Delta \zeta}\left[\left(\hat{\mathbf{H}}_{i, j, k+\frac{1}{2}}-\hat{\mathbf{H}}_{i, j, k-\frac{1}{2}}\right)-\left(\hat{\mathbf{H}}_{i, j, k+\frac{1}{2}}^{\mathrm{v}}-\hat{\mathbf{H}}_{i, j, k-\frac{1}{2}}^{\mathrm{v}}\right)\right].
\end{aligned}
\end{equation}

The above formulation is conservative throughout the computational domain since the residuals are computed based on fluxes at cell interfaces (half-locations). Even though the fluxes are estimated at the cell interfaces, the process involved in computing these fluxes ($\hat{\mathbf{F}}, \hat{\mathbf{G}}, \hat{\mathbf{H}}, \hat{\mathbf{F}}^{\mathrm{v}}, \hat{\mathbf{G}}^{\mathrm{v}}, \hat{\mathbf{H}}^{\mathrm{v}}$) only requires the cell-center solution information and uses finite difference approximations for reconstruction. Hence the approach is classified as a conservative finite difference approach. 

\begin{figure}[h!]
    \centering
    \includegraphics[width=130mm]{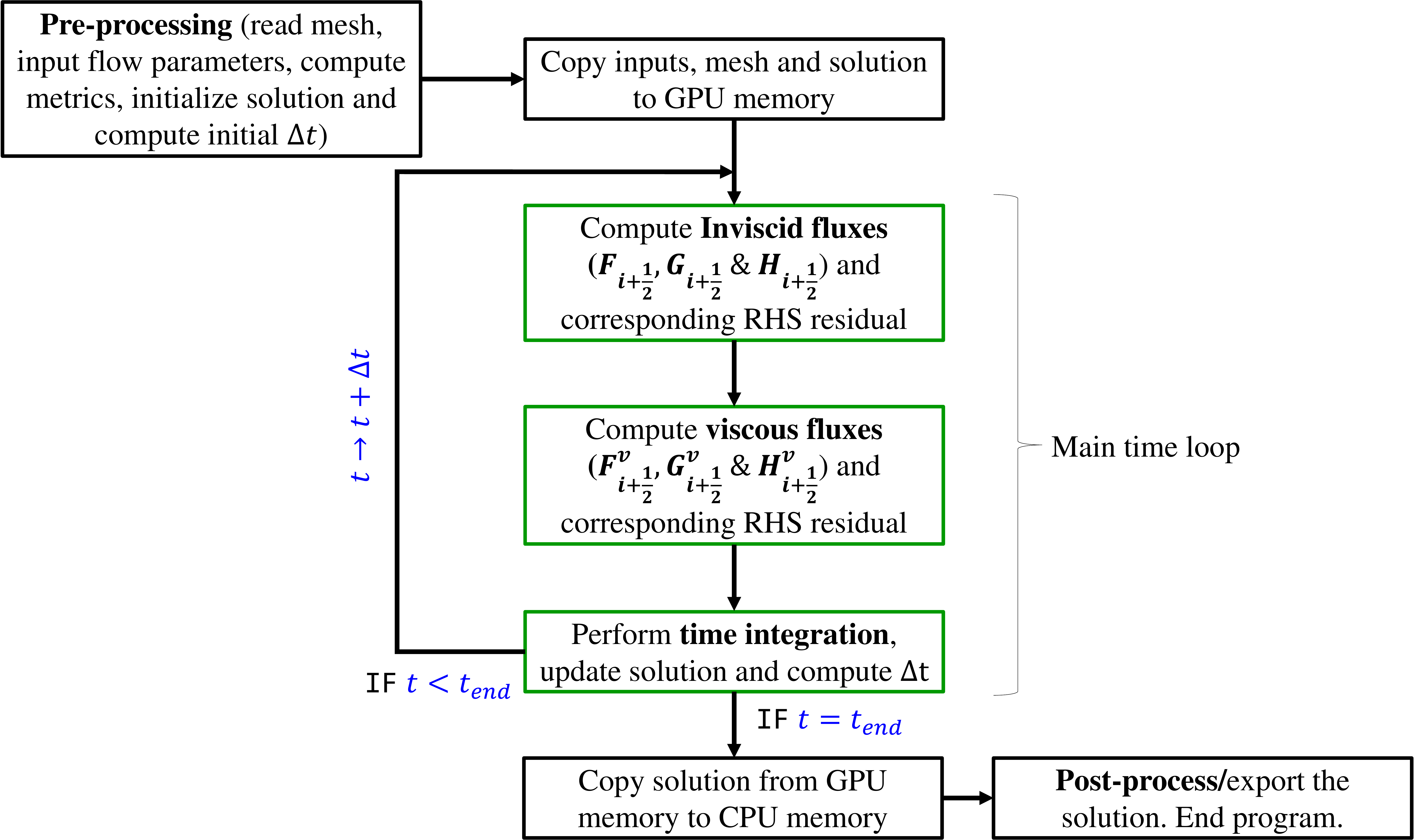}
    \caption{Flow chart of solver algorithm. The stages outlined in the green boxes are executed fully on the GPU.}
    \label{algo}
\end{figure}

Fig. \ref{algo} shows a flow chart view of key stages involved in the algorithm. The program starts with pre-processing steps and ends with post-processing. Before entering into the main time-loop, the necessary information required to execute the main time-loop is transferred to GPU memory, which will remain and update there without any communication with the CPU until the simulation ends at time $t_{end}$. This allows the computationally expensive parts of the program such as reconstruction, Riemann solvers, viscous fluxes, residuals, and time integration to be executed solely on GPUs. More details about the parallelization model employed and the achieved performance increase will be discussed in Section \ref{sec:gpu-accel}.\\

One of the key features of the present Navier-Stokes algorithm is gradient sharing. The gradients of primitive variables are utilized in multiple stages of the algorithm, enhancing efficiency and solution accuracy. This includes the use of gradients in inviscid fluxes (via gradient-based reconstruction - Sec. \ref{Inv-disc}) and viscous flux discretization (via $\alpha$-damping approach - Sec. \ref{sec:viscDisc}), as well as the improvement of shock-capturing through the improved MP-limiter. Additionally, the gradients of primitive variables are used in post-processing calculations such as enstrophy, vorticity, density gradient magnitudes, Q-criterion, and others. As a result, the approach is efficient and produces relatively accurate and well resolved flow solutions.

\section{Metric terms} \label{conser-metrics}
\subsection{Metric terms at cell centers and half locations}

Since the current framework stores the solution at cell centers and fluxes at half locations, the solver requires metric terms at both these locations. The metrics belonging to cell centers are used in the following stages of the algorithm:
\begin{enumerate}
    \item To impose boundary conditions (Section \ref{sec:mul-block})
    \item While computing the time-step based on the inviscid and viscous CFL condition (Section \ref{sec:time-int})
    \item For computing cell-interface metrics (Section \ref{sec:FP})
    \item While computing post-processing quantities such as enstrophy, vorticity, etc.
\end{enumerate}

\noindent On the other hand, the metrics corresponding to half locations ($i+\frac{1}{2},j+\frac{1}{2},k+\frac{1}{2}$) are used during the following stages of the algorithm:
\begin{enumerate}
    \item While transforming primitive variables in to characteristic space (to be discussed in Section \ref{Inv-disc})
    \item The approximate Riemann solver (Section \ref{sec:riemann-solver})
    \item While computing viscous fluxes (Section \ref{sec:viscDisc})
\end{enumerate}

The half location metrics required in the above mentioned steps are interpolated from cell-centered metrics employing the same scheme as that of the reconstruction scheme used for inviscid flux computation (Section \ref{Inv-disc}) in order to satisfy the geometric conservation law. A detailed description of the interpolation formula and it's effect on the flow simulations will be discussed in Section \ref{sec:FP}. Firstly, the formulations employed to compute cell-centered metrics will be presented.

\subsection{Conservative metrics for cell centered grid}
The following conservative form of grid metrics given by Thomas and Lombard \cite{thomas1979geometric} were used in this work to satisfy the Geometric Conservation Law (GCL).

\begin{equation} \label{GCL_metrics}
\begin{aligned}
&\hat{\xi}_{x}=\left(y_{\eta} z\right)_{\zeta}-\left(y_{\zeta} z\right)_{\eta}, \quad
 \hat{\xi}_{y}=\left(z_{\eta} x\right)_{\zeta}-\left(z_{\zeta} x\right)_{\eta}, \quad
 \hat{\xi}_{z}=\left(x_{\eta} y\right)_{\zeta}-\left(x_{\zeta} y\right)_{\eta}, \\
&\hat{\eta}_{x}=\left(y_{\zeta} z\right)_{\xi}-\left(y_{\xi} z\right)_{\zeta}, \quad 
 \hat{\eta}_{y}=\left(z_{\zeta} x\right)_{\xi}-\left(z_{\xi} x\right)_{\zeta}, \quad
 \hat{\eta}_{z}=\left(x_{\zeta} y\right)_{\xi}-\left(x_{\xi} y\right)_{\zeta}, \\
&\hat{\zeta}_{x}=\left(y_{\xi} z\right)_{\eta}-\left(y_{\eta} z\right)_{\xi}, \quad
 \hat{\zeta}_{y}=\left(z_{\xi} x\right)_{\eta}-\left(z_{\eta} x\right)_{\xi}, \quad
 \hat{\zeta}_{z}=\left(x_{\xi} y\right)_{\eta}-\left(x_{\eta} y\right)_{\xi}.
\end{aligned}
\end{equation}

By satisfying GCL, metric cancellation errors corresponding to the grid transformation are reduced to machine zero, thus achieving freestream and vortex preservation properties. Ensuring freestream preservation is particularly important while studying flows involving features such as transition, aero-acoustic feedback mechanisms, etc., which are significantly affected by the non-preserved freestream errors.

\noindent While the metric terms in Eqn. (\ref{GCL_metrics}) are only computed once per simulation for stationary grid cases, they are re-computed after each time-step for dynamic meshes. The chain rule is used to evaluate the temporal metric terms as follows,
\begin{equation}
\xi_{t}=-\left  (x_{t}\xi_{x}  +y_{t}\xi_{y}  +z_{t}\xi_{z}  \right), \quad \eta_{t}=-\left (x_{t}\eta_{x} +y_{t}\eta_{y} +z_{t}\eta_{z} \right), \quad \zeta_{t}=-\left(x_{t}\zeta_{x}+y_{t}\zeta_{y}+z_{t}\zeta_{z}\right).
\end{equation}

The grid velocities $x_t$, $y_t$, and $z_t$ could be either predetermined analytically or can also be estimated numerically on the fly as the grid deforms and changes with time. In the current work the test case considered uses predetermined analytical values for $x_t$, $y_t$, and $z_t$. For more details on the use of numerically estimated grid velocities and the associated errors, the reader is referred to  Ref. \cite{Visbal2002}.\\

\noindent The time derivative term in Eqn. (\ref{trans-eqn}), is split as follows,
\begin{equation} \label{time-split}
    \frac{\partial }{\partial t}\left(\frac{\textbf{Q}}{J}\right) = \frac{1}{J} \frac{\partial \textbf{Q}}{\partial t} + \left(\frac{1}{J}\right)_{t} \textbf{Q}
\end{equation}

The first term $1/J$ on the right hand side is the instantaneous Jacobian computed based on instantaneous grid cell positions as follows. 

\begin{equation} \label{Jac-matrix}
    \frac{1}{J}=\left|\begin{array}{lll}
    x_{\xi} & y_{\xi} & z_{\xi} \\
    x_{\eta} & y_{\eta} & z_{\eta} \\
    x_{\zeta} & y_{\zeta} & z_{\zeta}
    \end{array}\right|
\end{equation}

The second term with $\left(\frac{1}{J}\right)_{t}$ is only relevant to the dynamically changing meshes and is computed in accordance with GCL as follows,

\begin{equation} \label{Jacb2}
    \left(\frac{1}{J}\right)_{t}=-\left[(\xi_{t})_{\xi}+(\eta_{t})_{\eta}+(\zeta_{t})_{\zeta}\right]
\end{equation}


\section{Discretization}    \label{sec:disc}

In this section the discretization approaches used to compute inviscid and viscous fluxes are detailed. The MEG6/MIG4 algorithm of Chamarthi \cite{chamx}, the HLLC Riemann solver, and the $\alpha$-damping approach are explained in the context of generalized curvilinear coordinates with applicability to both stationary and moving grids.

\subsection{Inviscid flux discretization - MEG6 and MIG4 schemes}    \label{Inv-disc}
Monotonocity preserving Explicit sixth order Gradient scheme (MEG6) and the Monotonocity Preserving Implicit 4th order Gradient scheme (MIG4) proposed by Chamarthi \cite{chamx} are a class of gradient-based algorithms to discretize inviscid flux terms in the compressible Navier-Stokes equations. They are primarily driven by the gradients of primitive variables ($\mathbf{P}_{\xi},\mathbf{P}_{\eta}$, and $\mathbf{P}_{\zeta}$) and the Monotonicity Preserving limiter of Suresh and Huynh \cite{suresh1997accurate} for shock capturing. For interpolating the variables to the half locations, the following general form containing the functional values and their corresponding first and second derivatives at cell centers is employed \cite{van1977towards}:

\begin{equation}
   \phi(\xi)=\phi_i+\left( \frac{\partial \phi}{\partial \xi} \right)_i \left(\xi-\xi_i\right)+3 \kappa \left( \frac{\partial^2 \phi}{\partial \xi^2} \right)_i\left[\left(\xi-\xi_i\right)^2-\frac{\Delta \xi_i^2}{12}\right] \quad \text{for} \quad \xi_{i-\frac{1}{2}} \le \xi \le \xi_{i+\frac{1}{2}}
\end{equation}

where $\phi$ is the variable of interest to be interpolated and $\kappa=\frac{1}{3}$ \cite{van1977towards}. Since the above formulation contains functional values and its derivatives that are belonging to location $i$, by substituting $\xi = \pm \frac{\Delta \xi}{2}$ the left ($L$) and right ($R$) biased interpolation formulae for $\phi$ at locations $i+\frac{1}{2}$ and $i-\frac{1}{2}$ are obtained. Simplifying the resulting equations with $\Delta \xi =1$ (since unit grid spacing convention is followed for transformed coordinates) yields the following left and right biased states of $\phi$ at $i+\frac{1}{2}$ location.

\begin{equation} \label{vanleer-poly}
    \begin{aligned}
&\phi_{i+\frac{1}{2}}^L= \phi_i+\frac{1}{2} \left( \frac{\partial \phi}{\partial \xi} \right)_i+\frac{1}{12} \left( \frac{\partial^2 \phi}{\partial \xi^2} \right)_i \\
&\phi_{i+\frac{1}{2}}^R= \phi_{i+1}-\frac{1}{2} \left( \frac{\partial \phi}{\partial \xi} \right)_{i+1}+\frac{1}{12} \left( \frac{\partial^2 \phi}{\partial \xi^2} \right)_{i+1}
\end{aligned}
\end{equation}

The above two polynomials are used to reconstruct the variables in MEG6 and MIG4 schemes. The two methods MEG6 and MIG4 differ from each other only in the scheme that is used to compute the derivatives of primitive variables present in the above equations (Eqn. \ref{vanleer-poly}); the rest of the algorithm (which will be detailed soon) remains the same. Despite the different gradients employed, both MEG6 and MIG4 schemes yield fourth-order accurate estimates of inviscid flux derivatives $\frac{\partial \mathbf{\hat{F}}}{\partial \xi}, \frac{\partial \mathbf{\hat{G}}}{\partial \eta}$, and $\frac{\partial \mathbf{\hat{H}}}{\partial \zeta}$. The theoretical details regarding order of accuracy can be found in Ref. \cite{chamx}. The schemes however differ in their dispersion and dissipation properties. Readers are referred to Fig. 3 of Ref. \cite{chamx} which shows the spectral properties of MEG6 and MIG4 schemes in comparison with other well known discretization approaches.

\begin{figure}[h!]
    \centering
    \includegraphics[width=140mm]{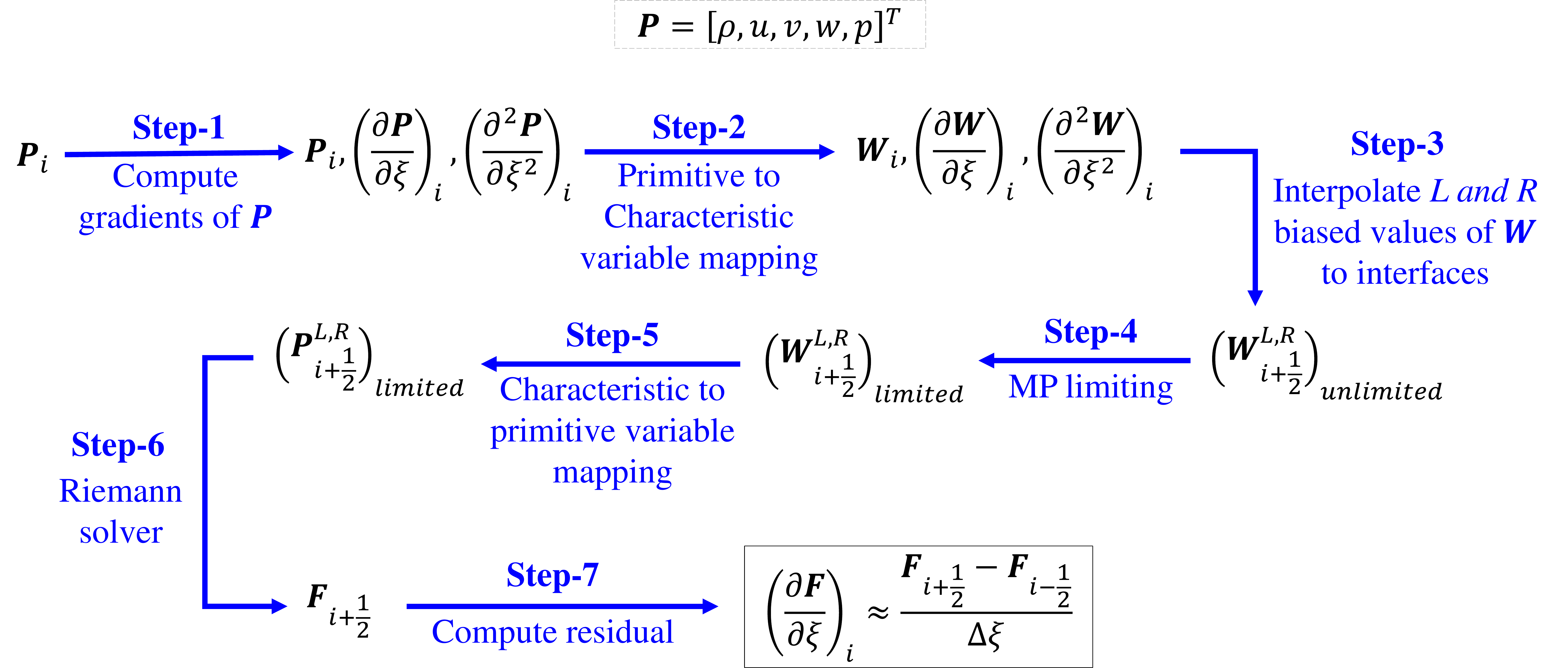}
    \caption{Various stages while estimating inviscid flux residual using the MEG/MIG schemes.}
    \label{inv_algo}
\end{figure}

Fig. \ref{inv_algo} depicts a flow chart view of different steps in computing the inviscid flux residual at half locations. A detailed step-by-step procedure is shown. For simplicity, the procedure is described for $\mathbf{F}_{i+\frac{1}{2}}$ fluxes only. Extending the same to other directions is straightforward. Also, for the sake of simplicity the `$j$' and `$k$' subscripts corresponding to $\eta$ and $\zeta$ directions are omitted in all the expressions to be presented now; the formulations remain the same and are independent of `$j$' and `$k$' indices. \\


\noindent \textbf{Step-1, Compute gradients:} Compute first and second gradients of primitive variables $\mathbf{P}=[\rho,u,v,w,p]^T$ at cell centers in the computational coordinate system using the following expressions:

\begin{equation}\label{E6_grads}
    \text{For MEG6:}\quad \left(\frac{\partial u}{\partial \xi}\right)_i = \frac{3}{4}\left( u_{i+1} + u_{i-1}\right) - \frac{3}{20}\left(u_{i+2} + u_{i-2}  \right) + \frac{1}{60}\left(u_{i+3} + u_{i-3}  \right)
\end{equation}

\begin{equation}\label{IG4_grads}
    \text{For MIG4:}\quad \alpha \left(\frac{\partial u}{\partial \xi}\right)_{i-1} + \left(\frac{\partial u}{\partial \xi}\right)_{i} + \alpha \left(\frac{\partial u}{\partial \xi}\right)_{i+1} = \frac{2 (2 + \alpha)}{6}\left( u_{i+1} + u_{i-1}\right) + \frac{-1+4 \alpha}{12}\left(u_{i+2} + u_{i-2} \right)
\end{equation}

\begin{equation}
    \text{For both MEG6 and MIG4:}\quad \left(\frac{\partial^{2} u}{\partial \xi^{2}}\right)_{i}=2\left(\hat{u}_{i+1}-2 \hat{u}_{i}+\hat{u}_{i-1}\right)-0.5\left(u_{i+1}^{\prime}-u_{i-1}^{\prime}\right)
\end{equation}

$\alpha=\frac{5}{14}$ is considered for the implicit scheme in Eqn.\ref{IG4_grads}. The above mentioned explicit sixth-order (E6) gradients are used for the MEG6 scheme, and implicit gradients of fourth-order (IG4) are used for the MIG4 scheme.\\

\noindent \textbf{Step-2, Perform characteristic transformation:} The cell center primitive variables $\mathbf{P}=[\rho,u,v,w,p]^T$, their first derivatives, $\frac{\partial \mathbf{P}}{\partial \xi}$ and second derivatives, $\frac{\partial^2 \mathbf{P}}{\partial \xi^2}$ are transformed into characteristic space $\mathbf{W}$, $\frac{\partial \mathbf{W}}{\partial \xi}$ and, $\frac{\partial^2 \mathbf{W}}{\partial \xi^2}$ by multiplying them with the left-eigenvectors of the flux Jacobian matrix, $\frac{\partial \mathbf{F}}{\partial \mathbf{P}}$, using the following relations Eqn. \ref{forward-projection}. Reconstructing the resulting characteristic variables will result in cleaner results without any oscillations near the discontinuities. This is since the Euler equations resemble linear wave equation in characteristic form for which the up-winding in the Riemann solver is designed for (will be discussed in Sec. \ref{sec:riemann-solver}).

\begin{subequations} \label{forward-projection}
    \begin{gather}
    \mathbf{W} = \overline{\mathbf{R}}_{\xi}^{-1} \mathbf{P}, \quad
    \frac{\partial \mathbf{W}}{\partial \xi}     = \overline{\mathbf{R}}_{\xi}^{-1}  \frac{\partial \mathbf{P}}{\partial \xi}, \quad \text{and }
    \frac{\partial^2 \mathbf{W}}{\partial \xi^2} = \overline{\mathbf{R}}_{\xi}^{-1} \frac{\partial^2 \mathbf{P}}{\partial \xi^2}
    \tag{\theequation a-\theequation c}
    \end{gather}
\end{subequations}

\noindent The eigenvector matrices employed in these equations are provided in \ref{app:Eig-structure}.\\

\begin{figure}[h!]
    \centering
    \includegraphics[width=50mm]{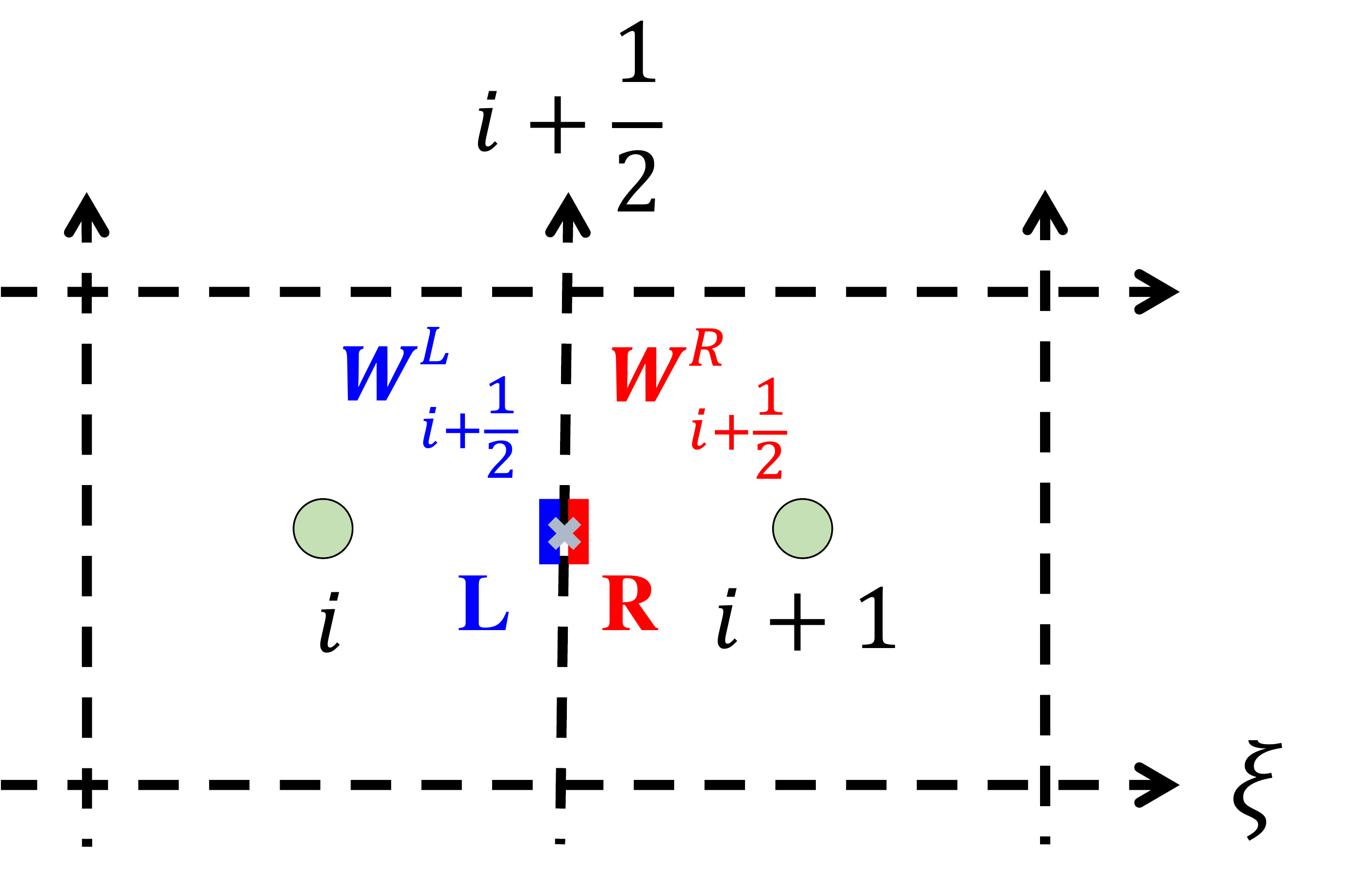}
    \caption{An illustration depicting left and right biased states of $W$ on the grid stencil.}
    \label{recon}
\end{figure}

\noindent \textbf{Step-3, Compute left and right biased states of $\mathbf{W}$:} Evaluate the unlimited left and right biased states of characteristic variables at half locations employing Eqns. \ref{vanleer-poly}. Fig. \ref{recon} illustrates the location of these quantities on the interpolation stencil.

\begin{eqnarray} \label{legendre_polys}
    \left(\mathbf{W}_{i+1/2}^L\right)_{unlimited} = \mathbf{W}_{i} + \frac{1}{2} \left(\frac{\partial \mathbf{W}}{\partial \xi}\right)_{i} + \frac{1}{12} \left(\frac{\partial^2 \mathbf{W}}{\partial \xi^2}\right)_{i} \\
    \left(\mathbf{W}_{i+1/2}^R\right)_{unlimited} = \mathbf{W}_{i+1} - \frac{1}{2} \left(\frac{\partial \mathbf{W}}{\partial \xi}\right)_{i+1} + \frac{1}{12} \left(\frac{\partial^2 \mathbf{W}}{\partial \xi^2}\right)_{i+1} 
\end{eqnarray}

\noindent \textbf{Step-4, Modify $\mathbf{W}_{i+1/2}^{L,R}$ through limiting:} An improved version of the Monotonicity Preserving fifth order limiting algorithm (MP5) \cite{suresh1997accurate} (see \ref{app:MP-limit}) proposed in Ref. \cite{chamx} is employed to check for discontinuities and limit the left and right reconstructed states of the characteristic variables ($\mathbf{W}_{i+1/2}^{L,R}$). This step essentially captures flow discontinuities by limiting the interpolated unlimited value from the previous step and avoids unphysical oscillations. One of the key favourable attributes of the MP5 limiter is that it allows an effective capturing of shock and acoustic waves simultaneously compared to the WENO based strategies \cite{Shu1997}. The limiting procedure is detailed in \ref{app:MP-limit}.


\noindent \textbf{Step-5, Map characteristic variables back to primitive variables:} The reconstructed left and right characteristic variable states computed in the previous step are now transformed back to primitive variables by multiplying them with the right-eigenvector matrices provided in \ref{app:Eig-structure}. 

\begin{subequations} \label{rev-projection}
    \begin{gather}
    (\mathbf{P}^L)_{i+1/2} = \overline{\mathbf{R}}_{\xi} (\mathbf{W}^L_{i+1/2})_{limited} \\
    (\mathbf{P}^R)_{i+1/2} = \overline{\mathbf{R}}_{\xi} (\mathbf{W}^R_{i+1/2})_{limited}
    \end{gather}
\end{subequations}

\noindent \textbf{Step-6, Compute interface fluxes:} Now that the flow states on both sides of the interface are known, the mathematical setting at each $i+\frac{1}{2}$ location reduces to a Riemann problem. In this step, the local one-dimensional Riemann problem is solved, and the resultant flux $\textbf{F}_{i+\frac{1}{2}}$ is estimated at each interface. The HLLC approximate riemann solver (Harten–Lax–van Leer with Contact restoration) \cite{toro2009riemann} is employed to estimate the fluxes in the present work. As shown in Eqn. \ref{hllc-function}, the only inputs required in order to compute the fluxes using the HLLC approach are the local left and right reconstructed states of the primitive variables and the interpolated metric terms at the $i+\frac{1}{2}$ locations. The required relations to compute the fluxes via the HLLC Riemann solver in the context of curvilinear coordinates is provided in a dedicated section on Riemann solvers in Section \ref{sec:riemann-solver}.

\begin{equation} \label{hllc-function}
    \hat{\mathbf{F}}_{i+\frac{1}{2}} = \mathbf{f}_{HLLC} \left\{ \mathbf{P}^L_{i+\frac{1}{2}}, \mathbf{P}^R_{i+\frac{1}{2}}, (\hat{\xi}_t)_{i+\frac{1}{2}}, (\hat{\xi}_x)_{i+\frac{1}{2}}, (\hat{\xi}_y)_{i+\frac{1}{2}}, (\hat{\xi}_z)_{i+\frac{1}{2}} \right\}
\end{equation}




\noindent \textbf{Step-7, Compute residual:} Using the interface fluxes estimated through the Riemann solver, the net flux entering into each cell is computed via Eqn. \ref{res-inv}. It should be noted that if the flow problem is linear in nature, the accuracy of this term theoretically yields fourth-order using both E6 and IG4 gradients in step-1 (Eqns. \ref{E6_grads} and \ref{IG4_grads}). More theoretical details including demonstrative tests regarding the order of accuracy can be found in \cite{chamx}. \\

\begin{equation} \label{res-inv}
    \left(\frac{\partial \boldsymbol{F}}{\partial \xi}\right)_i = \frac{ \boldsymbol{F}_{i+\frac{1}{2}}- \boldsymbol{F}_{i-\frac{1}{2}}}{\Delta \xi}
\end{equation}



\subsubsection{The HLLC Riemann solver} \label{sec:riemann-solver}
In this section, the HLLC Riemann solver is presented within the framework of curvilinear coordinates. The HLLC Riemann solver is a three-wave approximate solution that accounts for the presence of contact discontinuities. One of the notable advantages of the HLLC fluxes is their ability to minimize dissipation in comparison to other commonly used methods, such as HLL \cite{harten1983upstream} and Rusanov \cite{toro2009riemann}. As a result, HLLC is able to effectively resolve shocks, contact discontinuities, and shear layers. The inviscid flux $\mathbf{F}_{i+\frac{1}{2}}$ in the transformed coordinates through the HLLC approximate Riemann solver is obtained as follows:


\begin{equation}
\hat{\mathbf{F}}_{i+\frac{1}{2}}^{\mathrm{HLLC}}=\left\{\begin{array}{ll}
\mathbf{\hat{F}}^{\mathbf{L}} & \text { if } \quad \hat{S}^{L} \geq 0, \\
\mathbf{\hat{F}}^{* \mathbf{L}}=\mathbf{\hat{F}}^{\mathbf{L}}+\hat{S}^{L}\left(\mathbf{Q}^{* \mathbf{L}}-\mathbf{Q}^{\mathbf{L}}\right) & \text { if } \quad \hat{S}^{L} \leq 0 \leq \hat{S}^{*}, \\
\mathbf{\hat{F}}^{* \mathbf{R}}=\mathbf{\hat{F}}^{\mathbf{R}}+\hat{S}^{R}\left(\mathbf{Q}^{* \mathbf{R}}-\mathbf{Q}^{\mathbf{R}}\right) & \text { if } \quad \hat{S}^{*} \leq 0 \leq \hat{S}^{R}, \\
\mathbf{\hat{F}}^{\mathbf{R}} & \text { if } \quad \hat{S}^{R} \leq 0
\end{array}\right.
\end{equation}

The three wave speeds of the local one-dimensional problem are calculated as follows.
\begin{equation*}
    \hat{S}^{L}=\operatorname{Min}\left(\hat{U}^{L}-c^{L} \sqrt{\hat{\xi}_{x}^{2}+\hat{\xi}_{y}^{2}+\hat{\xi}_{z}^{2}}, \quad \hat{U}^{R}-c^{R} \sqrt{\hat{\xi}_{x}^{2}+\hat{\xi}_{y}^{2}+\hat{\xi}_{z}^{2}}\right) 
\end{equation*}

\begin{equation*}
    \hat{S}^{R}=\operatorname{Max}\left(\hat{U}^{L}+c^{L} \sqrt{\hat{\xi}_{x}^{2}+\hat{\xi}_{y}^{2}+\hat{\xi}_{z}^{2}}, \quad \hat{U}^{R}+c^{R} \sqrt{\hat{\xi}_{x}^{2}+\hat{\xi}_{y}^{2}+\hat{\xi}_{z}^{2}}\right)
\end{equation*}

\begin{equation*}
    \hat{S}^{*}=\frac{\rho^{R} \hat{U}^{R}\left(\hat{S}^{R}-\hat{U}^{R}\right)-\rho^{L} \hat{U}^{L}\left(\hat{S}^{L}-\hat{U}^{L}\right)+\left(P_{L}-P_{R}\right)\left(\hat{\xi}_{x}^{2}+\hat{\xi}_{y}^{2}+\hat{\xi}_{z}^{2}\right)}{\rho^{R}\left(\hat{S}^{R}-\hat{U}^{R}\right)-\rho^{L}\left(\hat{S}^{L}-\hat{U}^{L}\right)}
\end{equation*} 

The conservative variables in the star region $\mathbf{Q}^{* \mathbf{K}}$ for $\mathbf{K} = L,R$ are estimated as follows.

\begin{equation}
\mathbf{Q}^{* \mathbf{K}}=\rho^{\mathbf{K}}\left(\frac{\hat{S}^{\mathbf{K}}-\hat{U}^{\mathbf{K}}}{\hat{S}^{\mathbf{K}}-\hat{S}^{*}}\right)\begin{bmatrix}
    1 \\
\frac{\hat{\xi}_{x}(\hat{S}^{*}-\hat{\xi}_{t})+(\hat{\xi}_{y}^{2}+\hat{\xi}_{z}^{2}) u^{\mathbf{K}}-\hat{\xi}_{y} \hat{\xi}_{x} v^{\mathbf{K}}-\hat{\xi}_{z} \hat{\xi}_{x}w^{\mathbf{K}}}{\hat{\xi}_{x}^{2}+\hat{\xi}_{y}^{2}+\hat{\xi}_{z}^{2}} \\
\frac{\hat{\xi}_{y}(\hat{S}^{*}-\hat{\xi}_{t})-\hat{\xi}_{x} \hat{\xi}_{y} u^{\mathbf{K}}+(\hat{\xi}_{x}^{2}+\hat{\xi}_{z}^{2}) v^{\mathbf{K}}-\hat{\xi}_{z} \hat{\xi}_{y}w^{\mathbf{K}}}{\hat{\xi}_{x}^{2}+\hat{\xi}_{y}^{2}+\hat{\xi}_{z}^{2}} \\
\frac{\hat{\xi}_{z}(\hat{S}^{*}-\hat{\xi}_{t})-\hat{\xi}_{x} \hat{\xi}_{z} u^{\mathbf{K}}-\hat{\xi}_{y} \hat{\xi}_{z} v^{\mathbf{K}}+(\hat{\xi}_{x}^{2}+\hat{\xi}_{y}^{2})w^{\mathbf{K}}}{\hat{\xi}_{x}^{2}+\hat{\xi}_{y}^{2}+\hat{\xi}_{z}^{2}} \\
\frac{E^{\mathbf{K}}}{\rho^{\mathbf{K}}}+\left(\hat{S}^{*}-\hat{U}^{\mathbf{K}}\right)\left\{\frac{\hat{S}^{*}-\hat{\xi}_{t}}{\left(\hat{\xi}_{x}^{2}+\hat{\xi}_{y}^{2}+\hat{\xi}_{z}^{2}\right)}+\frac{p^{\mathbf{K}}}{\rho^{\mathbf{K}}(\hat{S}^{\mathbf{K}}-\hat{U}^{\mathbf{K}})}\right\}
  \end{bmatrix}
\end{equation}

The $i+\frac{1}{2}$ metrics involved in the above formulation are interpolated consistently with the inviscid flux discretization to prevent metric cancellation errors, which can accumulate in the flow solution. The corresponding interpolation formulae and demonstrative examples concerning metric cancellation errors and their effect on the flow solution will be discussed in Section \ref{sec:FP}. The above formulations only show the flux calculation in the $\xi$ direction. Extending these formulae to other directions is straightforward. For instance, 

\begin{equation}
    \hat{\mathbf{G}}_{i,j+\frac{1}{2},k} = \mathbf{f}_{HLLC} \left\{ \mathbf{P}^L_{i,j+\frac{1}{2},k}, \mathbf{P}^R_{i,j+\frac{1}{2},k}, (\hat{\eta}_t)_{i,j+\frac{1}{2},k}, (\hat{\eta}_x)_{i,j+\frac{1}{2},k}, (\hat{\eta}_y)_{i,j+\frac{1}{2},k}, (\hat{\eta}_z)_{i,j+\frac{1}{2},k} \right\}.
\end{equation}

\subsection{Viscous flux discretization - $\alpha$-damping scheme}  \label{sec:viscDisc}

\begin{figure}[h!]
    \centering
    \includegraphics[width=140mm]{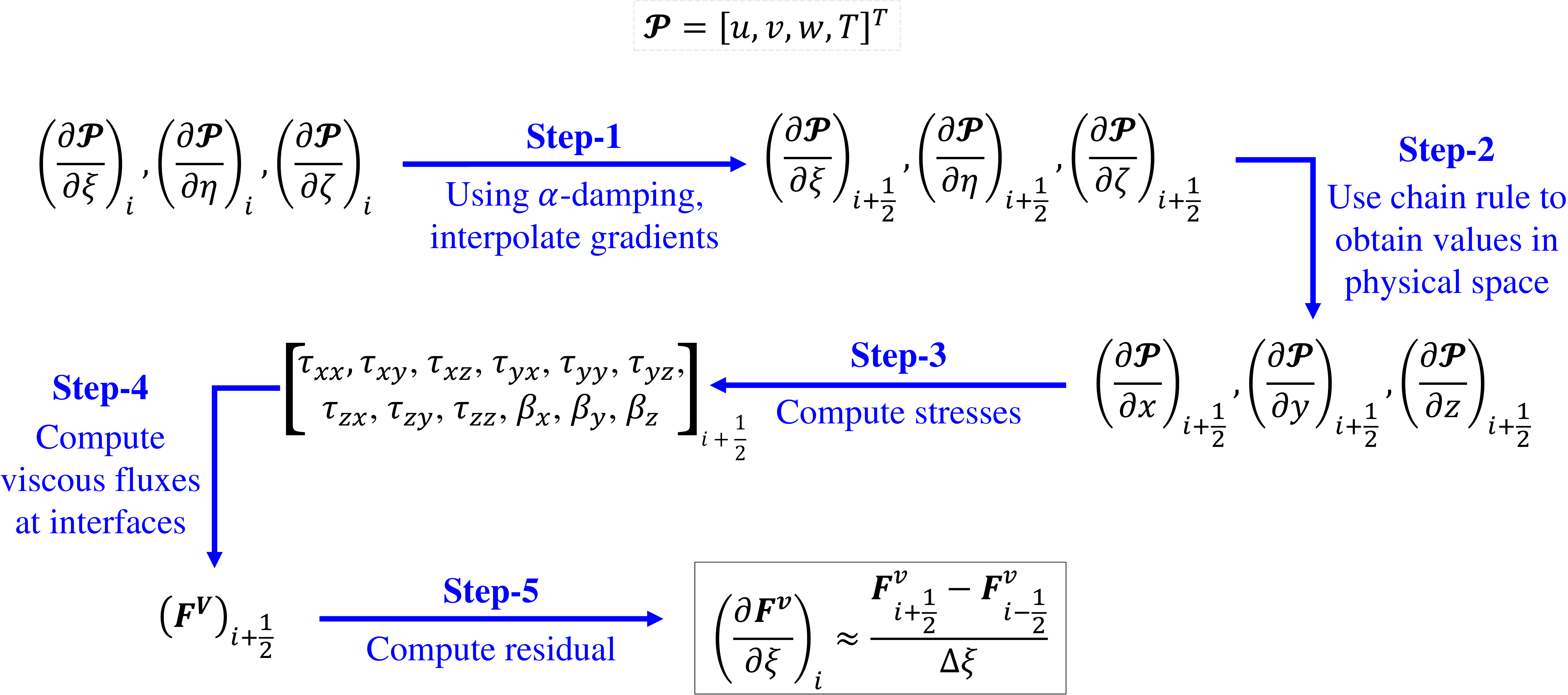}
    \caption{Various stages involved in estimating the viscous flux residual using the $\alpha$-damping scheme. Applies for both E6 and IG4 gradients of $\bm{\mathit{\mathscr{P}}}$, yielding 4th order accuracy for both \cite{chamx}.}
    \label{visc_algo}
\end{figure}

Many viscous schemes such as those proposed in \cite{Visbal2002,shen2010large} are prone to odd-even decoupling and inaccurate spectral representation particularly in the high wave-number region. In the recent past, efforts have been made in the development of numerical algorithms to compute diffusion fluxes without odd-even decoupling and good spectral properties over a broader wavenumber range \cite{Nishikawa2013, chamarthi2022importance,sainadh2022spectral}. The current work employs the fourth order accurate version of viscous fluxes proposed by Chamarthi et al. \cite{chamarthi2022importance}. The gradients required for the viscous fluxes are not re-computed here since they are already computed while estimating inviscid flux residuals (Eqns-\ref{E6_grads} and \ref{IG4_grads}). A step-by-step procedure to evaluate the interface viscous fluxes according to this algorithm in curvilinear coordinates is depicted in Fig. \ref{visc_algo}. For simplicity, the steps are described only for the $\xi$ direction but the same can be extrapolated to other directions, as well. \\

\noindent \textbf{Step-1, Compute interface gradients:} Interface gradients of $\bm{\mathit{\mathscr{P}}} = [u,v,w,T]^{T}$ in $\xi$ $\eta$ and $\zeta$ directions are first computed using Eqn. \ref{AD-eqn1}, where $\bm{\mathit{\mathscr{P}}}$ is a vector of variables required to evaluate the viscous fluxes. For simplicity, the equations are presented only for the variable $u$ here. The same should be repeated for other variables of $\bm{\mathit{\mathscr{P}}}$ as well.

\begin{equation} \label{AD-eqn1}
    \left(\frac{\partial u}{\partial \xi}\right)_{i+\frac{1}{2}}=\underbrace{\frac{1}{2}\left[\left(\frac{\partial u}{\partial \xi}\right)_{i}+\left(\frac{\partial u}{\partial \xi}\right)_{i+1}\right]}_{\text {Consistent term }}+\underbrace{\frac{\alpha^D}{2}\left(u_{R}-u_{L}\right)}_{\text {Damping term }},
\end{equation}

Where the left and right states of $u$ are defined using the following interpolation polynomial:
\begin{eqnarray} \label{AD-eqn2}
u_{L}=u_{i}+ 0.5 \left(\frac{\partial u}{\partial \xi}\right)_{i}+\beta^D\left(u_{i+1}-2 u_{i}+u_{i-1}\right) \\
\quad u_{R}=u_{i+1}- 0.5 \left(\frac{\partial u}{\partial \xi}\right)_{i+1}+\beta^D\left(u_{i+2}-2 u_{i+1}+u_{i}\right) .
\end{eqnarray}

$\alpha^D=4$ and $\beta^D=0$ are the scheme coefficients based on Ref. \cite{chamx} for both E6 and IG4 based gradients of $\bm{\mathit{\mathscr{P}}}$. These coefficients produce fourth order accuracy for both E6 and IG4 gradients. The temperature at the $i+\frac{1}{2}$ location is computed from an arithmetic average of $T_L$ and $T_R$ values. From this, the interface dynamic viscosity $\mu_{i+\frac{1}{2}}$ is estimated using the Sutherland's law of viscosity. \\

\noindent \textbf{Step-2, Compute the derivatives in Cartesian coordinates:} Using the $\xi$, $\eta$ and $\zeta$ direction gradients of $[u,v,w,T]$ computed in the previous step, the derivative chain rule is used to evaluate the $x$, $y$, and $z$ derivatives of the same variables. For instance, this is done as follows to compute the interface $x$ direction derivative:

\begin{equation} \label{AD-eqn3}
   \left(\frac{\partial u}{\partial x}\right)_{i+1 / 2} = (\xi_{x})_{i+\frac{1}{2}} \left(\frac{\partial u}{\partial \xi}\right)_{i+\frac{1}{2}} + (\eta_{x})_{i+\frac{1}{2}} \left(\frac{\partial u}{\partial \eta}\right)_{i+\frac{1}{2}} + (\zeta_{x})_{i+\frac{1}{2}} \left(\frac{\partial u}{\partial \zeta}\right)_{i+\frac{1}{2}}
\end{equation}

\noindent Note: Metric values at $i+\frac{1}{2}$ location are interpolated from the cell center metrics using the relations provided in Eqns. \ref{FP-interp} and \ref{FP-interp2}. \\

\noindent \textbf{Step-3, Compute stress terms:} The shear stress and energy diffusion terms $\tau_{ij}$, $\beta_i$ present in the viscous flux vector are computed at $i+\frac{1}{2}$ using Eqns. \ref{eqn:shear-str} and \ref{eqn:thermal-work} provided in Section \ref{sec:gov-eqns} and with the interface gradients computed in the previous step. The required interface velocities are computed via an arithmetic average of the left and right states of the velocities (e.g. $u_{i+\frac{1}{2}} = 0.5(u_L + u_R)$). The shear-stress term $\tau_{xx}$ at $i+\frac{1}{2}$ is evaluated as follows:

\begin{equation}
    (\tau_{xx})_{i+\frac{1}{2}} = 2 \mu_{i+\frac{1}{2}} \left(\frac{\partial u}{\partial x} \right)_{i+\frac{1}{2}} + \lambda_{i+\frac{1}{2}} \left[ \left(\frac{\partial u}{\partial x} \right)_{i+\frac{1}{2}} + 
    \left(\frac{\partial v}{\partial y} \right)_{i+\frac{1}{2}} +
    \left(\frac{\partial w}{\partial z} \right)_{i+\frac{1}{2}}\right]
\end{equation} \\

\noindent \textbf{Step-4, Estimate the interface fluxes:} The viscous fluxes at $i+\frac{1}{2}$ locations are computed using the relations provided in Eqn. \ref{visc-curvi}. For instance, the viscous flux vector in the $\xi$ direction is evaluated as follows:

\begin{equation}
    \hat{F}^{\text{V}}_{i+\frac{1}{2}}= \left[\begin{array}{c}
0 \\
\left(\hat{\xi}_{x}\right)_{i+\frac{1}{2}} (\tau_{x x})_{i+\frac{1}{2}}+\left(\hat{\xi}_{y}\right)_{i+\frac{1}{2}} (\tau_{x y})_{i+\frac{1}{2}}+\left(\hat{\xi}_{z}\right)_{i+\frac{1}{2}} (\tau_{x z})_{i+\frac{1}{2}} \\
\left(\hat{\xi}_{x}\right)_{i+\frac{1}{2}} (\tau_{y x})_{i+\frac{1}{2}}+\left(\hat{\xi}_{y}\right)_{i+\frac{1}{2}} (\tau_{y y})_{i+\frac{1}{2}}+\left(\hat{\xi}_{z}\right)_{i+\frac{1}{2}} (\tau_{y z})_{i+\frac{1}{2}} \\
\left(\hat{\xi}_{x}\right)_{i+\frac{1}{2}} (\tau_{z x})_{i+\frac{1}{2}}+\left(\hat{\xi}_{y}\right)_{i+\frac{1}{2}} (\tau_{z y})_{i+\frac{1}{2}}+\left(\hat{\xi}_{z}\right)_{i+\frac{1}{2}} (\tau_{z z})_{i+\frac{1}{2}} \\
\left(\hat{\xi}_{x}\right)_{i+\frac{1}{2}} (\beta_{x} )_{i+\frac{1}{2}}+\left(\hat{\xi}_{y}\right)_{i+\frac{1}{2}} (\beta_{y} )_{i+\frac{1}{2}}+\left(\hat{\xi}_{z}\right)_{i+\frac{1}{2}} (\beta_{z} )_{i+\frac{1}{2}}
\end{array}\right]
\end{equation}

\noindent \textbf{Step-5, Compute residual:} Finally, the viscous flux residual is computed using the following relation:

\begin{equation}
    \left(\frac{\partial \boldsymbol{F}^v}{\partial \xi}\right)_i \approx \frac{\boldsymbol{F}_{i+\frac{1}{2}}^v-\boldsymbol{F}_{i-\frac{1}{2}}^v}{\Delta \xi}
\end{equation}

After computing the inviscid and viscous flux residuals time marching is performed to compute the solution corresponding to next time-step. The details regarding the time marching scheme and the stable time-step evaluation for both inviscid and viscous flow simulations is detailed in \ref{sec:time-int}.
\section{Multi-block approach and boundary conditions} \label{sec:mul-block}
\begin{figure}[h!]
    \centering
    \includegraphics[width=150mm]{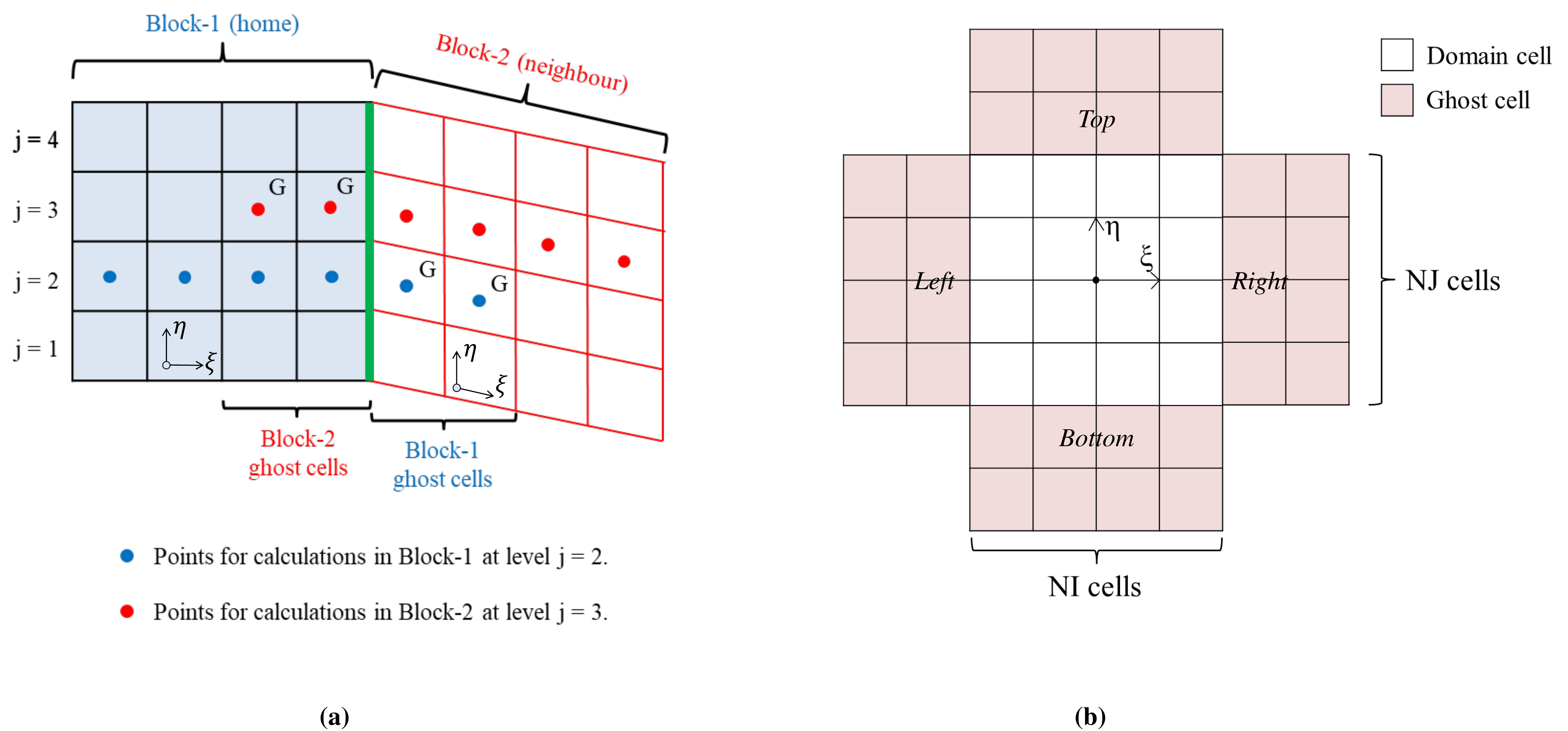}
    \caption{(a) Schematic of a block and its ghost points on left, right, bottom and top boundaries. (b) Schematic of ghost point sharing between blocks sharing interface along $\xi$ direction. The letter `G' in figure-b represents ghost cell region.}
    \label{multi-block}
\end{figure}

The multi-block approach enables geometric flexibility to simulate flows through and around domains of various atypical shapes. The multi-block approach adopted in the present work is similar to the approach proposed by Lien et al. \cite{lien1996multiblock}. Any block in the domain can interface one or more of it's boundaries (six in 3-D, four in 2-D) with the blocks surrounding it. Each block possesses its own coordinate system, essentially making the approach block unstructured. To ensure consistency of the block dimensions at the boundaries, the dimensions of the block interface plane should be identical on both home and neighbor blocks. A schematic of a multi-block layout depicting the arrangement of cells and the coordinate systems is illustrated in  Fig. \ref{multi-block}a. Ghost cells are adapted to introduce necessary inter-block information sharing and enable sufficient stencil size at block boundaries to compute high-order fluxes corresponding to MEG6, MIG4, and $\alpha$-damping.

Boundary conditions are imposed at each block boundary through ghost cells. The geometric location of the ghost points is arbitrary as they are physically non-existent. Therefore, metric terms are also non-existent and do not need to be computed in those locations. However, the lack of geometric information at boundaries poses difficulties in applying the usual high order approximations that are employed for cells inside the domain. \ref{app:one-sided} shows the special boundary formulae that were adopted in the present work.

A schematic of a two-dimensional block and the imaginary ghost cells surrounding it's four boundaries is illustrated in Fig. \ref{multi-block}. A total of five ghost cells were used for all the simulations in the current study. Suitable primitive variable values are specified in those ghost cells based on the type of boundary condition adapted at that boundary. A list of expressions to compute the primitive variables in ghost cells corresponding to various boundary conditions is provided in Table \ref{BCs}. All the formulations and concepts discussed in this section can be extrapolated to three dimensions.

\begin{table}[h!]
    \centering
    \includegraphics[width=\textwidth]{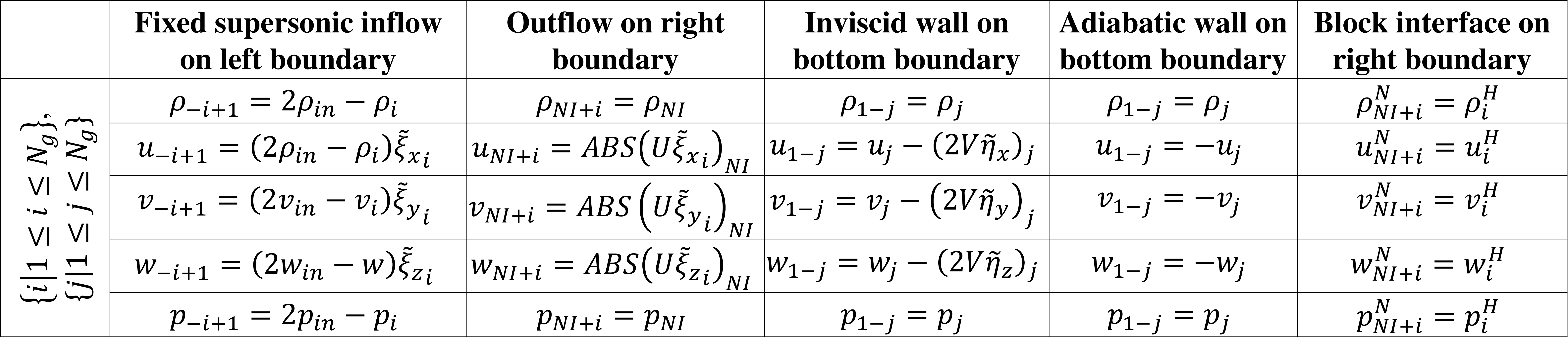}
    \caption{Expressions for primitive variable values in the ghost cells for various boundary conditions. Where, $N_g$ is number of ghost cells, prefix $(.)_{in}$ represents inflow conditions at the inflow boundary, $NI$ is number of points in $xi$ direction, $U$ is the contravariant velocity in $\xi$ direction, $V$ is the contravariant velocity in $\eta$ direction, Suffix `$H$' and `$N$' denote `Home' and `Neighbour' blocks respectively.}
    \label{BCs}
\end{table}


\section{Freestream preservation} \label{sec:FP}

In this section, firstly we layout the three essential steps to be followed to ensure freestream preservation for MEG6 and MIG4 schemes. Then we present a proof showing how the metric identities are satisfied with the use of MEG6 and MIG4 schemes to discretize and interpolate the conservative metric term formulae (Eqn. \ref{GCL_metrics}). The following are the three steps to be incorporated while computing metric terms in order to achieve freestream preservation:

\begin{enumerate}
    \item \textbf{Conservative metrics:} Conservative metrics presented in Eqn. \ref{GCL_metrics} should be used to ensure geometric and volume conservation \cite{thomas1979geometric}.
    
    \item \textbf{Consistent discretization using a central scheme}:  While computing the cell-centered metric terms, the same scheme as that of the inviscid flux discretization ($\frac{\partial F}{\partial \xi}$) has to be used to compute derivative terms involved in the conservative metric formulae. Furthermore, in order to estimate the metric terms in a linear fashion, the limiting steps (steps 4 and 5 in Fig. \ref{inv_algo}) are skipped and the left and right biased interpolated states should be averaged. 
    
    For instance, to compute the $(x_{\xi})_i$ term (which is required to compute metric terms), firstly, the values of $x$ at $i+\frac{1}{2}$ and $i-\frac{1}{2}$ should be computed by averaging the left and right biased interpolated values obtained using the formulation described in Eqn. \ref{vanleer-poly}. Then, using Eqn. \ref{FP-interp2} the derivative at location $i$ is computed.
    
    \begin{subequations} \label{FP-interp}
    \begin{gather}
    \begin{aligned}
         x_{i+\frac{1}{2}} &= 0.5\left(x_{i+\frac{1}{2}}^L + x_{i+\frac{1}{2}}^R\right) = 0.5 \left[\left(x_{i} + \frac{1}{2} x'_{i}+ \frac{1}{12} x''_{i} \right) + \left(x_{i+1} - \frac{1}{2} x'_{i+1}+ \frac{1}{12} x''_{i+1} \right) \right], \\
        x_{i-\frac{1}{2}} &= 0.5\left(x_{i-\frac{1}{2}}^L + x_{i-\frac{1}{2}}^R\right) = 0.5 \left[\left(x_{i-1} + \frac{1}{2} x'_{i-1}+ \frac{1}{12} x''_{i-1} \right) + \left(x_{i} - \frac{1}{2} x'_{i}+ \frac{1}{12} x''_{i} \right) \right] 
    \end{aligned}
    \tag{\theequation a-\theequation b}
    \end{gather}
    \end{subequations}
    
    \begin{equation} \label{FP-interp2}
        (x_{\xi})_i = \frac{x_{i+\frac{1}{2}} - x_{i-\frac{1}{2}}}{\Delta \xi}
    \end{equation}
    
    In the above equations, the $x'$ and $x''$ terms should be computed using the formulations described in Eqns. \ref{E6_grads} or \ref{IG4_grads} depending on whether the scheme being employed is MEG6 or MIG4. The procedure leads to linear interpolation and, consequently, a linear estimate of all the metric terms. It should be noted that though $x'$ and $x_{\xi}$ have mathematically the same meaning, they are different in the numerical sense.
    
    \item \textbf{Consistent interpolation using a central scheme}: The same formulation used in interpolating the left and right characteristic variable states in the inviscid flux algorithm (Eqns. \ref{legendre_polys}) should be used to interpolate metric terms from cell centers to cell interfaces. Furthermore, while interpolating metric terms of the form $\frac{\xi_x}{J}$ (which are used in the Riemann solver), the quantity should be interpolated as a whole rather than interpolating $\xi_x$ and $\frac{1}{J}$ separately and then multiplying them later. To make the interpolation central, the left and right states should be averaged. For instance, the following expression should be used to interpolate $\frac{\xi_x}{J}$ to the $i+\frac{1}{2}$ interface.
    
    
    \begin{equation}    \label{interp-formulae}
    \begin{aligned}
        \left(\frac{\xi_x}{J}\right)_{i+\frac{1}{2}} &= 0.5 \left\{\left(\frac{\xi_x}{J}\right)_{i+\frac{1}{2}}^L + \left(\frac{\xi_x}{J}\right)_{i+\frac{1}{2}}^R \right\}\\
        &= 0.5 \left\{\left[\left(\frac{\xi_x}{J}\right)_{i} + \frac{1}{2} \left(\frac{\xi_x}{J}\right)'_{i}+ \frac{1}{12} \left(\frac{\xi_x}{J}\right)''_{i} \right] + \left[\left(\frac{\xi_x}{J}\right)_{i+1} - \frac{1}{2} \left(\frac{\xi_x}{J}\right)'_{i+1}+ \frac{1}{12} \left(\frac{\xi_x}{J}\right)''_{i+1} \right] \right\}
    \end{aligned}
    \end{equation}
    
\end{enumerate}

Now that all the steps are detailed, next, we present a proof showing how freestream preservation is ensured by following these steps. To ensure freestream preservation, the interpolated metrics should satisfy the three metric identities listed in Eqn. \ref{metric-Idty} in the discrete sense \cite{thomas1979geometric,nonomura2010freestream}. These metric identities can be obtained by simplifying the momentum equations by considering the freestream flow assumption (i.e., constant $[\rho,u,v,w,p]$ values throughout the domain). For the sake of brevity, the proof is only presented for the metrics corresponding to MEG6. Nevertheless, following the essential steps, the same result can also be obtained even for metrics that are computed with the MIG4 scheme.
    
\begin{subequations}  \label{metric-Idty}
\begin{gather}
    \frac{\partial}{\partial \xi}\left(\hat{\xi_x}\right)+\frac{\partial}{\partial \eta}\left(\hat{\eta_x}\right)+\frac{\partial}{\partial \zeta}\left(\hat{\zeta_x}\right) =0, \\
    \frac{\partial}{\partial \xi}\left(\hat{\xi_y}\right)+\frac{\partial}{\partial \eta}\left(\hat{\eta_y}\right)+\frac{\partial}{\partial \zeta}\left(\hat{\zeta_y}\right) =0, \\
    \frac{\partial}{\partial \xi}\left(\hat{\xi_z}\right)+\frac{\partial}{\partial \eta}\left(\hat{\eta_z}\right)+\frac{\partial}{\partial \zeta}\left(\hat{\zeta_z}\right) =0
\end{gather}
\end{subequations}

\noindent \textbf{Statement:} The three metric identities (Eqn. \ref{metric-Idty}) will be satisfied in discrete sense if the conservative metrics are discretized and interpolated based on MEG6 or MIG4 schemes (as described in steps 2-3 above).

\noindent \textbf{Proof:} The metric terms $\hat{\xi}_{x}$, $\hat{\eta}_{x}$, and $\hat{\zeta}_{x}$ presented in Eqn. \ref{metric-Idty}a can be analytically expressed as,

\begin{equation}
\hat{\xi}_x=\left(y_\eta z\right)_\zeta-\left(y_\zeta z\right)_\eta, \quad \hat{\eta}_x=\left(y_\zeta z\right)_{\xi}-\left(y_{\xi} z\right)_\zeta, \quad \hat{\zeta}_x=\left(y_{\xi} z\right)_\eta-\left(y_\eta z\right)_{\xi}
\end{equation}

\noindent We proceed to discretize $\hat{\xi}_{x}$, $\hat{\eta}_{x}$ and $\hat{\zeta}_{x}$ at $i,j,k$ locations. Firstly the terms $y_\xi z$, $y_\eta z$, and $y_\zeta z$ are computed using the discretization scheme employed for the MEG6 scheme as follows:

\begin{equation}
    \begin{split}
    \left(y_\xi z\right)_{i,j,k} &= \frac{1}{\Delta \xi}\left(y_{i+\frac{1}{2},j,k} + y_{i-\frac{1}{2},j,k}\right) z_{i,j,k} \\
    &= \frac{1}{2 \Delta \xi}\left(y_{i+\frac{1}{2},j,k}^L+y_{i+\frac{1}{2},j,k}^R\right) z_{i,j,k} + \frac{1}{2 \Delta \xi}\left(y_{i-\frac{1}{2},j,k}^L+y_{i-\frac{1}{2},j,k}^R\right) z_{i,j,k} \\
    &= \frac{2282}{2880\Delta \xi}(y_{i + 1, j, k} - y_{i - 1, j, k})z_{i, j, k} + \frac{-546}{2880\Delta \xi}(y_{i + 2, j, k} - y_{i - 2, j, k})z_{i, j, k} \\
    &+ \frac{89}{2880\Delta \xi}(y_{i + 3, j, k} - y_{i - 3, j, k})z_{i, j, k} +\frac{-3}{2880\Delta \xi}(y_{i + 4, j, k} - y_{i - 4, j, k})z_{i, j, k} \\
    &+ \frac{-1}{2880\Delta \xi}(y_{i + 5, j, k} - y_{i - 5, j, k})z_{i, j, k}.
    \end{split}
\end{equation}

\noindent Similar expressions can be derived for $y_\eta z$ and $y_\zeta z$. Next we compute the terms $\left(y_\eta z\right)_\zeta$ and $\left(y_\zeta z\right)_\eta$ as follows.

\begin{equation}    \label{proof1}
    \begin{split}
    \left[\left(y_\eta z\right)_\zeta\right]_{i,j,k} &= \frac{1}{\Delta \zeta} \left[ \left(y_\eta z\right)_{i,j,k+\frac{1}{2}}-\left(y_\eta z\right)_{i,j,k-\frac{1}{2}} \right] \\
    &= \frac{1}{2 \Delta \zeta}\left[\left(y_\eta z\right)_{i,j,k+\frac{1}{2}}^L+\left(y_\eta z\right)_{i,j,k+\frac{1}{2}}^R\right] + \frac{1}{2 \Delta \zeta}\left[\left(y_\eta z\right)_{i,j,k-\frac{1}{2}}^L+\left(y_\eta z\right)_{i,j,k-\frac{1}{2}}^R\right] \\
    &= \frac{2282}{2880\Delta \zeta}\left[\left(y_\eta z\right)_{i , j, k+ 1} - \left(y_\eta z\right)_{i, j, k- 1}\right] + \frac{-546}{2880\Delta \zeta}\left[\left(y_\eta z\right)_{i, j, k + 2} - \left(y_\eta z\right)_{i, j, k - 2}\right] \\
    & + \frac{89}{2880\Delta \zeta}\left[\left(y_\eta z\right)_{i, j, k + 3} - \left(y_\eta z\right)_{i, j, k - 3}\right] + \frac{-3}{2880\Delta \zeta}\left[\left(y_\eta z\right)_{i, j, k + 4} - \left(y_\eta z\right)_{i, j, k - 4}\right] \\
    & + \frac{-1}{2880\Delta \zeta}\left[\left(y_\eta z\right)_{i, j, k + 5} - \left(y_\eta z\right)_{i, j, k - 5}\right].
    \end{split}
\end{equation}

\noindent After computing the term $\left[(y_{\zeta}z)_{\eta}\right]_{i,j,k}$ in a similar fashion, it can be subtracted from $\left[(y_\eta z)_\zeta\right]_{i,j,k}$ to obtain the metric term $\left(\xi_x\right)_{i,j,k}$. Similarly, other metric terms $\left(\eta_x\right)_{i,j,k}$ and $\left(\zeta_x\right)_{i,j,k}$ can also be computed. The next step is to interpolate $\xi_x$, $\eta_x$, and $\zeta_x$ terms from $(i,j,k)$ locations to the half locations and evaluate Eqn. \ref{metric-Idty}a. For instance, the first term of Eqn. \ref{metric-Idty}a is computed as follows. It should be noted that the grid spacing in the computational coordinates is uniform and taken equal to one unit, i.e., $\Delta \xi = \Delta \eta = \Delta \zeta = 1$.

\begin{equation}
    \begin{aligned}
    \frac{\partial}{\partial \xi}\left(\hat{\xi}_x\right) = \frac{(\xi_x)_{i+\frac{1}{2},j,k}-(\xi_x)_{i-\frac{1}{2},j,k}}{\Delta \xi} &= \frac{1}{2 \Delta \xi}\left[(\xi_x)_{i+\frac{1}{2},j,k}^L+(\xi_x)_{i+\frac{1}{2},j,k}^R\right] - \frac{1}{2 \Delta \xi}\left[(\xi_x)_{i-\frac{1}{2},j,k}^L+(\xi_x)_{i-\frac{1}{2},j,k}^R\right] \\
    &= \frac{1}{23887872000}[-y_{i-5,j-5,k-5}z_{i-5,j-5,k} -3y_{i-5,j-5,k-4}z_{i-5,j-5,k} \\
    & +\quad \cdots \quad  + 3y_{i+5,j+5,k+4}z_{i+5,j+5,k} + y_{i+5,j+5,k+5}z_{i+5,j+5,k}]
    \end{aligned}
\end{equation}

\noindent Similarly, the expressions $\frac{(\eta_x)_{i,j+\frac{1}{2},k}-(\eta_x)_{i,j-\frac{1}{2},k}}{\Delta \eta}$ and $\frac{(\zeta_x)_{i,j,k+\frac{1}{2}}-(\zeta_x)_{i,j,k-\frac{1}{2}}}{\Delta \zeta}$ can also be discretized. Adding up all the expressions will result in a sum equal to zero. The zero-sum proves that the first metric identity is satisfied in a discrete sense. The other two metric identities Eqns. \ref{metric-Idty}b-c can also be proved similarly. Due to the large number of terms involved, full length arithmetic operations are not showed here. A detailed, worked-out Mathematica notebook file has been provided in the supplementary section (Suppl. \ref{sec:supply}1) to support the proof. It was also found that the use of any other interpolation scheme for the present discretization approach (MEG4/MIG4), other than the one mentioned in step 3 above, will not satisfy the metric identities. 
\begin{equation}
     \frac{(\xi_x)_{i+\frac{1}{2},j,k}-(\xi_x)_{i-\frac{1}{2},j,k}}{\Delta \xi} +   \frac{(\eta_x)_{i,j+\frac{1}{2},k}-(\eta_x)_{i,j-\frac{1}{2},k}}{\Delta \eta} +  \frac{(\zeta_x)_{i,j,k+\frac{1}{2}}-(\zeta_x)_{i,j,k-\frac{1}{2}}}{\Delta \zeta} = 0
\end{equation}

\section{Results} \label{sec:results}
The results are organized into three parts. The first part (Sec. \ref{sec:FP-stationary}) presents demonstrative examples to show the freestream preserving nature of the MEG6/MIG4 schemes. The second part (Sec. \ref{sec:dynamic}) explores the efficacy of MEG6 and MIG4 schemes to simulate flows through dynamically deforming grids. Finally, in the third part (Sec. \ref{sec:JNapplication}), the resolving capability of MEG6 and MIG4 in capturing screech tones and unsteady aspects of an under-expanded supersonic jet \cite{ponton1997near} are shown employing coarse grid LES. The test cases used in each of the three parts are enumerated below.

\begin{enumerate}
    \item \textbf{Freestream preservation:} Three-dimensional uniform flow, convecting vortex, and Double Mach reflection.
    \item \textbf{Dynamically deforming meshes:} Three-dimensional uniform flow, stationary vortex, and double periodic shear layer.
    \item \textbf{Application to simulate supersonic jet noise:} Mach 1.35 choked under-expanded supersonic jet.
\end{enumerate}

\subsection{Freestream preservation: Demonstration} \label{sec:FP-stationary}

The test cases in this section were chosen to cover two-dimensional, three-dimensional, subsonic, and supersonic flow scenarios. To discern the effect of using a non-freestream preserving approach for computing metric terms, tests were also performed using a non-conservative metric term approach presented in \ref{app:non-FP}. The non-freestream preserving approach employs a non-conservative form of metrics and explicit fourth order central scheme for first and second derivative evaluation instead of Eqns. \ref{E6_grads} or \ref{IG4_grads}. The non-freestream preserving metrics are referred to with the names \textit{Non-Freestream Preserving MEG6} (Non-FP MEG6) and \textit{Non-Freestream Preserving MEG6} (Non-FP MIG4) when MEG6 and MIG4 schemes were used to compute the inviscid fluxes, respectively.

\subsubsection{Three-dimensional uniform flow} \label{3D-UF}

\begin{figure}[h!]
    \centering
    \includegraphics[width=90mm]{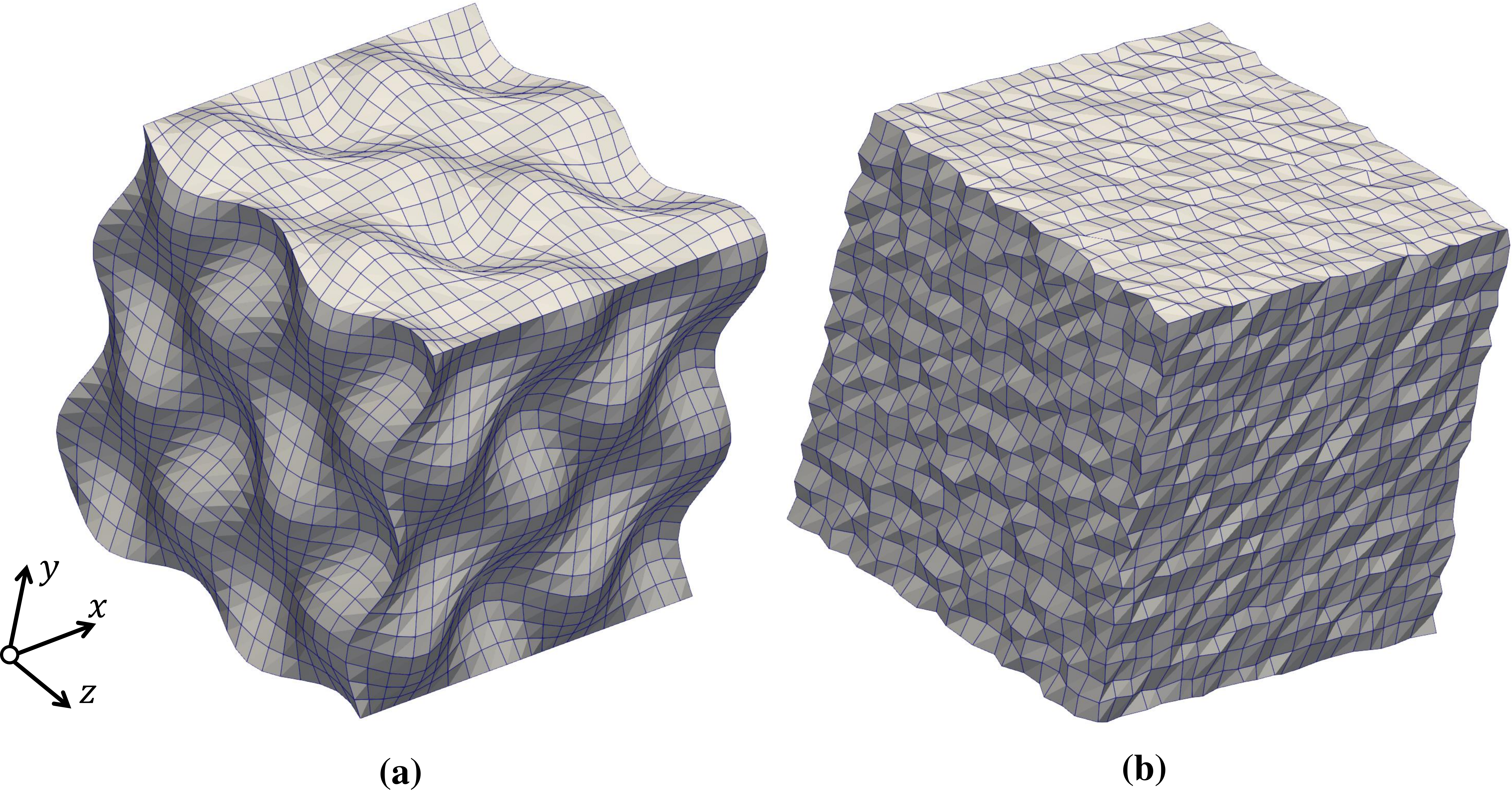}
    \caption{(a) Sinusoidal grid and (b) Randomized grid used to test freestream preservation nature of MEG6 and MIG4 schemes.}
    \label{freestream_grid}
\end{figure}

A fluid stream with uniform density, velocity, and pressure flowing through a periodic domain is simulated to test freestream preservation of MEG6 and MIG4 schemes in this test case. The effect of mesh non-uniformity on the solution is assessed by employing freestream preserving and non-freestream preserving metric terms. Two non-uniform meshes are employed in this test case; the first is a wavy sinusoidal domain, and the second is a randomized grid (Fig. \ref{freestream_grid}). The formulation used to generate the sinusoidal grid \cite{jiang2014free} is as follows:

\begin{equation}
   \begin{aligned}
    x_{i, j, k}=x_{\min }+&\Delta x_0\left[(i-1)+ \sin (\pi(j-1) \Delta y_0) \sin(\pi(k-1) \Delta z_0) \right], \\
    y_{i, j, k}=y_{\min }+&\Delta y_0\left[(j-1)+ \sin (\pi(k-1) \Delta z_0) \sin(\pi(i-1) \Delta x_0) \right], \\
    z_{i, j, k}=z_{\min }+&\Delta z_0\left[(k-1)+ \sin (\pi(i-1) \Delta x_0) \sin(\pi(j-1) \Delta y_0) \right]
    \end{aligned}
\end{equation}

where,

\begin{equation} \label{usual-pars}
    \begin{aligned}
        i=1,2, \cdots, NI, \quad & j=1,2, \cdots, NJ, \quad & k=1,2, \cdots, NK, \\
        \Delta x_0=\frac{L_x}{NI}, \quad & \Delta y_0=\frac{L_y}{NJ}, \quad & \Delta z_0=\frac{L_z}{NK}, \\
        x_{\min }=-\frac{L_x}{2}, \quad & y_{\min }=-\frac{L_y}{2}, \quad & z_{\min }=-\frac{L_z}{2}
    \end{aligned}
\end{equation}

The variables $NI$, $NJ$, and $NK$ represent the total number of cells along the $\xi$, $\eta$, and $\zeta$ directions, respectively; all of them are taken as $20$, thus making the grid resolution $20 \times 20 \times 20$. $L_x$, $L_y$, and $L_z$ are the length parameters taken as four units each. On the other hand, the randomized grid shown in Fig. \ref{freestream_grid}b is generated by randomly moving the uniformly distributed grid node positions by an amount of $20$ percent of the local grid spacing in each direction.

The initial conditions of the flow are: $(\rho, u, v, w, p)=(1,1,0,0,\frac{1}{\gamma M^2})$. The value of $\gamma$ is taken to as $1.4$. Inviscid calculations were performed with a Mach number of $0.5$. The initial conditions essentially dictate that the $y$ and $z$ velocities should remain zero everywhere in the domain at the end of the simulation. However, if the scheme is not freestream preserving, the errors associated with the metric cancellation will add up and grow. Consequently, they will be seen as non-zero values in the $y$ and $z$ direction velocities. 

The flow is simulated until a non-dimensional time of 10 units. Then, the $L^2$ norm errors associated with the $y$ and $z$ velocities are computed for both freestream and non-freestream preserving metric approaches. All the $L^2$ errors recorded in the simulations are tabulated in table \ref{freestream_table}. The errors associated with non-freestream preserving schemes presented in the first and second rows of the table can be seen to be significantly high. The errors are in the order of $10^{-3}$ and $10^{-1}$ in magnitude on sinusoidal and randomized grids, respectively. On the other hand, the errors associated with freestream preserving metrics are close to machine zero (of double-precision computations) on both sinusoidal and randomized grids; these errors can be treated as zeros. This clearly demonstrates that the current approach is freestream preserving concerning both MEG6 and MIG4 schemes. Experiments were also performed with conservative metrics computed using the fourth-order scheme (Eqn. \ref{4thorder-grads}), which also failed to preserve the freestream values, as expected.

\begin{table}[h!]
    \centering
    \includegraphics[width=110mm]{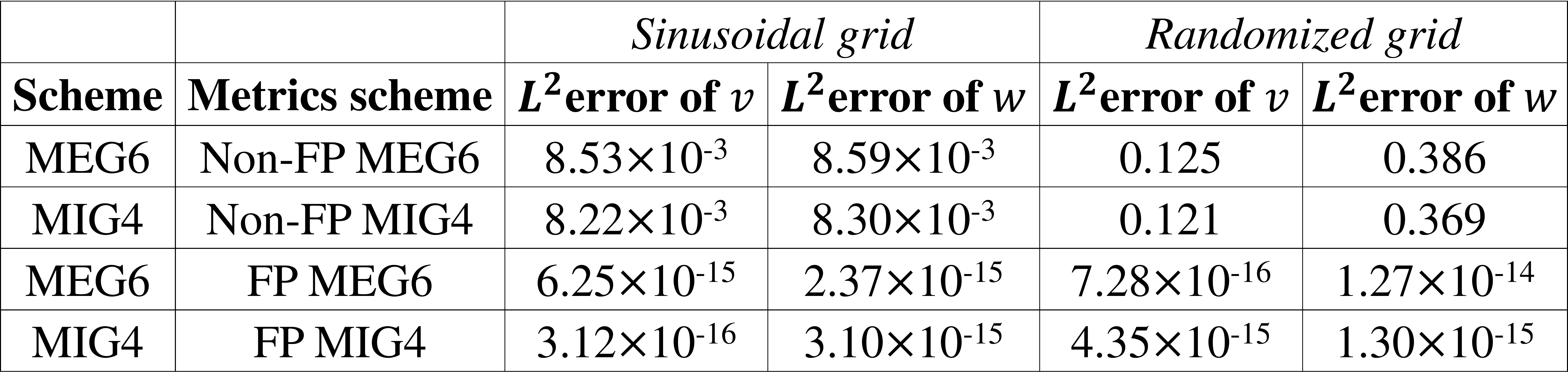}
    \caption{$L^2$ norm errors of $y$ and $z$ component velocity in the freestream preservation test on sinusoidal and randomized grids employing non-conservative and conservative metric term approaches.}
    \label{freestream_table}
\end{table}

\subsubsection{Two-dimensional moving vortex}

\begin{figure}[h!]
    \centering
    \includegraphics[width=135mm]{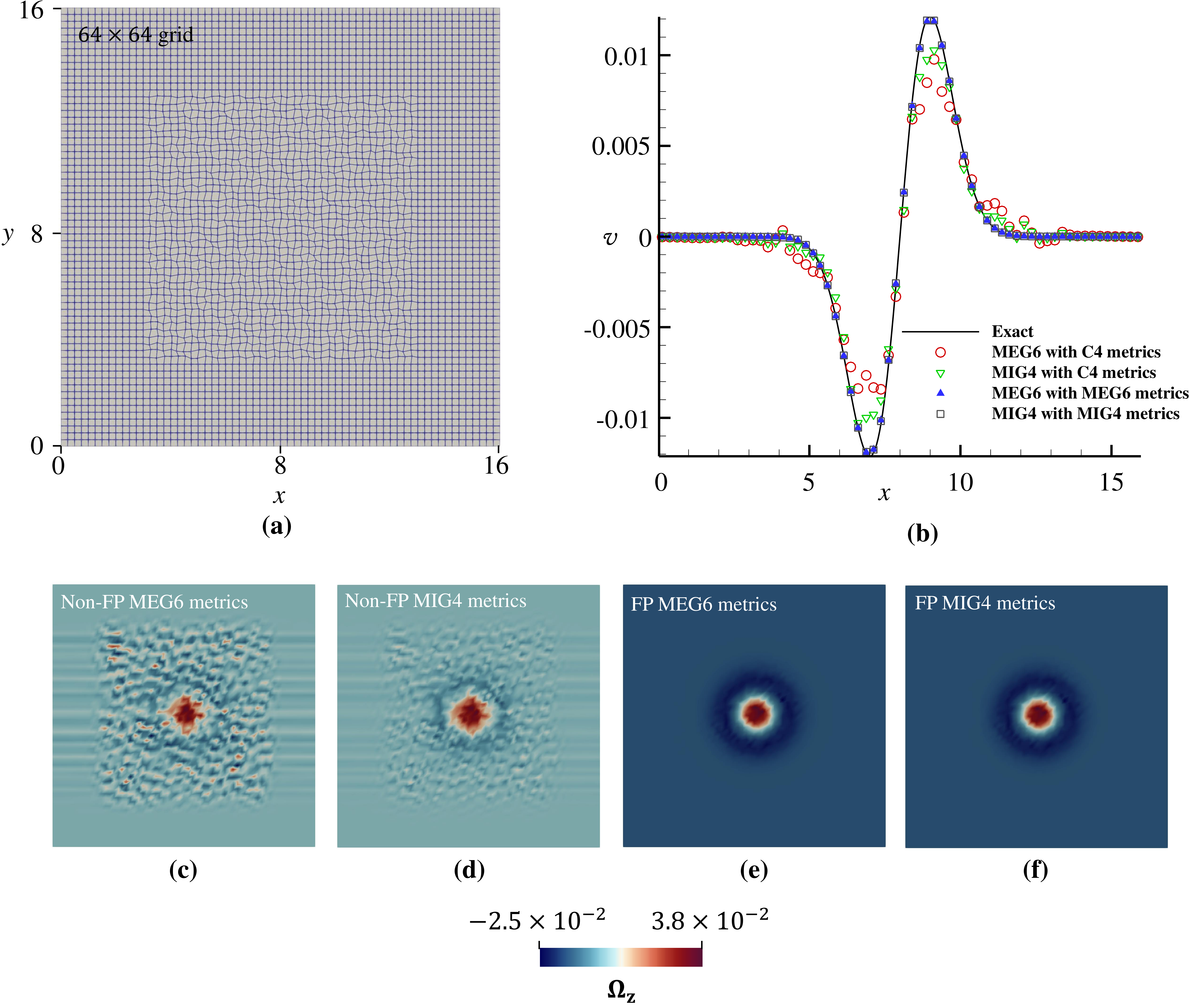}
    \caption{(a) $64\times64$ 2D grid used for the simulation, (b) Plots of swirl velocity along $y=8$ and, (c,d,e,f) Plots of z-vorticity ($\Omega_z$) at $t=16$ using conservative and non-conservative metrics for MEG6 and MIG4 schemes.}
    \label{freestream_COVO}
\end{figure}

In this test case, an isentropic inviscid vortex is advected through a distorted randomized grid of size $64 \times 64$ to test for the freestream and vortex preservation nature of the MEG6 and MIG4 schemes. The test is performed on a periodic domain spanning over square dimensions of $[0,16]\times[0,16]$. With the advection velocity chosen as one unit and the simulation time $16$ non-dimensional time units, the vortex is allowed to travel one full cycle to reach back to its initialized location. The grid randomization is only done in the central portion of the domain spanning over the region given by ${x \in [3,13] \cup y\in [3,13]}$ (Fig. \ref{freestream_COVO}a). Such an abrupt mesh non-uniformity (also an abrupt change in metric term values) is created intentionally to check the robustness of the schemes. The level of grid randomization in the current simulation is also considered twenty percent, similar to the previous case. The mesh resulting non-uniformities created a maximum included angle of about $160$ degrees in the domain. The initial conditions are given by:

\begin{equation}
\begin{aligned}
& \rho = 1 \\
& u =1-\frac{C\left(y-y_{c}\right)}{U_{\infty} R} \exp \left(-r^{2} / 2\right) \\
& v =\frac{C\left(x-x_{c}\right)}{U_{\infty} R^{2}} \exp \left(-r^{2} / 2\right) \\
& p =1-\gamma M^2 \frac{C^{2}}{U_{\infty}^2 R^{2}} \exp \left(-r^{2}\right) \\
& r^{2} =\left(x-x_{c}\right)^{2}+\left(y-y_{c}\right)^{2}
\end{aligned}
\end{equation}

where $(x_c,y_c) = (8,8)$ are the coordinates of vortex core center. $R$ denotes the vortex core radius. The value of non-dimensional vortex core strength $\frac{C}{U_{\infty}R}$ is taken as $0.02$. The freestream Mach number $M$ is set to $0.1$.

Fig. \ref{freestream_COVO}b shows the swirl velocity profiles (y-direction velocity) along $y=8$ computed using freestream preserving and non-freestream preserving metric schemes in comparison to the exact solution. The blue triangles and the black circles can be noted to follow close to the exact solution, unlike the solution corresponding to the non freestream scheme. A similar conclusion can be drawn from the vorticity contours presented in Fig \ref{freestream_COVO}c-d. While MEG6 and MIG4 conservative approaches preserve the vortex shape and its surrounding field, the solution computed using non-conservative compact fourth-order schemes yields an unphysical solution. This test has successfully demonstrated the vortex-preserving nature of the MEG6 and MIG4 conservative metric approaches.

\subsubsection{Double Mach reflection} \label{DMR-case}

\begin{figure}[h!]
    \centering
    \includegraphics[width=135mm]{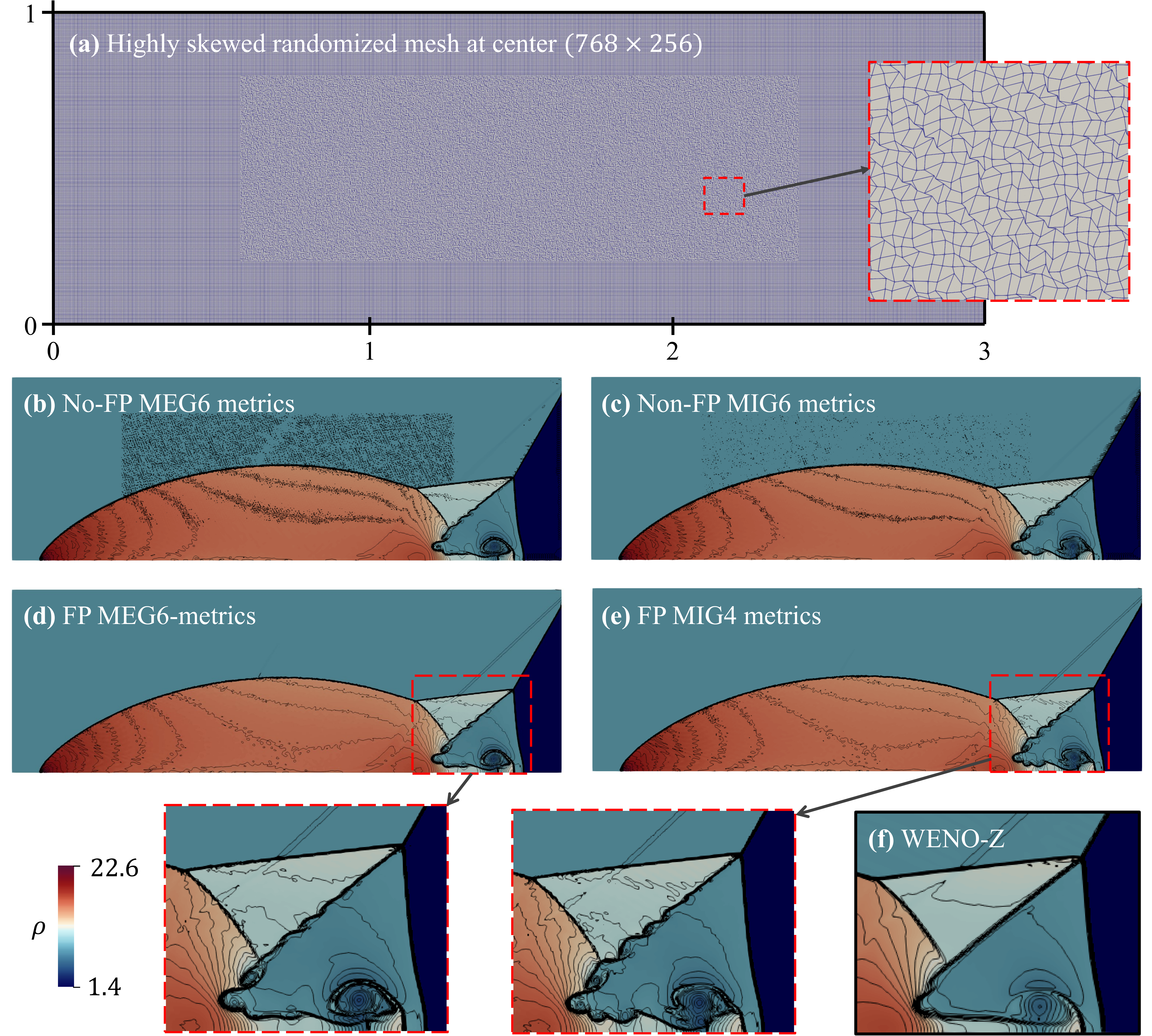}
    \caption{(a) $768\times256$ 2D grid used for the simulation, (b,c,d,e) Plots of density contours using conservative and non-conservative metrics for MEG6 and MIG4 schemes. Blow-up view of the shear layer corresponding to (d,e) conservative metrics approach are also shown at the bottom of the image. (f) Close-up view of the shear layer region, corresponding to the results obtained from the WENO-Z simulation employing uniform grid.}
    \label{freestream_DMR}
\end{figure}

Through this test case, the freestream preservation property of the MEG6 and MIG4 schemes is further accessed in the presence of shock waves and shear layers in the solution field. The problem comprises a planar shock wave moving towards a compression ramp. The ramp is inclined at 30 degrees relative to the shock. The moving shock wave interacts with the wall to form two triple points, two reflected shocks, two Mach stems, and a slip-stream. In order to simplify the computational domain and grid, the planar shock wave is inclined while keeping the ramp wall horizontal; this consequently makes the domain rectangular. Simulations are performed on a domain size of $[0,3]\times[0,1]$ at a resolution of $768\times256$. The grid is randomized in the central portion of the domain by randomly moving the grid nodes in both $x$ and $y$ directions by a maximum distance of fifty percent of the uniform grid spacing. The maximum included angle of the resulting mesh in the entire domain is close to $200$ degrees, making the grid highly skewed and non-uniform. A closeup view of the randomized grid is shown in Fig. \ref{freestream_DMR}a. The initial conditions for this test case are:

\begin{equation}
    (\rho, u, v, p)= \begin{cases}\left(8,8.25 \cos 30^{0},-8.25 \sin 30^{0}, 116.5\right), & x<1 / 6+\frac{y}{\tan 60^{0}} \\ (1.4,0,0,1), & x>1 / 6+\frac{y}{\tan 60^{0}}\end{cases}
\end{equation}

The boundary conditions are as follows. Post-shock flow conditions are set at the left boundary, and zero-gradient conditions are imposed on the right boundary. At the bottom boundary, reflecting boundary conditions are used from $x = 1/6$ to $x = 3.0$ and post-shock conditions for rest of the region i.e., $x \in [0, 1/6]$. A time-varying boundary condition is imposed on the top with pre- and post-shock conditions to the left and right of the shock, respectively, to accommodate the shock motion. The flow is simulated until an end time of $t=0.2$. 

Fig. \ref{freestream_DMR}b-d shows density contours at $t=0.2$ obtained using MEG6 and MIG4 schemes with freestream-preserving and non-freestream preserving metrics. The leftover errors corresponding to the non-freestream preserving schemes can be seen in the central region of the solution as rough patches in Figs. \ref{freestream_DMR}b-c. Interestingly the errors corresponding to MIG4 are minimal and significantly less compared to the MEG6 Non-FP approach (Fig. \ref{freestream_DMR}c). This might be attributed to the accuracy of implicit gradient values used in estimating the metric terms of the MIG4-based derivatives and interpolations. On the other hand, with conservative metric terms evaluated using MEG6 and MIG4 schemes, the solution in the central region can be seen unaffected by the grid distortion. Furthermore, the original solution is noted to be preserved even in the presence of shocks. In addition, the bottom right of the figure illustrates the outcome of a WENO-Z simulation for comparison with the current schemes. The simulation was conducted utilizing a uniform grid at the same resolution. The variations in the shear layer characteristics between the current schemes and the WENO-Z scheme demonstrate the enhanced resolving capabilities of the current schemes. This concludes the discussion on demonstrating the freestream and vortex-preserving nature of the MEG6 and MIG4 schemes. The use of these schemes on dynamically deforming meshes is explored next.\\


\subsection{Application to dynamically deforming meshes} \label{sec:dynamic}
In this section, we examine the usage of MEG6 and MIG4 schemes to simulate flows through dynamically deforming mesh environments using three canonical test cases, namely `3-D uniform flow', `stationary vortex,' and `doubly periodic shear layer.' The time derivative is split into two parts using the chain rule described in Eqn. \ref{time-split}. To ensure the geometric conservation law (GCL) which is a necessary condition for freestream preservation, the term $\left(\frac{1}{J}\right)_t$ is computed using Eqn. \ref{Jacb2}. The mesh velocities are computed analytically for all the test cases presented in this section. The corresponding expressions are presented while describing the case.


\subsubsection{Three-dimensional uniform flow on deforming sinusoidal grid} \label{3D-moving-FP}
Firstly, we start with the trivial case of three-dimensional uniform flow. The flow is simulated in a dynamically moving mesh environment with periodic boundary conditions on all sides. The initial and boundary conditions for this test case are the same as that of the conditions described in section \ref{3D-UF}. Inviscid calculations are performed using both freestream and non-freestream preserving metric approaches. The time-dependent location of the grid nodes for the simulation is given by: 

\begin{equation}
    \begin{aligned}
        x_{i, j, k}(t)=& x_{\min }+\Delta x_o\left[i+A_x \sin (2 \pi \omega t) \sin \frac{n_{x y} \pi j \Delta y_o}{L_y} \sin \frac{n_{x z} \pi k \Delta z_o}{L_z}\right] \\
        y_{i, j, k}(t)=& y_{\min }+\Delta y_o\left[j+A_y \sin (2 \pi \omega t) \sin \frac{n_{y x} \pi i \Delta x_o}{L_x} \sin \frac{n_{y z} \pi k \Delta z_o}{L_z}\right] \\
        z_{i, j, k}(t)=& z_{\min }+\Delta z_o\left[k+A_z \sin (2 \pi \omega t) \sin \frac{n_{z x} \pi i \Delta x_o}{L_x} \sin \frac{n_{z y} \pi j \Delta y_o}{L_y}\right]
    \end{aligned}
\end{equation}  

Taking the analytical derivative of the above expressions, the following grid time-dependent deformation velocity expressions are obtained:

\begin{equation}
    \begin{aligned}
        x_{i, j, k}^{'}(t)=&  2 \pi \omega \Delta x_o A_x \cos (2 \pi \omega t) \sin \frac{n_{x y} \pi j \Delta y_o}{L_y} \sin \frac{n_{x z} \pi k \Delta z_o}{L_z}, \\
        y_{i, j, k}^{'}(t)=&  2 \pi \omega \Delta y_o A_y \cos (2 \pi \omega t) \sin \frac{n_{y x} \pi i \Delta x_o}{L_x} \sin \frac{n_{y z} \pi k \Delta z_o}{L_z}, \\
        z_{i, j, k}^{'}(t)=&  2 \pi \omega \Delta z_o A_z \cos (2 \pi \omega t) \sin \frac{n_{z x} \pi i \Delta x_o}{L_x} \sin \frac{n_{z y} \pi j \Delta y_o}{L_y},
    \end{aligned}
\end{equation}

where, $i$, $j$, $k$, $\Delta x_o$, $\Delta y_o$, $\Delta z_o$, $x_{\min}$, $y_{\min}$, $z_{\min}$ assume their usual meanings described in Eqn. \ref{usual-pars}. The grid resolution for the simulation is set to $32^3$. The values of the rest of the parameters are: $A_x=A_y=A_z=1.5$, $L_x=L_y=L_z=12$, and $n_{xy}=n_{yz}= \cdots = 4$. The grid deformation oscillation frequency $\omega$ is set to $1$. As the mesh undergoes deformation, the maximum included angle in the domain reaches a maximum value of $\approx 160$.

\begin{figure}[h!]
    \centering
    \includegraphics[width=150mm]{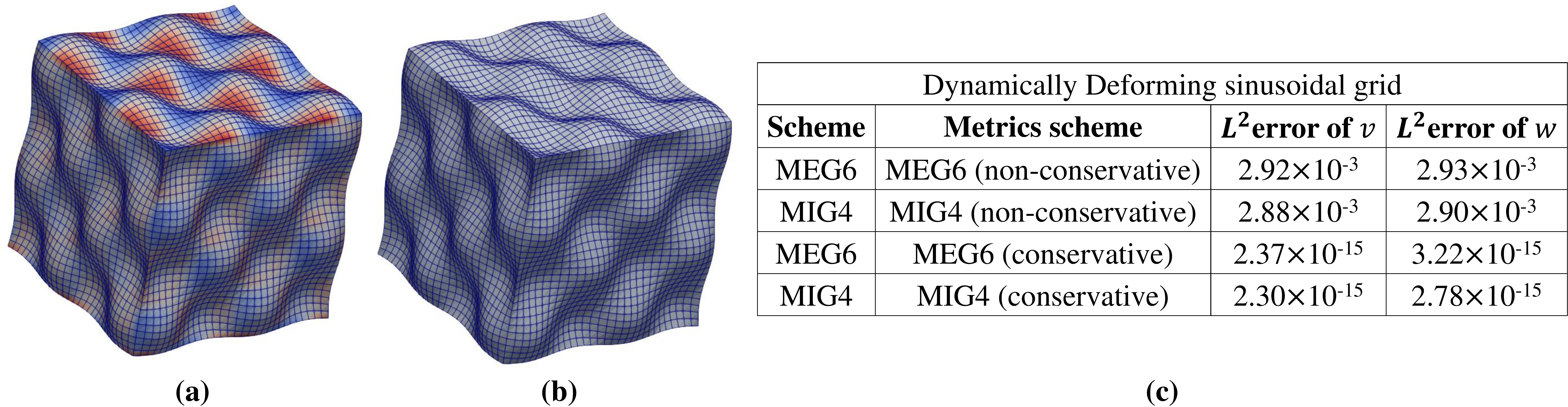}
    \caption{Snapshots of $20\times20$ three dimensional deforming sinusoidal grid with y-velocity contours on the surface using (a) non conservative (non-FP) metrics and (b) conservative (FP) metrics for MEG6 and MIG4 schemes. (c) $L^2$ norm error in y and z velocities using different different metric schemes.}
    \label{dynamic1}
\end{figure}

Contours of $y$-velocity and the corresponding grid at $t=5.25$ are shown in Fig. \ref{dynamic1}. The contours clearly indicate that the non-conservative metrics (non-FP) approach feeds error into the solution while the conservative metrics (FP) approach does not. Fig. \ref{dynamic1}c shows the $L^2$ norm errors computed using MEG6 and MIG4 schemes. The errors corresponding to the conservative metric formulation can be noted to stay close to machine zero, while the non-conservative approach does not. This clearly demonstrates that the present schemes satisfy the geometric conservation law and are thus freestream preserving even on dynamically deforming meshes.

\subsubsection{Stationary vortex on a three-dimensional deforming mesh}
In this test case, the vortex preservation nature of MEG6 and MIG4 schemes is shown in a moving grid environment. A stationary columnar vortex is initialized at the center of a three-dimensional, cubical domain of dimensions $[-8,8]\times[-8,8]\times[-8,8]$. In each direction, thirty-two cells were used. The grid is set to deform in time according to the same equations presented in the previous test case on \textit{three-dimensional uniform flow} (section \ref{3D-moving-FP}). The oscillation frequency of the grid is set to $\omega=1$.

\begin{figure}[h!]
    \centering
    \includegraphics[width=140mm]{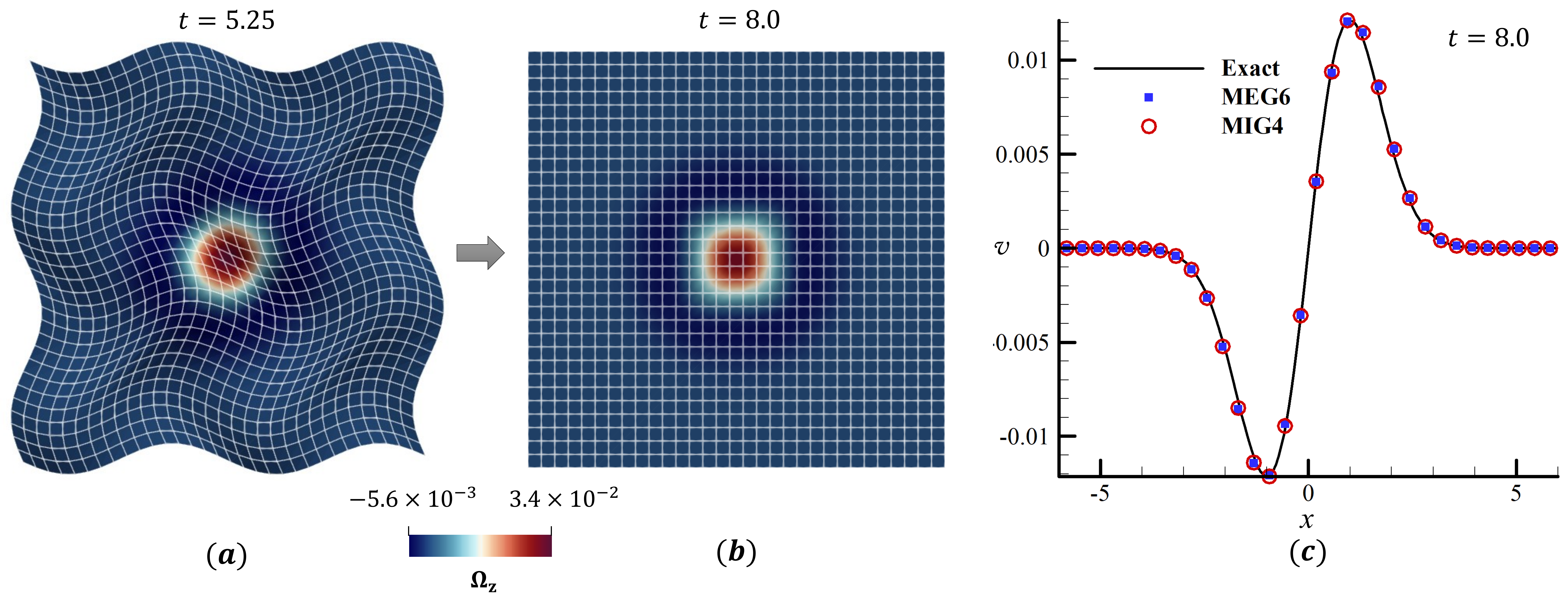}
    \caption{(a,b) Instantaneous snapshots of vorticity contours collected on the surface corresponding cells on the 16th grid index in the $\zeta$-direction. (c) Swirl velocity plot along NZ/2 and y = 8 at $t=8$.}
    \label{dynamic2}
\end{figure}

Inviscid calculations were performed until an end-time of $t_{end}=8$ using MEG6 and MIG4 schemes. The grid undergoes eight deformation cycles before the end time. Fig. \ref{dynamic2}a-b shows the grid and $z$-vorticity contours computed using the MIG4 scheme (with conservative metrics approach) at two different time instances. The vortex shape can be noted to stay intact and unaffected by the grid motion throughout. Fig. \ref{dynamic2}c shows the computed and exact swirl velocity profiles along the mid-horizontal line. Both MEG6 and MIG4 approaches can be noted to be in excellent agreement with the theoretical solution.

\subsubsection{Double periodic shear-layer on dynamic mesh}
In this test case, the freestream preserving nature of the MEG6 and MIG4 schemes in the presence of viscous stresses are tested. The case consists of two shear layers initially parallel to each other that evolve to produce two large vortices at $t=1$. Viscous forces are dominant near the shear layer in this test case. The non-dimensional parameters concerning the flow are, $\mathrm{Re} = 1 \times 10^4$ and $\mathrm{M} = 0.1$. The value of $\gamma$ is taken as $5/3$. The initial conditions for the flow are:

\begin{subequations}
    \begin{align}
        \rho &= \frac{1}{\gamma \mathrm{M}^2}, \\
        u &= 
        \begin{cases}
            \tanh (80 \times(y-0.25)), & \text{ if } (y \leq 0.5), \\
            \tanh (80 \times(0.75-y)), & \text{ if } (y > 0.5),
        \end{cases} \\
        v &= 0.05 \times \sin (2 \pi(x+0.25)) \\
        T &= 1.
    \end{align}
\end{subequations}

Simulations are performed on a two-dimensional deforming grid of resolution $360 \times 360$ using MEG6 and MIG4 schemes with the freestream preserving metrics. The spatial positions of mesh nodes and their respective velocities in time are given by Eqn. \ref{DPSL-eqn1} and Eqn. \ref{DPSL-eqn2} respectively, as described in Ref-\cite{achu2021entropically}. 

\begin{equation} \label{DPSL-eqn1}
    \begin{aligned}
x_{i, j}(\tau)=x_{\min }+\Delta x_0&\left[(i-1)+A_x \sin (2 \pi \omega \tau) \sin \left(\frac{n_x \pi(j-1) \Delta y_0}{L_y}+\frac{i \phi_x}{I L-1}\right)\right] \\
y_{i, j}(\tau)=y_{\min }+\Delta y_0&\left[(j-1)+A_y \sin (2 \pi \omega \tau) \sin \left(\frac{n_y \pi(i-1) \Delta x_0}{L_x}+\frac{i \phi_y}{J L-1}\right)\right]
\end{aligned}
\end{equation}

\begin{equation} \label{DPSL-eqn2}
    \begin{aligned}
&x_\tau=2 \pi \omega \Delta x_0 A_x \cos (2 \pi \omega \tau) \sin \left(\frac{n_x \pi(j-1) \Delta y_0}{L_y}+\frac{i \phi_x}{I L-1}\right) \\
&y_\tau=2 \pi \omega \Delta y_0 A_y \cos (2 \pi \omega \tau) \sin \left(\frac{n_y \pi(i-1) \Delta x_0}{L_x}+\frac{i \phi_y}{J L-1}\right)
\end{aligned}
\end{equation}

The length parameters $L_x$ and $L_y$ are considered $1$ unit each. The mesh oscillation frequency $\omega$ is set to 1, enabling the mesh to deform through one full cycle during the simulation ($t_{end}=1$).

\begin{equation} \label{KE-Ens}
    \begin{gathered}
E_k=\frac{1}{L_x L_y} \int_0^{L y} \int_0^{L x} \rho \frac{u^2+v^2}{2} \mathrm{~d} x \mathrm{~d} y \\
\mathcal{E}=\frac{1}{L_x L_y} \int_0^{L y} \int_0^{L x} \rho \frac{\boldsymbol{\Omega} \cdot \boldsymbol{\Omega}}{2} \mathrm{~d} x \mathrm{~d} y
\end{gathered}
\end{equation}

\begin{figure}[h!]
    \centering
    \includegraphics[width=150mm]{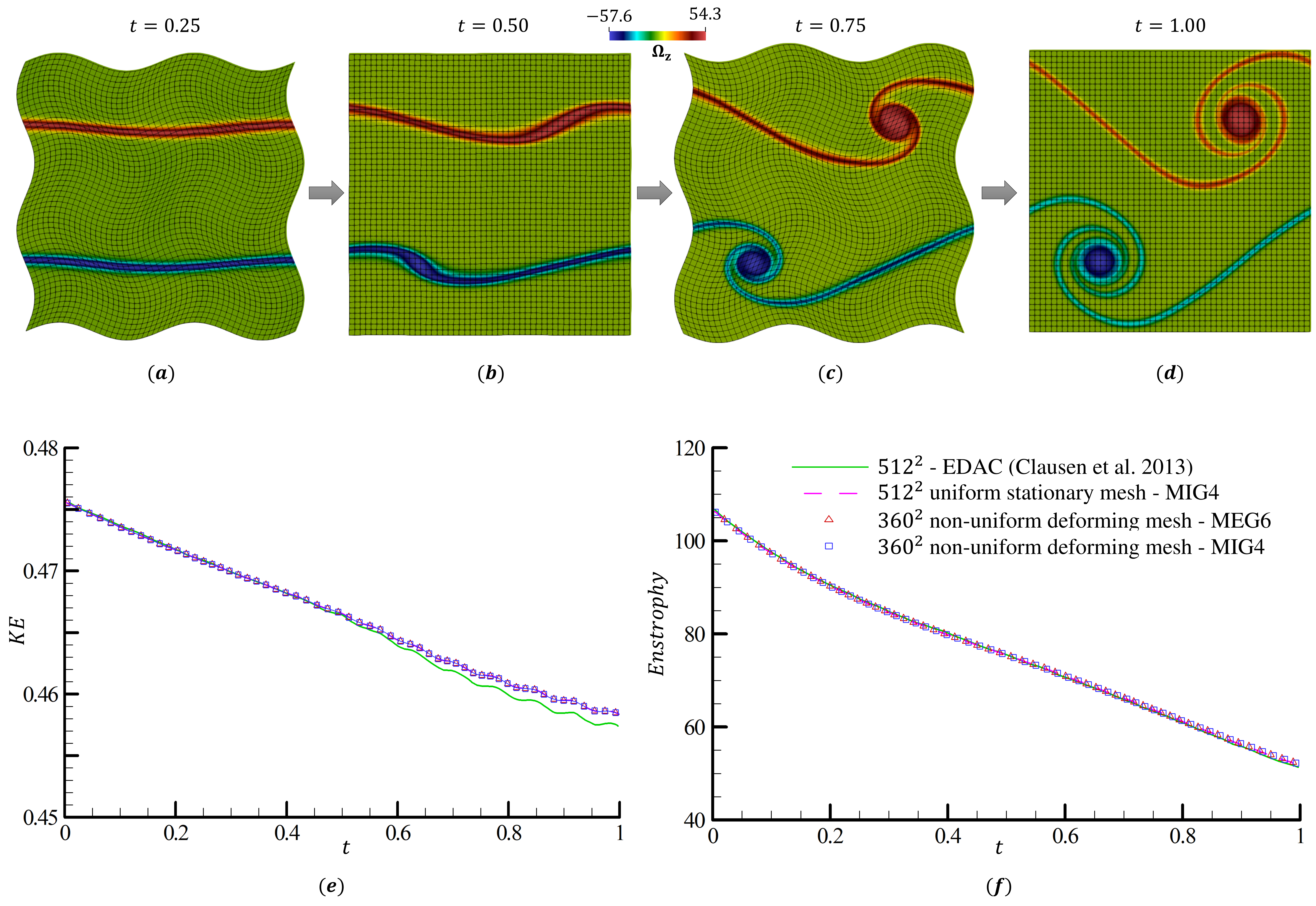}
    \caption{(a,b,c,d) Snapshots of instantaneous vorticity contours collected at four different time instances with every 10th point of the grid overlain on the solution computed using MIG4 scheme. Evolution of (e) Kinetic energy (KE) and (f) Enstrophy computed using MEG6 and MIG4 schemes compared against the incompressible solution of Clausen \cite{clausen2013entropically}.}
    \label{dynamic3}
\end{figure}

Fig. \ref{dynamic3} shows the $z$-vorticity field at four equally spaced time instances during the simulation. Two large vortices of opposing angular velocity can be seen to evolve as time progresses. The dynamic mesh motion can be seen to not show any effect on the flow contours (for comparison against the stationary grid DNS solution, please refer to \cite{sainadh2022spectral}). The evolution of Kinetic Energy (KE) and Enstrophy ($\mathcal{E}$) in the entire computational domain is tracked in time for both MEG6 and MIG4 schemes. The expressions to compute these quantities are provided in Eqn. \ref{KE-Ens}. Evolution plots of these quantities are shown in Figs. \ref{dynamic3}e-f in comparison to the solutions corresponding to a stationary grid simulation and the incompressible flow solution of Clausen \cite{clausen2013entropically}, both simulated on a uniform grid with a resolution of on $512 \times 512$. An excellent agreement between the reference solutions and the current calculations can be noted in both KE and Enstrophy plots. This demonstrates that a Navier-Stokes solution computed using the higher accuracy schemes, MEG6 and MIG4, will resist the errors associated with the dynamic grid motion.

\subsection{Mach 1.35 under-expanded supersonic round jet} \label{sec:JNapplication}
Shock containing supersonic jets emit intense aeroacoustic noise at distinct frequencies due to the phenomenon termed `screech'. The nearfield noise levels due to screech in a free single jet configuration typically range between 120dB-180dB. Although the physical mechanism of screech tone production is still not completely understood, it is a well-established fact that it occurs as the consequence of aeroacoustic resonance \cite{edgington2019aeroacoustic,edgington2021generation}. According to the classical feedback loop theory of Powell \cite{powell1953mechanism,raman1999supersonic}, firstly, a series of small flow disturbances are initiated at the nozzle lip region. These disturbances travel downstream (in the form of Kelvin-Helmholtz instabilities) towards the shock cells to interact with the oblique shocks. As the disturbances interact with the shocks (generally after the second or third shock cells), they undergo strong amplification and radiate intense acoustic waves, which drive the entire jet plume to resonate and consequently oscillate. Some of the produced acoustic waves travel upstream towards the nozzle lip through the atmosphere outside the jet to close the feedback loop. The screech is sustained with the continuous excitation of the shear layer (receptivity process) from these emission waves. Therefore, since the screech depends on several flow features, such as shock cells, shear layer, vortical structures, and acoustic waves, it is crucial to capture all these features with sufficient fidelity in order to simulate its effect through computations. The main objective of this section is to test the efficacy of MEG6 and MIG4 schemes in resolving `screech' and `unsteady jet oscillation mode' of an under-expanded supersonic jet. Simulations are also performed using WENO-Z \cite{borges2008improved} scheme for comparison against the present scheme.

\begin{figure}[h!]
    \centering
    \includegraphics[width=160mm]{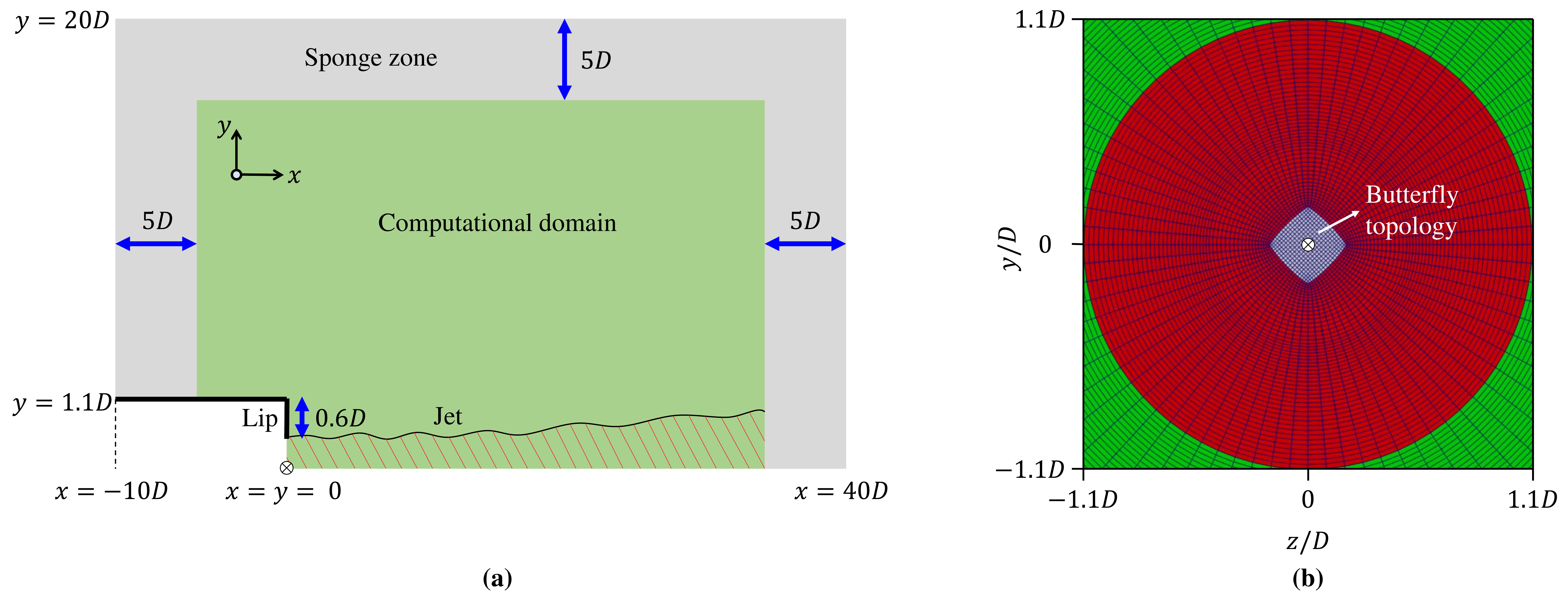}
    \caption{(a) Computational domain used, (b) Butterfly mesh topology adapted to avoid singularity at the axis, visualized on $x=2D$ plane.}
    \label{jet-domain-mesh}
\end{figure}

An under-expanded supersonic jet ejecting from a converging, choked nozzle operating at a Nozzle Pressure Ratio (NPR) of $2.97$ is considered as the test case. The flow simulated in the present computations is compared with the nearfield microphone data of Ponton et al. \cite{ponton1997near}. The experimental setup used in their work consists of a settling chamber attached to an axisymmetric converging nozzle with an exit diameter (D) equal to one inch. The lip thickness of the nozzle exit is $0.6D$. The flow diagnostics include a microphone placed on the nozzle exit plane at a radial location of two nozzle diameters away from the nozzle axis. Fig. \ref{jet-domain-mesh}a shows the dimensions of the computational domain used to simulate the flow. The domain is axisymmetric and consists of four blocks. The jet Reynolds number based on the nozzle diameter and the jet exit conditions is $Re_D=1.144 \times 10^6$. The flow is non-dimensionalized based on the ambient atmospheric conditions ($T_{\infty}=293\text{ K}$, $p_{\infty} = 101325\text{ Pa}$). The nozzle exit conditions are theoretically computed using Eqns. \ref{ini-cons}. The computed inlet conditions are directly provided as an inlet boundary condition on the nozzle exit plane. The effect of the boundary layer and turbulent fluctuations developed inside the annular region are not considered. Although this does not replicate the experimental conditions accurately, the effect of annular fluctuations in the present case was believed to show little effect on the screech feedback loop based on the study by \cite{ahn2021numerical,ahn2018supersonic}. To prevent spurious reflections from entering in to the computational domain, sponge zones are employed at the far-field boundaries in the regions marked in Fig. \ref{jet-domain-mesh}. These zones are implemented by incorporating a dissipative source term into the governing equations as outlined in the methodology of Bodony \cite{bodony2006analysis}.

\begin{equation} \label{ini-cons}
    \begin{aligned}
    &\frac{p_e}{p_{\infty}}=\frac{1}{\gamma}\left[\frac{2+(\gamma-1) M_j^2}{\gamma+1}\right]^{\frac{\gamma}{\gamma-1}} \\
    &\frac{\rho_e}{\rho_{\infty}}=\frac{\gamma(\gamma+1) p_e}{2 (T_n/T_{\infty})} \\
    &\frac{u_e}{a_{\infty}}=\sqrt{\frac{2 (T_n/T_{\infty})}{\gamma+1}}, \quad v_e = w_e = 0
\end{aligned}
\end{equation}

In this test case, three grids were utilized with resolutions of $10$, $13$, and $20$ million cells to examine the independence of the results from the choice of grid. To prevent a singularity at the nozzle axis, the butterfly topology was employed, as illustrated in Fig. \ref{jet-domain-mesh}b. The grid was clustered towards the axis and exit of the nozzle, where the key flow features are located. Along the axial direction of the geometry, a minimum grid spacing of $\Delta x = 0.0085D$, $0.0075D$, and $0.0065D$ was maintained for the $10$, $13$, and $20$ million grids, respectively. The grid spacing was linearly increased along the $x$-direction, with the maximum grid spacing not exceeding $0.1D$ in all three grids. The grid was uniformly distributed along the azimuthal direction, with 80, 100, and 120 divisions for the 10, 13, and 20 million grids, respectively. The grid distribution along the radial direction was consistent across all three meshes used in the study. Along the radial direction, a minimum spacing of $0.0075D$ was used close to the axis, with a growth rate of $\approx 4\%$, resulting in a total of 120 cells. Computations were performed with a $\Delta t a_{\infty}/D = 2\times10^{-3}$ until an end-time of $t a_{\infty}/D = 800$ ($400,000$ iterations). All the statistics were collected after the flow reached a statistically steady state, which is after $t a_{\infty}/D = 400$. All the simulations were run in parallel on a single Nvidia A100 GPU. The simulations corresponding to 13 million grid were completed in a span of $34.0$ and $41.7$ hours using MEG6 and MIG4 schemes, respectively. More details about the GPU acceleration will be discussed in the next section.

Given the high Reynolds number in the present case, Large Eddy Simulations (LES) were conducted on course grids. In conventional Large Eddy Simulation (LES) approaches, the effect of unresolved sub-grid-scale (SGS) flow features is modeled through an explicit SGS model while the larger scales are resolved by the grid \cite{germano1991dynamic}. However, in the present simulations, we solve the unfiltered Navier-Stokes equations directly and utilize the inherent numerical dissipation of the MEG6/MIG4 schemes to implicitly mimic the effects of SGS models. This approach is commonly referred to as ``implicit-LES'' or, specifically in the present case, ``Monotonically integrated Implicit LES (MILES)'' \cite{fureby1999monotonically} because of the use of the Monotonocity Preserving limiter in the inviscid flux discretization and its high spectral resolution. A similar LES strategy was also employed in the study performed by Ahn et al. \cite{ahn2021numerical} on twin-jet configurations.

\subsubsection{Instantaneous and average flow features}
\begin{figure}[h!]
    \centering
    \includegraphics[width=150mm]{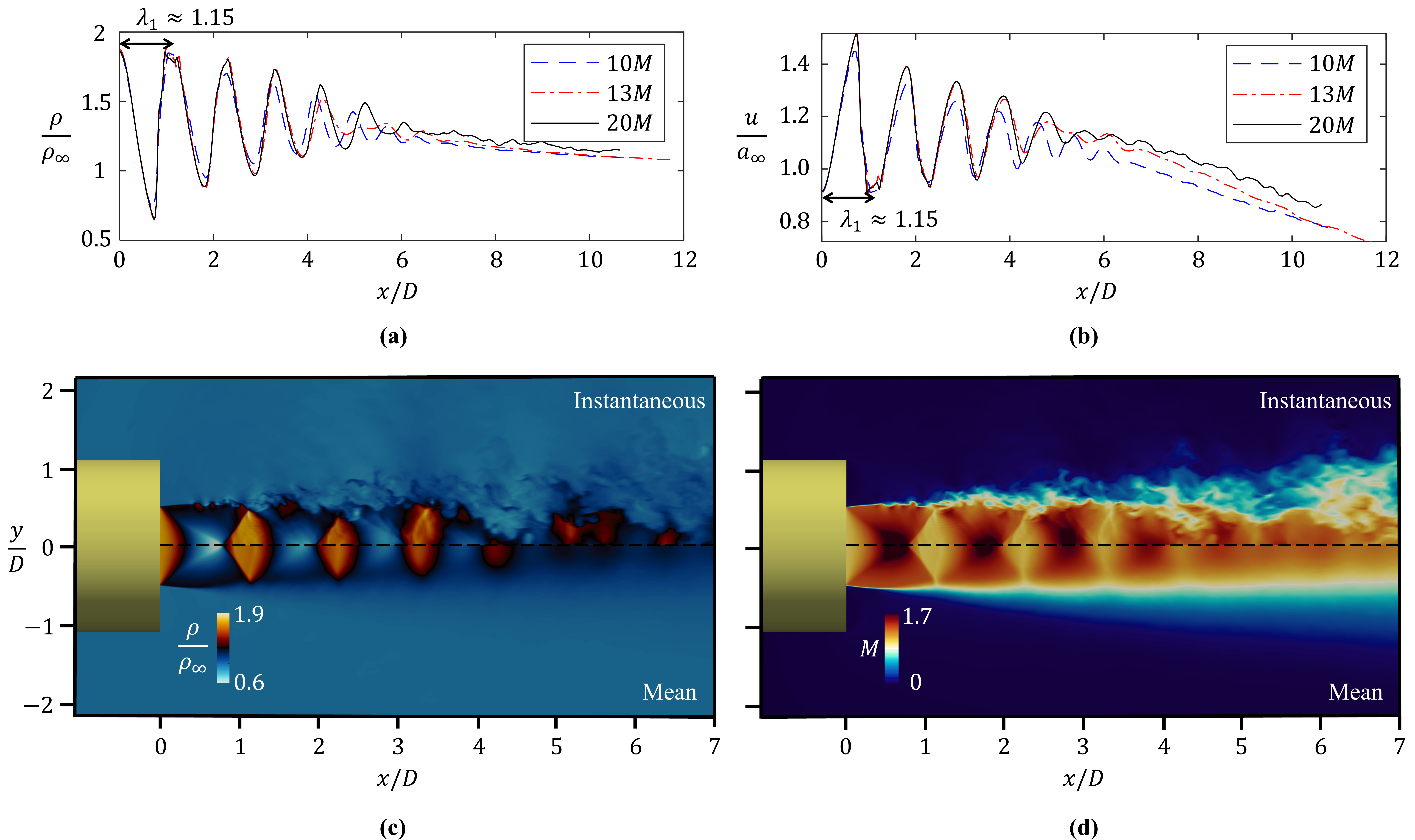}
    \caption{Time averaged axial profiles of density (a) and x-velocity (b) at various grid resolutions. The length of first shock cell spacing $\lambda_1$ is also indicated in the figures (a) and (b)}. Instantaneous and mean density contours (c) and local Mach number (d) contours of $M_{j}=1.35$ jet solution plotted on $z/D = 0$ plane employing a grid resolution of 20M. All the plots shown here are based on MIG4 scheme.
    \label{jet-inst-ave}
\end{figure}

Fig. \ref{jet-inst-ave}a-b shows the time-averaged axial density and $x$-velocity profiles on three different grid resolutions. The peaks and troughs in the profiles suggest the presence of shock cells. The solution has converged sufficiently at a grid resolution of $13$ million cells, especially in the region corresponding to the first four shock cells. It can be noted that the solution corresponding to the 10 million grid fails to capture the peaks and troughs of the density and velocity profiles between the range of $x/D=0$ to $4$, which is critical for the screech phenomenon. Thus, the results corresponding to $10$ million grid are not considered for the analysis. Fig. \ref{jet-inst-ave}c-d shows the instantaneous and mean contours of density and Mach number. Various flow features such as shocks, expansion fans, shear-layer, vortices, and induced density fluctuations can be noted in the pictures. The first two shock cells appear clearly defined, but progressively downstream the structure of shock cells become less distinct due to the growth of instability waves and turbulent mixing. The non-dimensional shock cell spacing in the jet is a crucial length scale in supersonic jets as it dictates the location of effective screech noise source \cite{powell1953mechanism}. The first shock cell spacing denoted by $\lambda_1$, can be calculated theoretically using the formulation proposed by Pack \cite{pack1950note} through Eqn. \ref{firstShockSpace} approximately.

\begin{equation}\label{firstShockSpace}
    \lambda_1=2.695 \sqrt{\left(\mathrm{NPR}^{0.291}-1.205\right)}
\end{equation}

Plugging in NPR=$2.97$ in to the equation, results in $\lambda_1 \approx 1.1$. On the other hand in the present simulations, as can be seen in the time-averaged density and velocity plots shown in Fig. \ref{jet-inst-ave}(a,b), the first shock cell spacing was obtained to be $\lambda_1 = 1.15$. A good agreement is observed between the theoretical calculations and the numerical results. Subsequently the shock cell spacing downstream was noted to slowly decrease with  $\lambda_2=1.15$, $\lambda_3=1.0$, $\lambda_4=1.0$, and $\lambda_5=0.9$. In Fig. \ref{jet-inst-ave}(c,d) the shear layer instabilities initiated near the nozzle lip can be seen to grow progressively in scale as they move downstream to interact with the shock cells. Consequently, local acoustic disturbances are produced, which travel upstream to close the feedback loop causing screech, as explained before. To better visualize this phenomenon, movies of density field are provided in the supplementary materials section \ref{sec:supply}. From the time-resolved data shown in Movie-1 of Sec. \ref{sec:supply}, the evolution of Kelvin-Helmholtz instabilities, the interaction of shear layer vortices with the shock cells, emitted acoustic radiation, and the consequent jet oscillation behavior can be remarked.

\begin{figure}[h!]
    \centering
    \includegraphics[width=\textwidth]{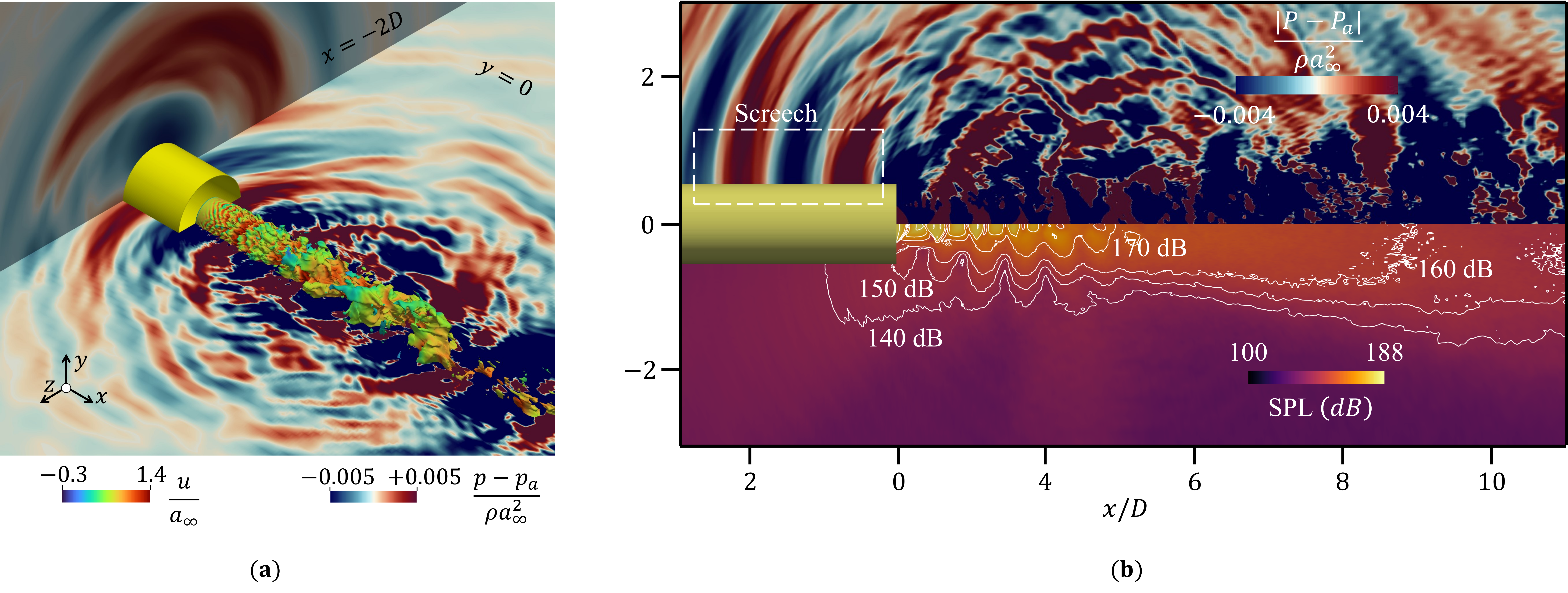}
    \caption{(a) Visualization of the jet surface using an iso-surface of density corresponding to $\rho/\rho_{\infty}=1.2$. The figure also illustrates the circular pressure ripples surrounding the jet planes $y=0$ and $x=-2D$. (b) Instantaneous pressure fluctuations (upper half) and overall sound pressure levels (lower half) are shown.}
    \label{jet-spl}
\end{figure}

A supersonic jet has multiple noise sources \cite{tam1995supersonic,bailly2016high,edgington2019aeroacoustic}. The above-mentioned acoustic resonance, shock shear layer interaction, and the jet mixing are part of the generation mechanisms behind different the various noise components. The tones corresponding to screech noise are predominant and can be characterized in the upstream region of the jet \cite{gojon2019antisymmetric,powell1953mechanism,davies1962tones,westley1969near}. Fig. \ref{jet-spl}(a) presents a snapshot of the instantaneous 3-D jet surface visualized through density iso-surface, along with the pressure fluctuation field surrounding it. The figure illustrates the progressive amplification of three-dimensional hydrodynamic disturbances on the jet surface close to the nozzle exit. In Fig. \ref{jet-spl}(b), the instantaneous pressure fluctuation contours and the mean Sound Pressure Levels (SPL) are displayed on the $z=0$ plane. The pressure fluctuation field immediately close to the jet surface appears choppy and seemingly random; however, a clear sinusoidal-type pattern can be observed in the upstream region of the jet just above the nozzle, as indicated by the dashed white box in Fig. \ref{jet-spl}. Given the pronounced sinusoidal nature of the wave pattern and the prevalence of screech in the upstream location, it can be inferred that these fluctuations correspond to screech. The lower half of Fig. \ref{jet-spl} displays isolines of SPL, with SPL near the lip of the nozzle observed to be close to $150$dB. The solutions presented in Fig. \ref{jet-inst-ave} correspond to the MIG4 scheme; however, similar results were also observed with the MEG6 scheme.

\subsubsection{Comparison of acoustic data with experiments} \label{freqs}

\begin{figure}[h!]
    \centering
    \includegraphics[width=\textwidth]{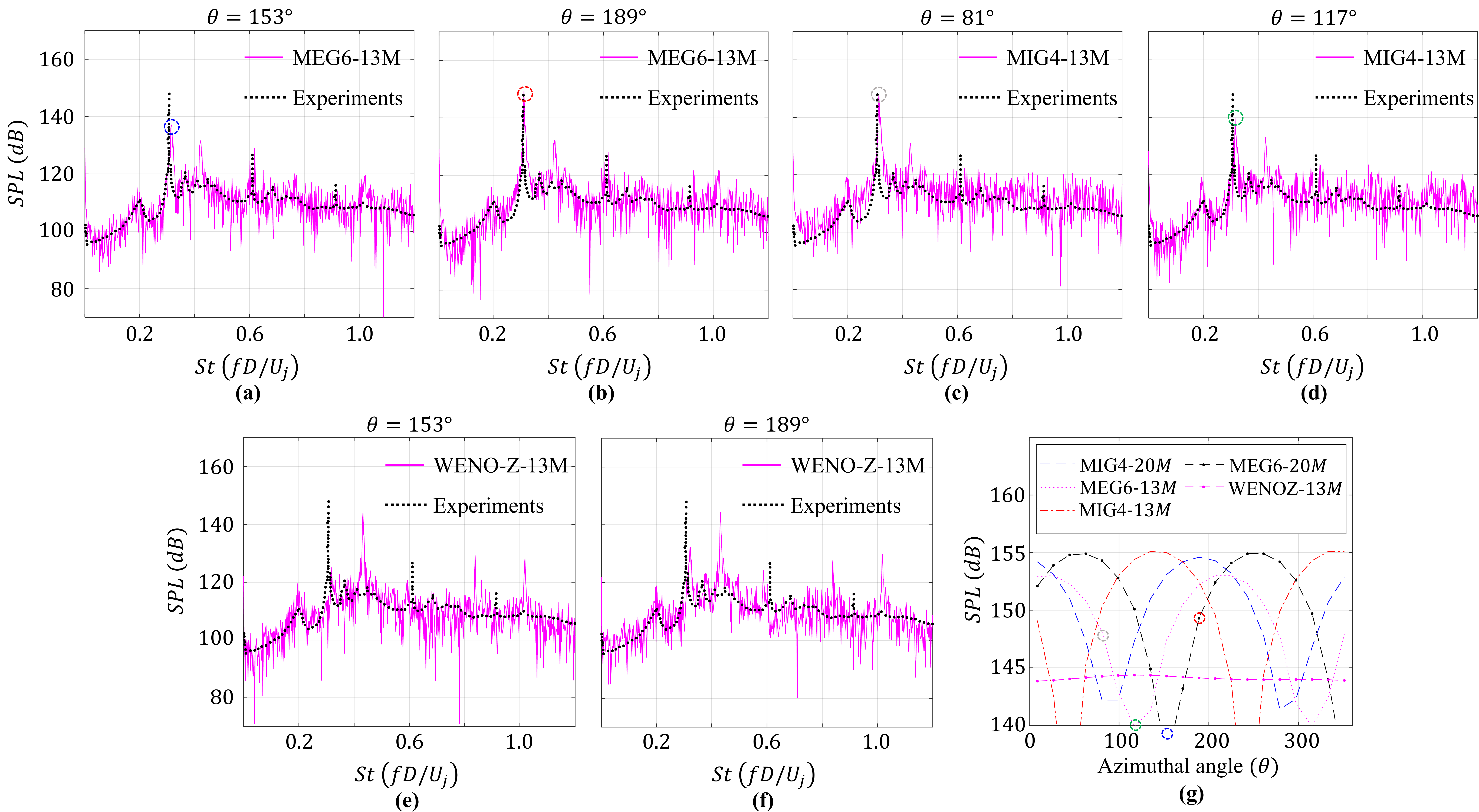}
    \caption{Comparison of pressure spectra of probe data collected at various azimuthal angles on the nozzle exit plane $2D$ away from the nozzle axis, using (a,b) MEG6, (c,d) MIG4, and (e,f) WENO-Z scheme, with the experimental data of Ponton et al. \cite{ponton1997near}. The figure also includes (g) a plot illustrating the variation of the amplitude of the fundamental tone with azimuthal angle, using data from twenty probes and the various schemes.}
    \label{jet-freqs}
\end{figure}

To evaluate the screech tones and amplitudes captured in the flowfield (Fig. \ref{jet-spl}), a Fourier analysis of the nearfield pressure data is conducted and compared to the experiments of Ponton et al. \cite{ponton1997near}. Given the non-axisymmetric nature of screech pressure fluctuations at the present jet Mach number as previously reported in studies \cite{ahn2021numerical,gojon2019antisymmetric}, time-series data is gathered at multiple azimuthal locations on the jet exit plane. A total of 20 equally spaced pressure probes (computational) are placed at two nozzle diameters from the nozzle axis on the jet exit plane, as shown in Fig. \ref{jet-freqs}e. In contrast, the experiments of Ponton et al. \cite{ponton1997near} only utilized one microphone placed at an arbitrary azimuthal location two diameters away from the nozzle axis on the jet exit plane. The Fourier analysis of the time-series data collected at various azimuthal locations is then compared to the pressure spectra obtained in the experiments \cite{ponton1997near}.

Fig. \ref{jet-freqs}(a-d) presents pressure spectra of probe data using the MEG6, MIG4, and WENO-Z schemes on a 13 million cell grid at various azimuthal locations. The accuracy of capturing the fundamental screech frequency and the first harmonic at all azimuthal locations is observed in computations performed on both 13 and 20 million cell grids. As an illustration, four sample pressure spectra are displayed at different azimuthal angles, with the peak of the fundamental tone indicated by a dashed circle. The Strouhal number remains constant across all azimuthal angles, as summarized in Table \ref{jet-table} for the fundamental screech tone at different grid resolutions using the MEG6 and MIG4 schemes, showing a good agreement with experimental results. In addition to the screech, the plots also reveal the presence of the first harmonic at around St$\approx 0.62$ and mixing noise component at St$\approx 0.2$. However, the WENO-Z scheme in Fig. \ref{jet-freqs}(e,f) exhibited a fundamental tone at St$=0.43$ which is significantly off from the experimental value. The inaccuracy can be attributed to the relatively poor spectral properties of the scheme.

Despite the constant Strouhal number across various azimuthal probe locations, variations in amplitude can be observed in Fig. \ref{jet-freqs}a-d. The screech amplitudes recorded at different azimuthal angles ($\theta$) using the MEG6, MIG4, and WENO-Z schemes are presented in Fig. \ref{jet-freqs}g. The amplitude of the WENO-Z scheme maintains a consistent level, while the MEG6 and MIG4 schemes exhibit a fluctuating wavy pattern with a difference in amplitude between peaks and troughs greater than 15dB. The physical reasoning behind this observation is due to the flapping mode of the jet which will be discussed next in Sec. \ref{sec:jet-oscill}. Table \ref{jet-table} presents the azimuthal average of screech amplitudes for different grid resolutions using the MEG6 and MIG4 schemes, and a good agreement with experimental values is observed. This supports the effectiveness of both MEG6 and MIG4 schemes in resolving supersonic jet screech `frequencies' and `amplitudes'.


\begin{table}[h!]
    \centering
    \includegraphics[width=120mm]{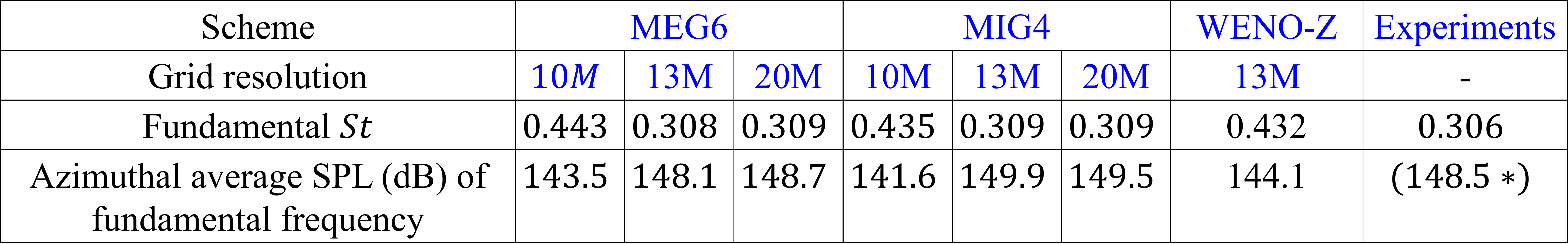}
    \caption{Summary of Strouhal numbers and average amplitudes of fundamental screech tones obtained at various grid resolutions employing MEG6, MIG4 and WENO-Z (only at 13M cell count) schemes. The screech amplitude corresponding to experiments presented in the table does not correspond to azimuthal average since only one microphone is employed in the experiments.}
    \label{jet-table}
\end{table}

\subsubsection{Unsteady jet oscillation behaviour} \label{sec:jet-oscill}

\begin{figure}[h!]
    \centering
    \includegraphics[width=\textwidth]{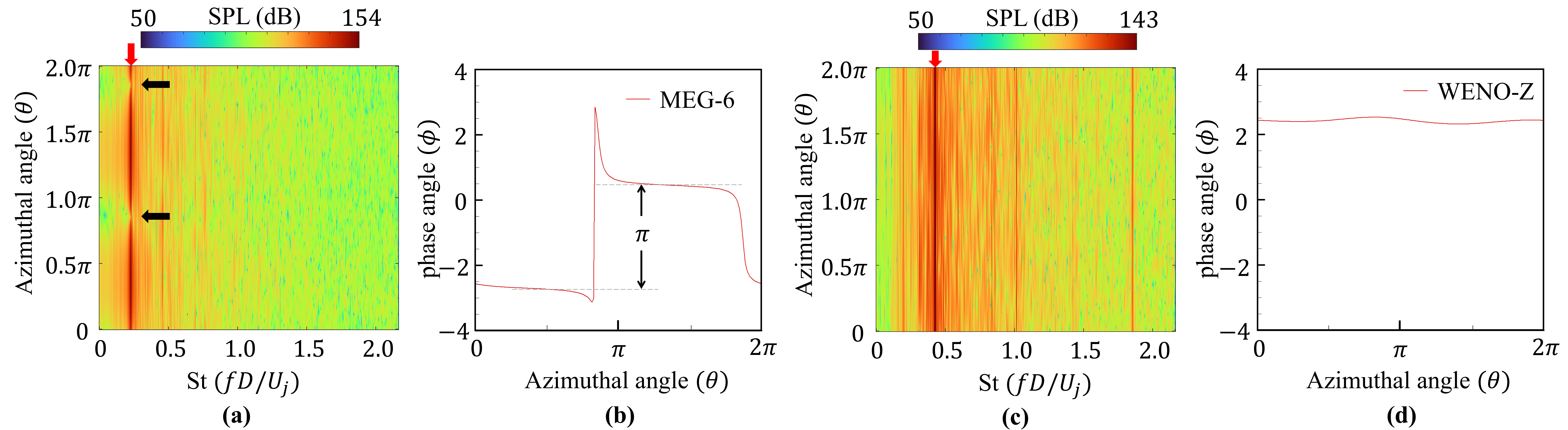}
    \caption{Pressure spectra (a,c) and phase variation of fundamental frequency $St=0.308$ (b,d) along a circular line two diameters away from the nozzle axis on the $x=-2D$ plane using MEG6 and WENO-Z schemes. The location of the fundamental frequency is marked with a red arrow on the top in figures (a) and (c). The phase shift of $\pi$ in figure (b) illustrates the flapping nature of the jet as captured by the MEG6 scheme and in the experiments of Ponton et al. \cite{ponton1997near}. Black arrows in figure (a) indicate the azimuthal location at which this phase shift occurs.}
    \label{new1}
\end{figure}

Previously in Sec \ref{freqs}, the amplitude variation depicted in Fig. \ref{jet-freqs} indicated the presence of azimuthally asymmetric jet oscillations. However, the specific oscillation mode cannot be determined from those plots alone as they correspond to a data from single point. Previous research by Powell et al. \cite{powell1992observations} has determined that the jet should exhibit flapping B mode oscillations at the current Nozzle Pressure Ratio of 2.97. In order to accurately identify the jet oscillation mode in the present simulations, Fourier analysis was conducted on two-dimensional nearfield pressure data on plane $x=-2D$, which is a jet normal plane located two diameters upstream of the nozzle exit. The results from the MEG6 and WENO-Z methods are compared to each other along with experimental observations \cite{ponton1997near,powell1992observations}.

In Fig. \ref{new1}(a,c), the frequency and amplitude of pressure data collected along a circular line located two diameters away from the nozzle axis on the $x=-2D$ plane are presented, utilizing the MEG6 and WENO-Z schemes. The ordinate of the plot represents the azimuthal angle $\theta$ along the circle. The data reveals the presence of fundamental tones (marked with red arrows on top) at Strouhal numbers of $0.308$ and $0.432$, which are consistent with the observations made in the Fourier analysis presented in Sec. \ref{freqs}. The spectrum for the WENO-Z scheme in Fig. \ref{new1}(c) appears fairly uniform at all azimuthal angles, whereas the MEG6 results in Fig. \ref{new1}(a) exhibit two notches separated by an angular distance of $\pi$ radians which are indicated by two black arrows. This suggests the presence of a plane of symmetry on the $x=-2D$ surface. The phase angle (the argument of the Fourier transformation) at the fundamental tone for the MEG6 and WENO-Z schemes is plotted in Fig. \ref{new1}(b,c). The MEG6 results show a step-like distribution with a step height equal to $\pi$ radians, which is indicative of sinuous flapping behavior of the jet, where pressure fluctuations are separated by a phase difference of $\pi$ radians about a plane of symmetry. In contrast, the WENO-Z results shown in Fig. \ref{new1}(d) exhibited a constant phase value at all azimuthal locations, indicating an axisymmetric or toroidal oscillation mode (or A1 mode) of the jet, which ideally leads to axisymmetric pressure fluctuations. To better understand these aspects noted from the 1D line data, the Fourier amplitude and phase fields of 2D data from the plane at $x = -2D$ are presented next.

\begin{figure}[h!]
    \centering
    \includegraphics[width=\textwidth]{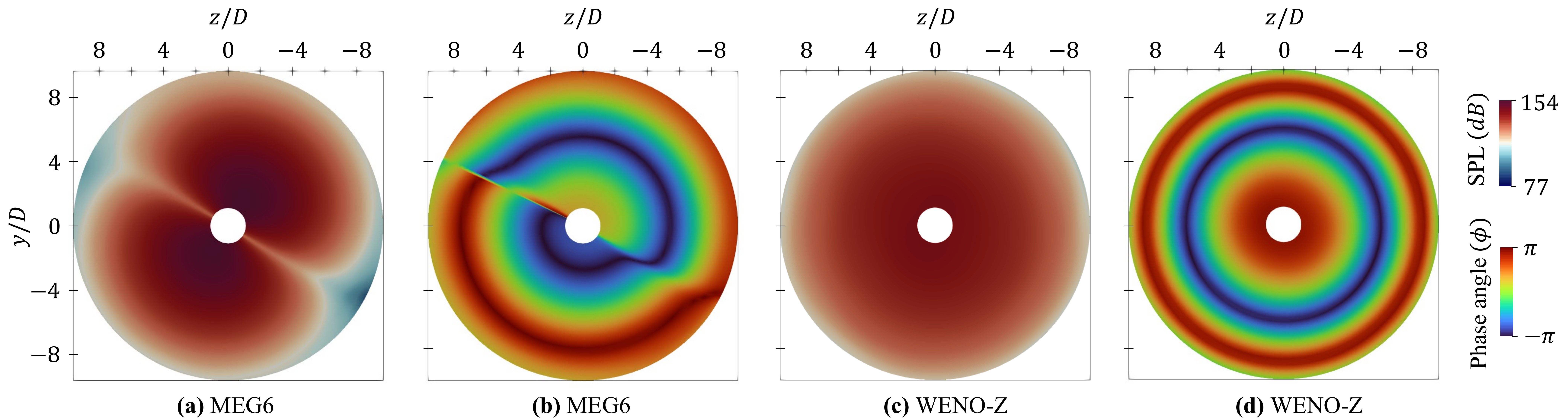}
    \caption{Fourier amplitude fields (a,c) and Fourier phase fields (b,d), corresponding to the fundamental frequency $St=0.308$ on $x=-2D$ plane using MEG6 and WENO-Z schemes. The contour distributions depicted in these figures indicate the presence of the flapping B-mode and A1-toroidal modes, captured by the MEG6 and WENO-Z schemes, respectively.}
    \label{new2}
\end{figure}

Fig. \ref{new2} illustrates the Fourier amplitudes and phase values computed on the $x=-2D$ plane at the fundamental tone frequency, as determined by the MEG6 and WENO-Z schemes. It can be observed that the results obtained using the WENO-Z scheme exhibit a clear axisymmetric distribution of both amplitude and phase, which is not consistent with the flapping behavior observed in experiments \cite{powell1992observations}. In contrast, the amplitude levels for the MEG6 scheme, as shown in Fig. \ref{new2}(a), displays a plane of symmetry, about which the jet flapping occurs. Additionally, a spiraling phase distribution, as seen in Fig. \ref{new2}(b) for the MEG6 scheme, is a clear indication of the sinuous flapping mode. This conclusively confirms the efficacy of the MEG6 scheme in resolving the Flapping B mode observed by Powell et al. \cite{powell1992observations} in their experiments.

\subsubsection{Proper Orthogonal Decomposition of density field}

Proper Orthogonal Decomposition (POD) was performed on the fluctuations of the 2D density field collected on the $xy$-plane, within a window of $x/D=[0,8]$ and $y/D=[-4,4]$, to identify the dominant unsteady coherent structures within the jet that contribute to the screech. The POD analysis in this Sec. is based on the snapshot technique described by Weiss in Ref-\cite{weiss2019tutorial} and follows a similar approach to the analysis conducted by Chandravamsi et al. \cite{chandravamsi2023control}. Firstly, a time series of `$N$' ($N=1600$ in the present case) evenly spaced instantaneous 2D density fluctuation field data (separated by $\Delta t a_{\infty}/D=0.015$) is collected and reshaped to form a 2D solution matrix named $\mathbf{A} = [\rho^{'}(x,y,t_1), \rho^{'}(x,y,t_2), .... \rho^{'}(x,y,t_N)]$. The data is reshaped in such a way that each column in $\mathbf{A}$ corresponds to a `time-instance' and each row `spatial location'. The covariance matrix $\mathbf{C}=\mathbf{A}^{T}\mathbf{A}$ was then subjected to eigenvalue decomposition to find its eigenvalues `$\Lambda_i$' and eigenvectors $\mathbf{E_i}$, forming an orthogonal basis in the eigenspace where each eigenvalue represents the variance along the corresponding eigenvector. Finally, the POD modes ($\mathbf{\Phi}_i$) and their corresponding time coefficients ($\Tilde{\mathbf{a}}$) are evaluated using the eigenvectors and input flow data ($\rho^{'}(x,y,t)$) using the following relations:


\begin{equation}
    \boldsymbol{\Phi}_i=\frac{\sum_{j=1}^N E_i^j \rho^{'}(x,y,t_j)}{\left\|\sum_{j=1}^N E_i^j \rho^{'}(x,y,t_j)\right\|}, \quad \text{and} \quad \Tilde{\mathbf{a}}_i = (\mathbf{\Phi}_i)^{T}\mathbf{A}, \quad i=1, 2 \ldots, N.
\end{equation}

\begin{figure}[h!]
    \centering
    \includegraphics[width=\textwidth]{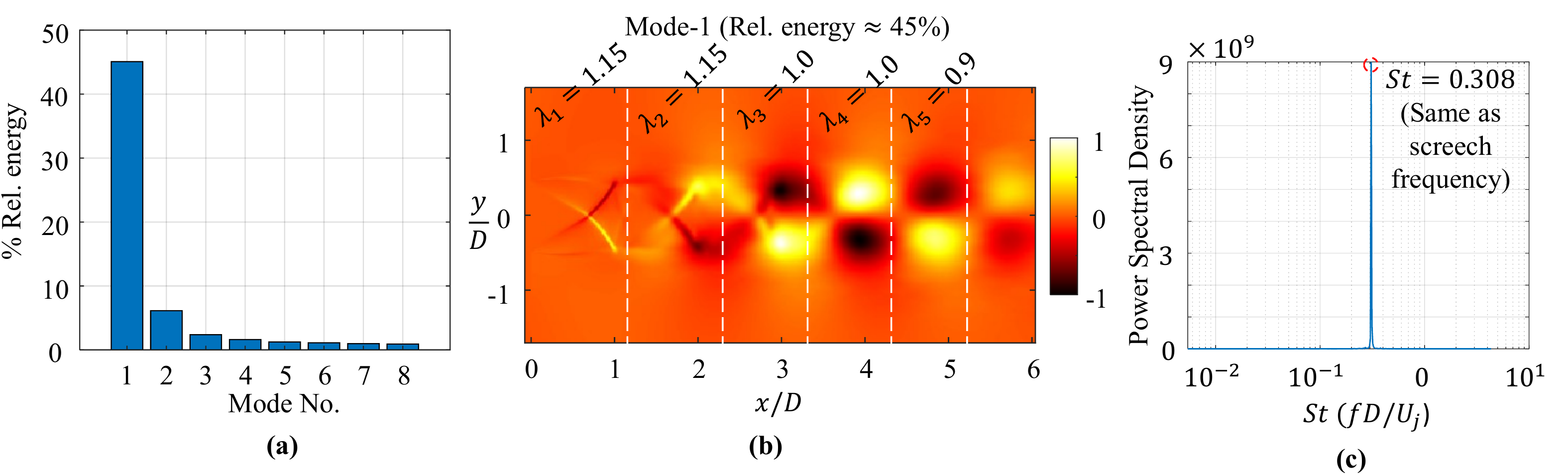}
    \caption{(a) Mode energies of first eight POD modes, (b) Mode-1 shape with shock cell locations and spacing indicated, and (c) Power Spectral Density of mode-1 POD
coefficients.}
    \label{POD}
\end{figure}

The index `$i$' here represents the mode number. The modes are ordered based on their relative mode energy, calculated as $RE_i = \Lambda_i \times \left(\sum_{i=1}^N \Lambda_i\right)^{-1}$. As depicted in Fig. \ref{POD}(a), amongst the first eight dominant modes displayed, the first mode itself accounts for $\approx 45\%$ of the total energy. A noticeable decrease in energy can be observed following the first mode, implying that only a single dominant mode exists, and the majority of the jet's unsteady dynamics are contained within mode-1. The normalized mode-1 shape is displayed in Fig. \ref{POD}(b), along with annotations describing the locations of shock cells by white dashed lines. The shock cell spacings ($\lambda$), as determined from Fig. \ref{jet-inst-ave}(b), are indicated at the top. The mode shape exhibits symmetry about the jet axis, with a reversal in sign, indicating that the jet oscillations of the most energetic portion are oriented along the $y$-axis. The dominant fluctuations are observed in the region corresponding to the third, fourth, and fifth shock cells, suggesting that the shear layer structures along the jet surface experience significant amplification after their interaction with the third shock cell. Therefore, the effective source of the screech noise can be considered to be situated between the third and fifth shock cells. As depicted in Fig. \ref{POD}(c), the Fourier analysis of the mode-1 time coefficients revealed a ``pure tone'' that precisely matches the screech frequency ($St = 0.308$) determined from the acoustic data in Sec. \ref{freqs}. This suggests that the oscillation frequency of the oblique shocks within the jet plume and the turbulent structures between the third and fifth shock cells are all operating at the same frequency as the screech.

\newpage
\section{GPU acceleration model and speedup analysis} \label{sec:gpu-accel}

This section discusses the GPU parallelization model employed in the present work to implement the MEG6 and MIG4 schemes along with the performance results achieved on various GPUs. In presenting the performance details, a particular emphasis was laid on subjects such as `the contribution of various functions to the overall computational time', `the effect of cell count', `memory occupancy versus cell count', and `the effect of working precision'.

In order to specify the parallelism on GPUs, the directive based `OpenACC' programming language \cite{openacc-web1} is used in the present work. The use of OpenACC directives over the more popular `CUDA' programming language is motivated by less code development time associated with OpenACC while still being able to achieve beneficial speedup results that are comparable to CUDA, as will be demonstrated soon. The details of OpenACC implementation are being skipped here for brevity. Interested readers are refered to the OpenACC API guide \cite{openacc-web2} and other resources available in their website \cite{openacc-web1}.

\begin{figure}[h!]
    \centering
    \includegraphics[width=\textwidth]{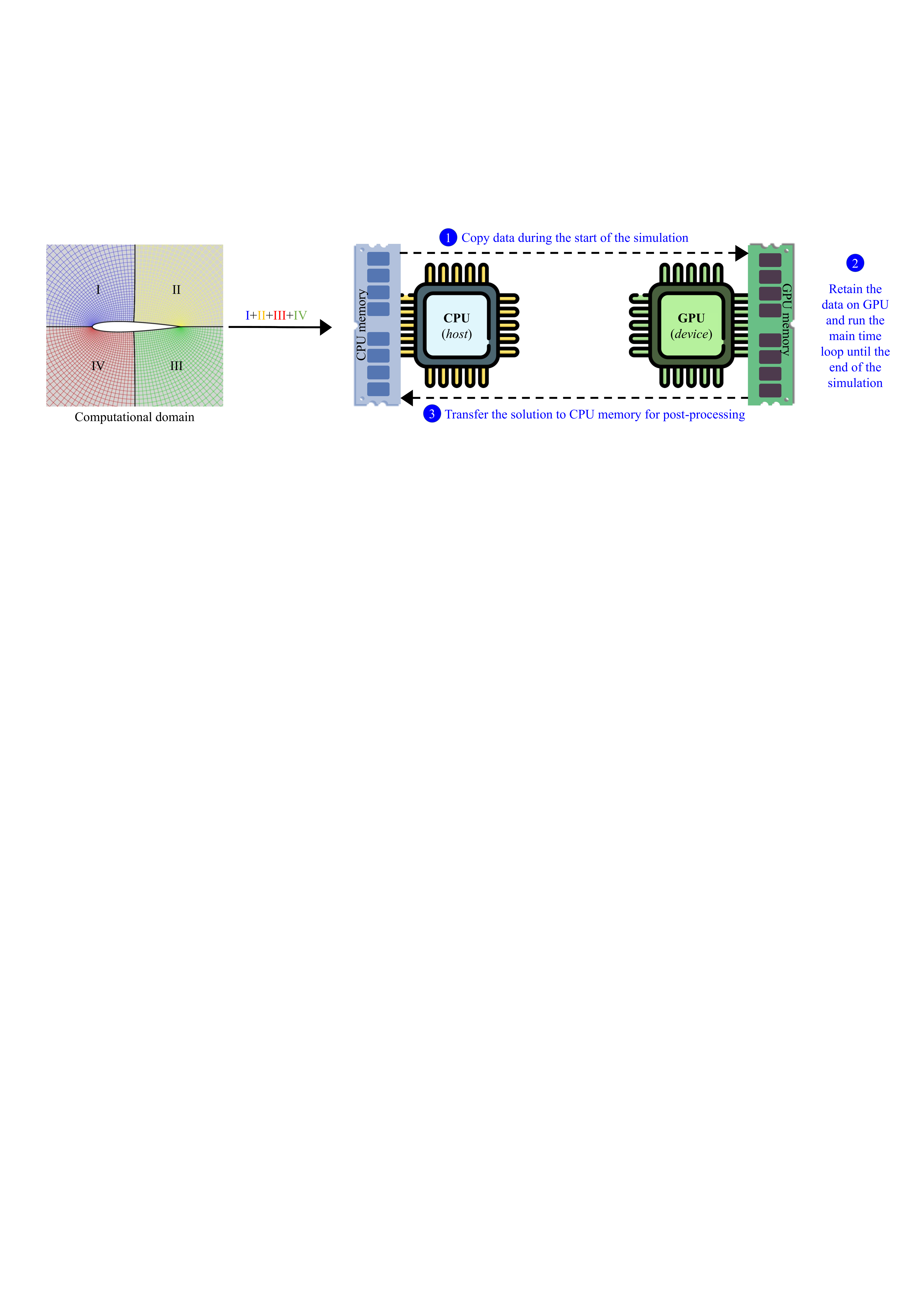}
    \caption{Schematic diagram of the parallelization model employed for single device GPU computations. Shown in the left of the figure, is an example multi-block computational domain (of an airfoil) whose geometrical and initial conditions data is distributed to the host memory. The three steps involved in the parallelization model are shown in blue circles with numbers.}
    \label{gpu-strategy1}
\end{figure}

A high-level view of the computing model adapted for single GPU computations is pictorially illustrated in Fig. \ref{gpu-strategy1}. It consists of a CPU-GPU pair connected via a PCI Express bus or an NVLink interconnect; the latter option which is an NVIDIA proprietary bus is designed to offer a higher bandwidth. The program is hosted by the CPU (or called `host'), which initializes the compute environment, reads the geometry, mesh, and other input variables. The compute routines of the program are off-loaded onto the GPU; GPU is refereed to as `device'. Fig. \ref{gpu-strategy1}, shows the three elemental steps involved in performing a single device GPU simulation. The first step before starting the main time loop involves creating a copy of all the necessary data, such as mesh, initial conditions, and the flow parameters on the device memory. Since scientific computing based programs such as present are mainly bandwidth bounded, this step is done as a one time act to restrict the communication between the host and device during the simulation. It is followed by executing all the routines that are part of the main time loop listed in Fig. \ref{algo} until the simulation reaches its end-time. Once the computations inside the main time loop are completed, the solution is transferred to the host memory (step-3 in Fig. \ref{gpu-strategy1}), from where it can be viewed and post-processed in the end. But often in many cases, the user requires to store the solution corresponding to intermediate time-steps for temporal analysis. In such a scenario, only the primitive variable data is transferred to the host whenever necessary. Such communication during the middle of a simulation, if carried out just once per hundred iterations or more, was noted to add only a negligible amount of computational time to the overall simulation.

Now that the parallelization strategy is outlined, the performance details achieved employing both MEG6 and MIG4 schemes will be presented. All the computations are performed using double precision by default unless any other working precision is specified. The high performance `Intel(R) Xeon(R) Gold 5115 CPU @ 2.40GHz' is used for all the CPU computations. For GPU computations, three generations of data center GPUs, namely, NVIDIA Quadro A6000, NVIDIA V100, and NVIDIA A100 are employed to present a comparative study of their efficiencies. These GPUs are referred to as A6000, V100 and A100 for short in this paper. The acceleration offered by a GPU is mainly dependent on three key hardware attributes which are listed in table \ref{gpuspecs}.


\begin{table}[h!]
    \centering
    \includegraphics[width=120mm]{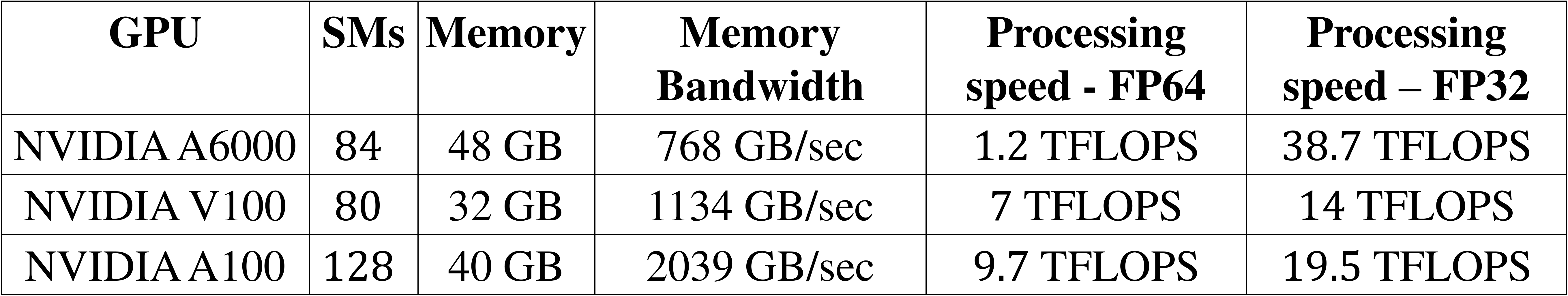}
    \caption{Key hardware specifications of GPUs used in the current study.}
    \label{gpuspecs}
\end{table}


\subsection{Speedup comparison to single core CPU} 
Tables \ref{gpu_table1} and \ref{gpu_table2} show the elapsed simulation times per hundred iterations on the CPU and various GPUs. To access the performance, three test cases are employed. The first is the two-dimensional Inviscid Double Mach Reflection (DMR) case (described in section \ref{DMR-case}) consisting of 3 million cells. The second is the three-dimensional Viscous Taylor Greene Vortex (TGV) case (details of the test case can be found in \cite{sainadh2022spectral}) consisting of 27 million cells. The relative difference in the grid resolution between the two cases and the fact that the viscous TGV test case has extra routines to be executed (thus higher exposed parallelism) is planned intentionally to note the performance variation between two such contrasting compute scenarios. To also understand the speedup on a practically relevant multi-block test case, the supersonic jet noise case discussed in Sec. \ref{sec:jet} is simulated at a grid resolution of 13 million.

\begin{table}[h!]
    \centering
    \includegraphics[width=155mm]{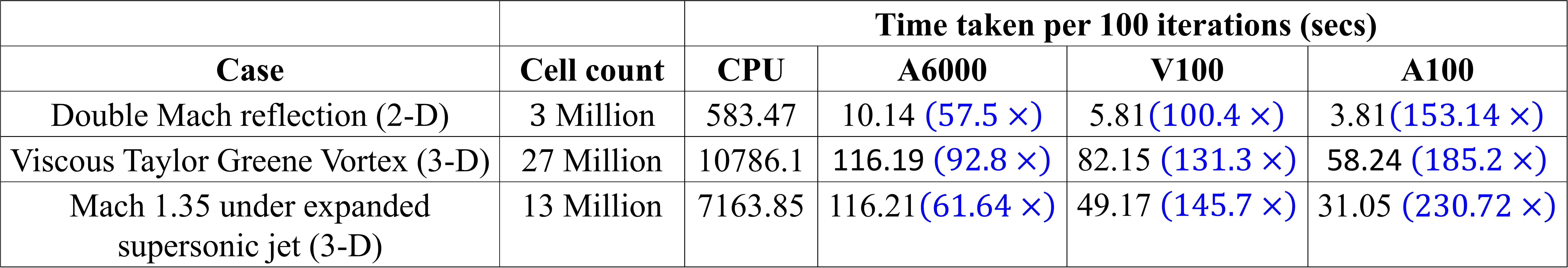}
    \caption{Comparison of time taken per 100 iterations on CPU versus various Nvidia GPUs employing MEG6 scheme.}
    \label{gpu_table1}
\end{table}

\begin{table}[h!]
    \centering
    \includegraphics[width=155mm]{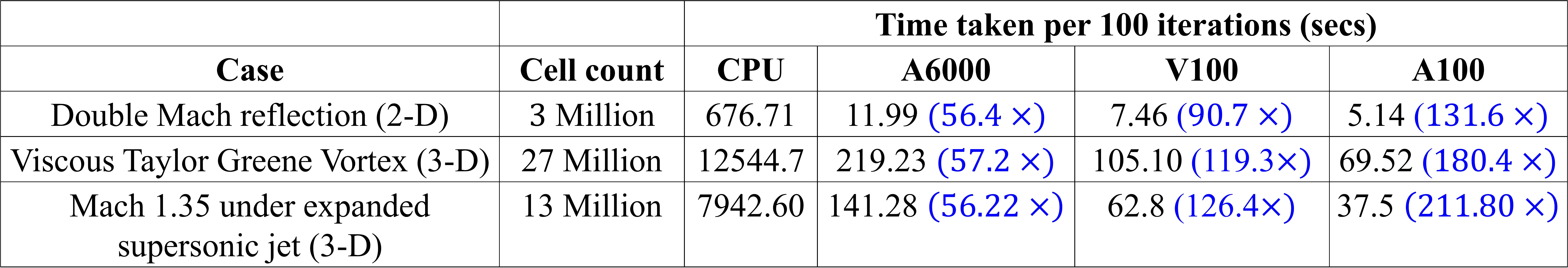}
    \caption{Comparison of time taken per 100 iterations on CPU versus various Nvidia GPUs employing MIG4 scheme.}
    \label{gpu_table2}
\end{table}

\begin{table}[h!]
    \centering
    \includegraphics[width=155mm]{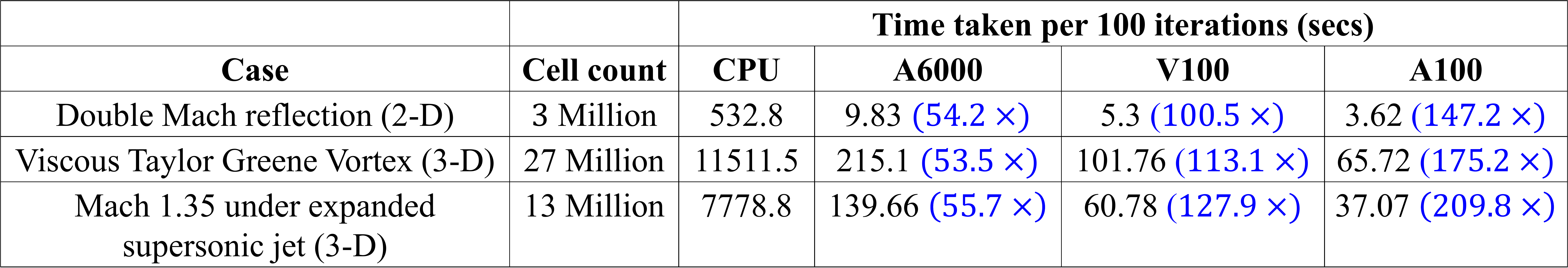}
    \caption{Comparison of time taken per 100 iterations on CPU versus various Nvidia GPUs employing WENO-Z scheme.}
    \label{gpu_table3}
\end{table}

The results presented in Tables \ref{gpu_table1}, \ref{gpu_table2}, and \ref{gpu_table3} suggest that the computations performed on V100 and A100 are about two orders of magnitude faster than the calculations performed on CPU. These speedup results can be noted to be on par with the CUDA based parallelization models implemented in other studies, for instance Refs-\cite{cernetic2022high,crespo2015dualsphysics,laufer2022gpu}. Amongst the GPUs, the A100 can be noted to perform the best by producing a maximum speedup of greater than $200\times$ for both MEG6 and MIG4 schemes. V100 stands next in terms of performance, followed by A6000 in the last position. This is consistent with the hardware specifications presented in table \ref{gpuspecs}; the double precision processing speed and memory bandwidth of Nvidia A100 is the highest which is followed by V100 and A6000. The effect of superior single precision speed corresponding to A6000 on the simulation times will be discussed later in section \ref{effect-WP}. 

The results also indicate that the speedup of the simulation is influenced by its parallelism and size. The viscous TGV simulation at 27 million cell count was found to run efficiently compared to the 3 million cell count inviscid DMR test case. The MEG6 scheme was the most efficient among the three schemes tested, due to its explicit derivative computations as opposed to implicit derivatives being used in MIG4 and WENO-Z tests. Since WENO-Z does not use gradients for inviscid flux calculations, for the inviscid DMR case, the WENO-Z scheme was the fastest. However, for the viscous cases (viscous TGV and supersonic jet), the computational time of WENO-Z and MIG4 were similar, with MIG4 being slightly more efficient. Although the computational time of WENO-Z and MIG4 are close for viscous cases, the solution resolution is significantly different, with MIG4 being superior.


\subsection{Contribution of various routines to the overall computational time}
\begin{figure}[h!]
    \centering
    \includegraphics[width=\textwidth]{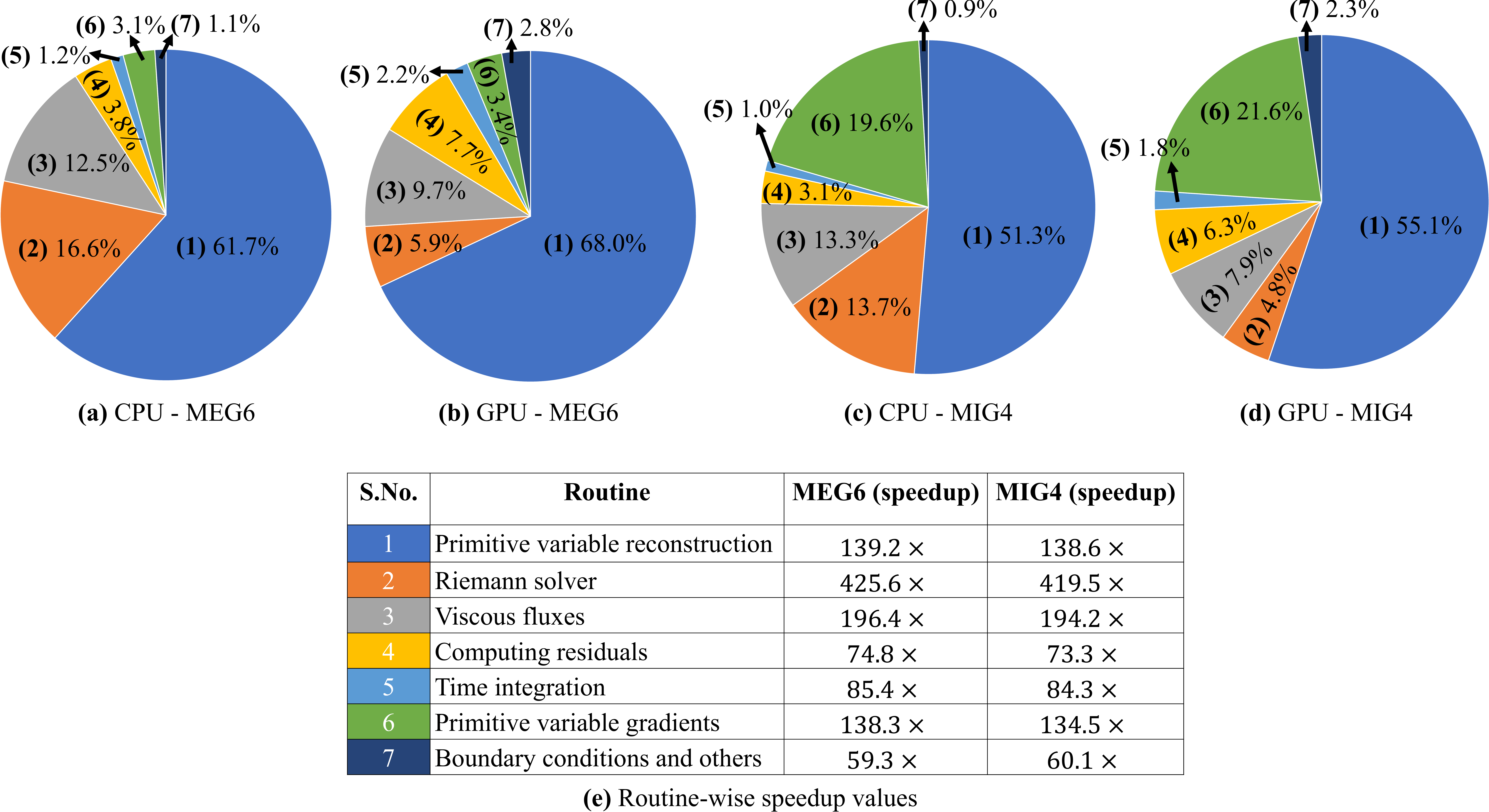}
    \caption{Contribution of various routines to the total computational time while executing one RK-time step, corresponding to (a) MEG-6, (b) MIG-4 schemes tested using Viscous Taylor Greene Vortex test case ($256\times256\times256$) on Nvidia A100 GPU.}
    \label{gpu1}
\end{figure}

Fig. \ref{gpu1}a-d shows a pie-chart view representing the contribution of various routines to the overall computational time in performing one RK-iteration of MEG6 and MIG4 algorithms on both CPU (Intel Xeon Gold 5115) and GPU (Nvidia A100). All the tests were run using the Viscous Taylor-Greene-Vortex case at a grid resolution of $256^3$. The plots suggest that more than 50\% of the time on both CPU and GPU is spent just performing the `reconstruction routine' (indicated with number `1') in all the scenarios. This is mainly because of the multiple sub-steps involved in the reconstruction routine described in \ref{inv_algo}. Furthermore, the contribution of primitive variable gradients (indicated with the number `6') can be noted to occupy a more significant portion of the pie in the MIG4 scheme compared to MEG6. This is due to the extra tri-diagonal matrix inversion step involved in the MIG4 algorithm. Apart from reconstruction and gradient computation routines, the other routines can be noted to occupy a relatively smaller portion of the pie due to the relatively simple nature of the corresponding kernels.

To understand the efficiency of parallelization corresponding to each routine, the individual speedup values for each routine achieved for both MEG6 and MIG4 are computed and tabulated in Fig. \ref{gpu1}e. Interestingly, the speedup values corresponding to MEG6 and MIG4 are nearly similar for all the routines, including the `primitive variable gradient computation'. The `Riemann solver' routine can be noted to run with highest efficiency achieving a speedup of about $\approx 420 \times$ in both MEG6 and MIG4 approaches. On the other hand, the least speedup is noted for the residual computations, time integration, and the boundary condition routines.

\subsection{Effect of cell count on the speedup}
\begin{figure}[h!]
    \centering
    \includegraphics[width=100mm]{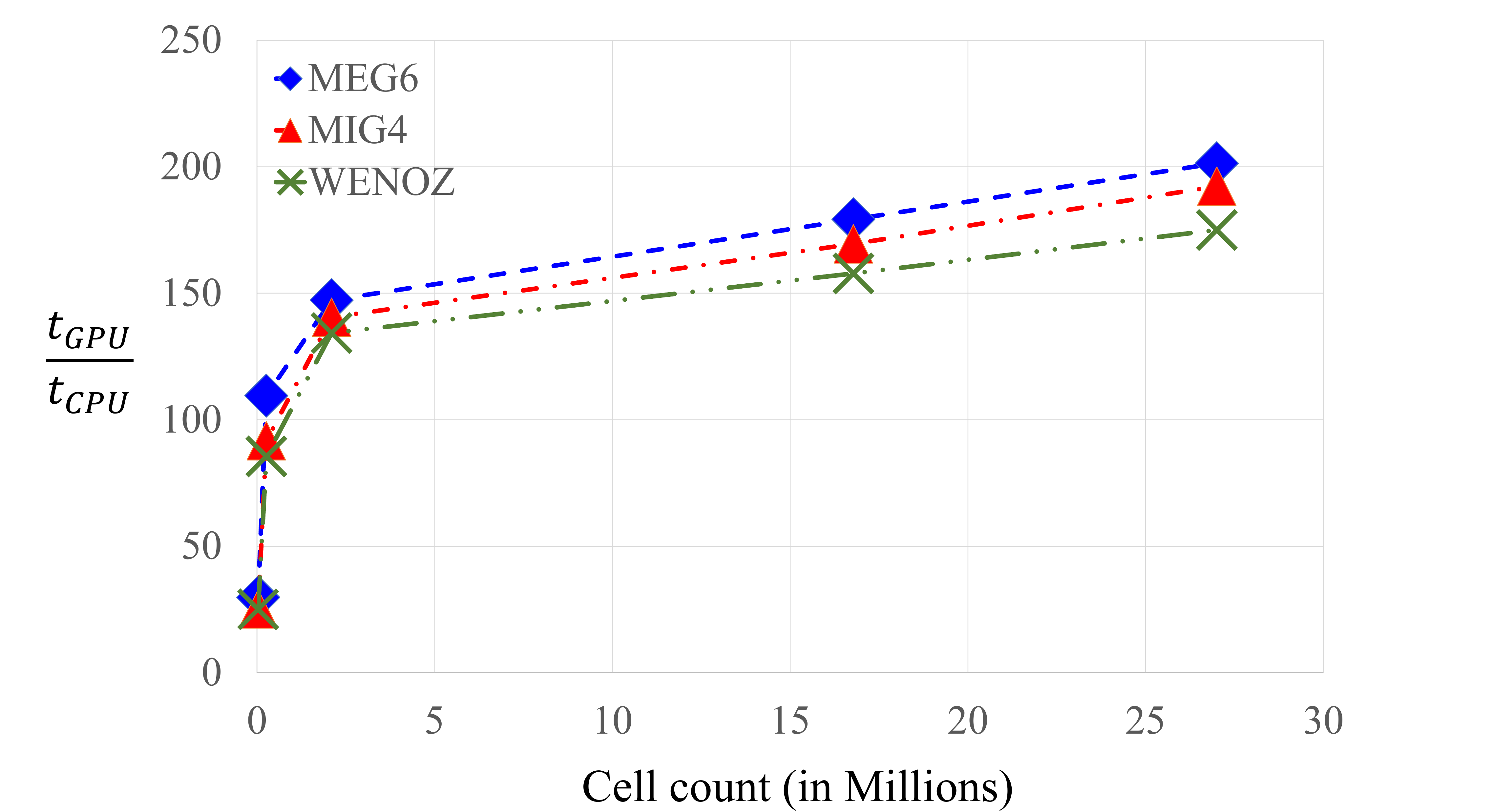}
    \caption{Speedup efficiencies of MEG6, MIG4, and WENO-Z schemes with increasing grid size on the Nvidia A100 GPU with reference to single-core CPU.}
    \label{gpu3}
\end{figure}

Variation of speedup with increasing cell count is studied relative to the computations on a single core CPU. The test runs were performed on the Nvidia A100 using the Viscous Taylor Greene Vortex test case. The speedup was measured by taking the ratio of time taken per hundred iterations on a single GPU to that of a single core CPU. To avoid GPU memory over-subscription (40GB - A100), the simulations were not carried out beyond 27 Million cells. Fig. \ref{gpu3} shows a monotonically increasing trend in the efficiency for all the three schemes tested. However, due to the implicit nature of the derivative computation involved in the MIG4 and WENO-Z schemes, their parallel efficiency turned out to be slightly lower compared to MEG6. In comparison, WENO-Z was noted to produce the least amount of efficiency, which can be acknowledged by comparing the curves in Fig. \ref{gpu3} near the 27 million range.


The curves shown in Fig. \ref{gpu3} suggest that the rate of rise in speedup is considerably high in the initial range of $0$ to $2.5$ million cells. Although the speedup still increases beyond the $0-2.5$ million range, the curve gets progressively plateaued with increasing cell count. This trend can be explained as follows. Initially, when the cell count is increased, the threads in the GPU are progressively occupied, resulting in a steady and almost linear rise in the speedup until all the GPU threads are fully occupied. This maximum GPU occupancy reaches at close to two million cells in the specific case studied here. However, when the cell count is further increased, the kernel divides the program loops into multiple sets to run each of them in parallel resulting in multiple computation cycles within the kernel. This effect manifests as a slow rise in the parallel efficiency after a certain cell count limit. Despite this plateauing effect, keeping the GPU unit as occupied as possible is always a good practice to extract the highest possible efficiency, especially keeping in view the net power consumption when multiple GPUs are employed.

\subsection{Memory occupancy of the solver on GPU} \label{effect-WP}

\begin{figure}[h!]
    \centering
    \includegraphics[width=92mm]{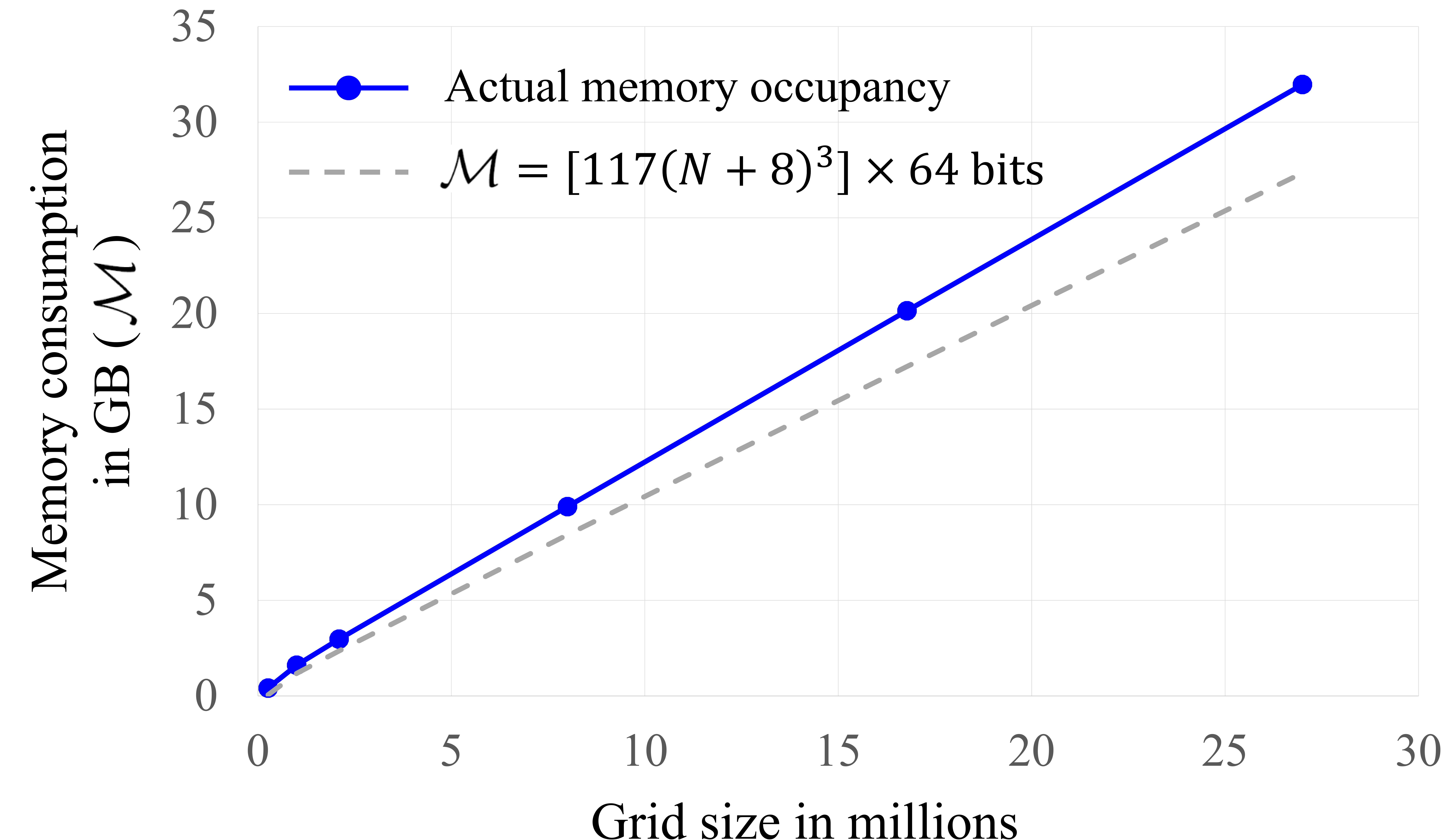}
    \caption{The GPU memory consumption with respect to variation in the grid size of the simulation, as demonstrated by tests performed on an NVIDIA A100 with 40GB RAM using the Viscous Taylor-Green Vortex test case. The gray dashed line corresponds to the approximate estimate of memory consumption based on the number of global variables defined in the solver.}
    \label{mem_cons}
\end{figure}

The capability to run a simulation of given size on a GPU is largely determined by the available RAM on the device. Memory utilization is a crucial factor for finite difference algorithms based on curvilinear coordinates, as they require multiple variables representing metric terms to be defined at each cell center and interface. The memory footprint of a three-dimensional solver scales approximately as $O(N^3)$, where $N$ represents the number of cells in each direction, and for simplicity, it is assumed to be equal in all directions.

Fig. \ref{mem_cons} displays the memory occupancy ($\mathcal{M}$) of the flow solver at different grid resolutions. The plot also includes a rough estimate of the expected memory consumption ($\mathcal{M}=\left[117(N+8)^3\right] \times 64 \text { bits }$) derived based on the number of variables in the solver. The solver uses $117$ global 3D double-precision arrays, each of size $(N+8)^3$ (including ghost cells), excluding small scalars and arrays. The discrepancy between the blue line and gray dashed line in Fig. \ref{mem_cons} is due to arrays/variables defined in each function that consume memory on GPU RAM. For the present flow solver (employing MEG6 or MIG4), tests show that a Nvidia A100 (40GB) can handle simulations with up to 34 million cells without any decrease in performance or memory over-subscription issues. Among the various data variables, metric terms consume about $60\%$ of the total memory as they are needed at cell centers and six-cell interfaces surrounding each cell. $71$ out of a total $117$ variables belong to the metric terms such as $(\xi_{x})_i$, $(\xi_{x})_{i+\frac{1}{2}}$, $(\hat{\xi}_{x})_{i+\frac{1}{2}}$, and $(\Tilde{\xi}_{x})_{i+\frac{1}{2}}$.

\subsection{Effect of working precision on the speedup} \label{effect-WP}

The floating point representation in computer calculations can influence the computational time as it can directly affect the number of arithmetic operations and the overall memory movement required to complete a task. A 32-bit/single-precision memory allocation (FP32) is generally used for tasks where the accuracy of calculations is not critical and the tasks are bounded by limitations in allocated memory. On the other hand, a 64-bit/double-precision (FP64) allocation is usually employed for tasks like scientific calculations where accuracy is needed. Although FP64 is widely used for CFD simulations, a few studies show that the working precision does not significantly influence the solution and the corresponding physics. One can also improve the simulation turnaround times by employing FP32 for the transient phase of simulation (where the flowfield is not of interest to the user) and switching back to FP64 for the rest of the flowfield calculations when the accuracy of the solution is essential.

\begin{figure}[h!]
    \centering
    \includegraphics[width=150mm]{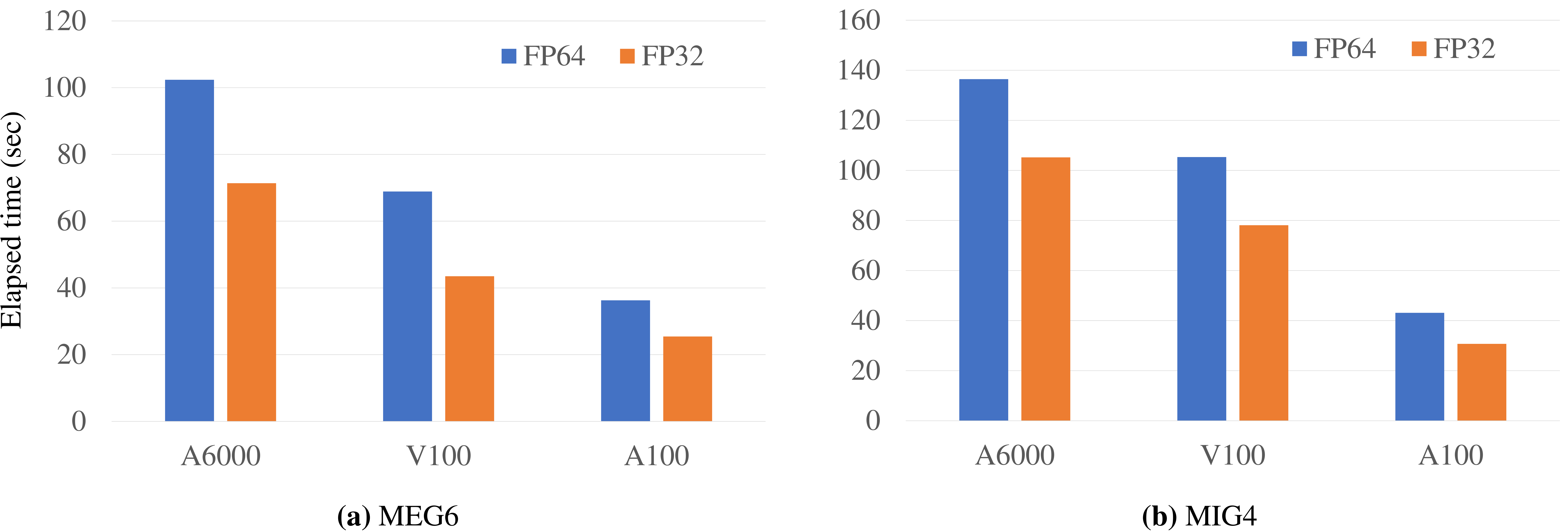}
    \caption{Comparison of elapsed times using Single Precision (FP32) and Double Precision (FP64) data variables.}
    \label{gpu-SP-DP}
\end{figure}

Fig. \ref{gpu-SP-DP} compares FP32 and FP64 computational times for both MEG6 and MIG4 schemes on various GPUs. The tests were performed by running one hundred iterations of the Viscous Taylor Greene vortex test case at $256^3$ resolution. The simulations that use FP32 can be noted to run faster than FP64 by a factor of $\approx 1.3\times$ to $1.4\times$ contrary to the anticipated $2\times$ speedup that is suggestive from Table \ref{gpuspecs}. There is a high discrepancy between the FP32 processing speed of A6000 displayed in table \ref{gpuspecs} and the FP32 speedup achieved here. Furthermore, the FP32 calculations on A6000 can be noted to only run almost as fast as FP64 calculations on the V100. This suggests that rather than the processing speed, the computations are primarily bounded by the memory bandwidth of GPU. Which means that the overall simulation time is not only spent in performing the floating point operations but also on other essential tasks such as initializing the kernels, moving data across GPU RAM and various levels of cache memory storage, etc. It is worth noting that scientific computing applications like the present solver are generally bounded by memory bandwidth as opposed to arithmetic operations. Therefore, the working precision does not hugely affect the performance, as demonstrated here. Nevertheless, the speedup from the employment of FP32 can still be used to accelerate the simulations.

\section{Conclusions}

Various aspects of extending the `MEG6/MIG4' and `$\alpha$-damping' schemes (proposed by Chamarthi (2022) in Ref. \cite{chamx}) to simulate flows over curvilinear, multi-block meshes are presented in this paper. The first part of this paper discusses the extension of MEG6/MIG4 and $\alpha$-damping schemes to the generalized curvilinear coordinates. Various steps required to achieve freestream and vortex preservation on static and dynamic low-quality stretched and skewed grids are presented with demonstrative examples. The theoretical and numerical results showed that the freestream preservation property was satisfied by the use of central schemes to compute metrics that were derived consistently from the corresponding upwind schemes MEG6 and MIG4 (averaging left and right states). The efficacy of MEG6 and MIG4 schemes to run simulations on dynamically deforming meshes is also illustrated using both inviscid and viscous test cases. The results show that employing conservative metric terms that are computed and interpolated consistently using the same scheme as that of the inviscid flux discretization will ensure freestream preservation on both stationary and moving grids. \\



The second part demonstrates the efficacy of MEG6 and MIG4 schemes in resolving supersonic jet screech. Both the schemes were noted to perform well in terms of resolving the essential flow features necessary to simulate the screech. The predicted screech frequencies and screech amplitudes of a round Mach $1.35$ under-expanded jet are noted to be in great agreement with the experimental measurements. Furthermore, aspects concerning the jet oscillation mode and screech amplitude variation along the azimuthal direction are elucidated. In the particular case studied, it was noted that the nearfield screech amplitude can differ by almost $15$dB (varies between $155$dB to $140$dB) depending on the azimuthal location of the observer. \\

The third part of the paper discusses the GPU acceleration model employed for single GPU computations. The benchmark speedup tests conducted on different GPUs show that the simulations can be run up to $\approx 200$ times faster on a single NVIDIA A100 GPU card compared to a single-core Intel Xeon Gold CPU. The achieved speedup statistics were noted be on par with some of the CUDA based implementations reported in the literature. In the context of the MEG6/MIG4 schemes, this study also investigates several aspects of GPU acceleration through the use of example test cases, including the `speedup efficiencies of different GPU hardware,' `compute loads of various functions,' `memory occupancy,' `the impact of cell-count on speedup,' and the `influence of working precision.' The supersonic jet noise simulation (13 million cells, 400,000 iterations, MEG6 scheme) presented in the paper was completed in a reasonable turnaround time of 34.5 hours on a single A100 GPU card.

The current approach however has a limitation in its inviscid scheme which only achieves second-order accuracy for non-linear test cases, as depicted in Ref. \cite{chamx}. Despite efforts to improve the MP limiter, it may still get activated frequently in the regions that doses not correspond to shocks. The ongoing work to address these issues will be reported in a separate publication. The readers should note that, despite its low-order nature, the present method outperforms a truly fifth-order scheme, as shown in Ref. \cite{chamx}.

\section*{Supplementary material} \label{sec:supply}
\begin{enumerate}
    \item \textbf{Notebook-1:} \href{https://drive.google.com/file/d/1KlICDIN3MvDogqhGxzNNohNTtJfqIW41/view?usp=sharing}{\underline{link}} - Mathametica notebook proving the metric identities through symbolic manipulation employing MEG6 based derivatives and interpolation.
    \item \textbf{Movie-1:} \href{https://drive.google.com/file/d/1LRFdi6fLINZTarDmMU-yvyAG6zTGXr81/view?usp=sharing}{\underline{link}} - Density contours of Mach 1.35 under-expanded supersonic jet corresponding to Fig. \ref{jet-inst-ave}c.
\end{enumerate}

\appendix

\section{Primitive variables form of 3-D Euler equations and its eigen Structure} \label{app:Eig-structure}
The eigenvectors used for characteristic transformation employed during the reconstruction procedure in the MEG6 and MIG4 schemes (section \ref{Inv-disc}) are presented in this section. Firstly, removing the viscous terms from the Navier-Stokes equations (Eqn. \ref{trans-eqn}) we obtain the following inviscid counterpart.

\begin{equation}
    \frac{\partial}{\partial t}\left(\frac{\mathbf{Q}}{J}\right)+\frac{\partial \hat{\mathbf{F}}}{\partial \xi}+\frac{\partial \hat{\mathbf{G}}}{\partial \eta}+\frac{\partial \hat{\mathbf{H}}}{\partial \zeta}=0
\end{equation}\label{EulerGC}

Following a series of algebraic manipulations, the above equation can be transformed into its primitive variables form with $\mathbf{P}=[\rho,u,v,w,p]^T$ present in the time, and spatial derivative terms \cite{masatsuka2013like}.

\begin{equation}\label{prim-form}
    \frac{\partial \mathbf{P}}{\partial t}+\overline{\mathbf{B}}_{\xi} \frac{\partial \mathbf{P}}{\partial \xi}+\overline{\mathbf{B}}_{\eta} \frac{\partial \mathbf{P}}{\partial \eta}+\overline{\mathbf{B}}_{\zeta} \frac{\partial \mathbf{P}}{\partial \zeta}=0
\end{equation}

where the coefficient matrices $\overline{\mathbf{B}}_{\xi}$, $\overline{\mathbf{B}}_{\eta}$ and $\overline{\mathbf{B}}_{\zeta}$ are,

\begin{equation*}
\overline{\mathbf{B}}_{\xi}=\frac{1}{J}\left[\begin{array}{ccccc}
U & \rho \xi_{x} & \rho \xi_{y} & \rho \xi_{z} & 0 \\
0 & U & 0 & 0 & \frac{\xi_{x}}{\rho} \\
0 & 0 & U & 0 & \frac{\xi_{y}}{\rho} \\
0 & 0 & 0 & U & \frac{\xi_{z}}{\rho} \\
0 & \gamma p \xi_{x} & \gamma p \xi_{y} & \gamma p \xi_{z} & U
\end{array}\right], \quad
\overline{\mathbf{B}}_{\eta}=\frac{1}{J}\left[\begin{array}{ccccc}
V & \rho \eta_{x} & \rho \eta_{y} & \rho \eta_{z} & 0 \\
0 & V & 0 & 0 & \frac{\eta_{x}}{\rho} \\
0 & 0 & V & 0 & \frac{\eta_{y}}{\rho} \\
0 & 0 & 0 & V & \frac{\eta_{z}}{\rho} \\
0 & \gamma p \eta_{x} & \gamma p \eta_{y} & \gamma p \eta_{z} & V
\end{array}\right] \text{and}
\end{equation*}

\begin{equation*}
\overline{\mathbf{B}}_{\zeta}=\frac{1}{J}\left[\begin{array}{ccccc}
W & \rho \zeta_{x} & \rho \zeta_{y} & \rho \zeta_{z} & 0 \\
0 & W & 0 & 0 & \frac{\zeta_{x}}{\rho} \\
0 & 0 & W & 0 & \frac{\zeta_{y}}{\rho} \\
0 & 0 & 0 & W & \frac{\zeta_{z}}{\rho} \\
0 & \gamma p \zeta_{x} & \gamma p \zeta_{y} & \gamma p \zeta_{z} & W
\end{array}\right] .
\end{equation*}

For the characteristic variable projection used in the reconstruction procedure of the inviscid flux computation algorithm (Eqns. \ref{forward-projection} and \ref{rev-projection}), the eigenvectors of the coefficient matrices $\overline{\mathbf{B}}_{\xi}$, $\overline{\mathbf{B}}_{\eta}$ and $\overline{\mathbf{B}}_{\zeta}$ are used. The eigen structure of the coefficient matrices can be found by splitting them using eigen decomposition. The following three eigenvalue problems must be solved to find the corresponding eigen matrices. 

\begin{subequations}
    \begin{gather}
        \overline{\mathbf{B}}_{\xi} \overline{\mathbf{R}}_{\xi}=\overline{\mathbf{R}}_{\xi} \bar{\Lambda}_{\xi}, \quad \overline{\mathbf{B}}_{\eta} \overline{\mathbf{R}}_{\eta}=\overline{\mathbf{R}}_{\eta} \bar{\Lambda}_{\eta}, \text{ and} \quad \overline{\mathbf{B}}_{\zeta} \overline{\mathbf{R}}_{\zeta}=\overline{\mathbf{R}}_{\zeta} \bar{\Lambda}_{\zeta}.
        \tag{\theequation a-\theequation c}
    \end{gather}
\end{subequations}

Where, $\overline{\mathbf{R}}_{\xi}$, $\overline{\mathbf{R}}_{\eta}$, and $\overline{\mathbf{R}}_{\zeta}$ are the right eigen vector matrices and $\bar{\Lambda}_{\xi}$, $\bar{\Lambda}_{\eta}$, and $\bar{\Lambda}_{\zeta}$ are the eigen value matrices which are block diagonal in nature.

%

Solving the above mentioned eigen value problems, the following left and right eigen vector and eigen value matrices are obtained.

\begin{subequations}
    \begin{gather}
        \overline{\boldsymbol{\Lambda}}_{\xi}=\left[\begin{array}{ccccc}
        U-a_{\xi} & 0 & 0 & 0 & 0 \\
        0 & U & 0 & 0 & 0 \\
        0 & 0 & U & 0 & 0 \\
        0 & 0 & 0 & U & 0 \\
        0 & 0 & 0 & 0 & U+a_{\xi}
        \end{array}\right], \quad  \overline{\mathbf{R}}_{\xi}=\left[\begin{array}{ccccc}
        \frac{1}{2 a^{2}} & \frac{\tilde{\xi}_{x}}{a^{2}} & \frac{\tilde{\xi}_{z}}{a^{2}} & -\frac{\tilde{\xi}_{y}}{a^{2}} & \frac{1}{2 a^{2}} \\
        -\frac{\tilde{\xi}_{x}}{2 \rho a} & 0 & -\tilde{\xi}_{y} & -\tilde{\xi}_{z} & \frac{\tilde{\xi}_{x}}{2 \rho a} \\
        -\frac{\tilde{\xi}_{y}}{2 \rho a} & -\tilde{\xi}_{z} & \tilde{\xi}_{x} & 0 & \frac{\tilde{\xi}_{y}}{2 \rho a} \\
        -\frac{\tilde{\xi}_{z}}{2 \rho a} & \tilde{\xi}_{y} & 0 & \tilde{\xi}_{x} & \frac{\tilde{\xi}_{z}}{2 \rho a} \\
        \frac{1}{2} & 0 & 0 & 0 & \frac{1}{2}
        \end{array}\right] \tag{\theequation a-\theequation b} \\
        \overline{\boldsymbol{\Lambda}}_{\eta}=\left[\begin{array}{ccccc}
        V-a_{\eta} & 0 & 0 & 0 & 0 \\
        0 & V & 0 & 0 & 0 \\
        0 & 0 & V & 0 & 0 \\
        0 & 0 & 0 & V & 0 \\
        0 & 0 & 0 & 0 & V+a_{\eta}
        \end{array}\right], \quad \overline{\mathbf{R}}_{\eta}=\left[\begin{array}{ccccc}
        \frac{1}{2 a^{2}} & \frac{\tilde{\eta}_{x}}{a^{2}} & \frac{\tilde{\eta}_{z}}{a^{2}} & -\frac{\tilde{\eta}_{y}}{a^{2}} & \frac{1}{2 a^{2}} \\
        -\frac{\tilde{\eta}_{x}}{2 \rho a} & 0 & -\tilde{\eta}_{y} & -\tilde{\eta}_{z} & \frac{\tilde{\eta}_{x}}{2 \rho a} \\
        -\frac{\tilde{\eta}_{y}}{2 \rho a} & -\tilde{\eta}_{z} & \tilde{\eta}_{x} & 0 & \frac{\tilde{\eta}_{y}}{2 \rho a} \\
        -\frac{\tilde{\eta}_{z}}{2 \rho a} & \tilde{\eta}_{y} & 0 & \tilde{\eta}_{x} & \frac{\tilde{\eta}_{z}}{2 \rho a} \\
        \frac{1}{2} & 0 & 0 & 0 & \frac{1}{2}
        \end{array}\right] \tag{\theequation c-\theequation d} \\
        \overline{\boldsymbol{\Lambda}}_{\zeta}=\left[\begin{array}{ccccc}
        W-a_{\zeta} & 0 & 0 & 0 & 0 \\
        0 & W & 0 & 0 & 0 \\
        0 & 0 & W & 0 & 0 \\
        0 & 0 & 0 & W & 0 \\
        0 & 0 & 0 & 0 & W+a_{\zeta}
        \end{array}\right] , \quad
        \overline{\mathbf{R}}_{\zeta}=\left[\begin{array}{ccccc}
        \frac{1}{2 a^{2}} & \frac{\tilde{\zeta}_{x}}{a^{2}} & \frac{\tilde{\zeta}_{z}}{a^{2}} & -\frac{\tilde{\zeta}_{y}}{a^{2}} & \frac{1}{2 a^{2}} \\
        -\frac{\tilde{\zeta}_{x}}{2 \rho a} & 0 & -\tilde{\zeta}_{y} & -\tilde{\zeta}_{z} & \frac{\tilde{\zeta}_{x}}{2 \rho a} \\
        -\frac{\tilde{\zeta}_{y}}{2 \rho a} & -\tilde{\zeta}_{z} & \tilde{\zeta}_{x} & 0 & \frac{\tilde{\zeta}_{y}}{2 \rho a} \\
        -\frac{\tilde{\zeta}_{z}}{2 \rho a} & \tilde{\zeta}_{y} & 0 & \tilde{\zeta}_{x} & \frac{\tilde{\zeta}_{z}}{2 \rho a} \\
        \frac{1}{2} & 0 & 0 & 0 & \frac{1}{2}
        \end{array}\right] \tag{\theequation e-\theequation f} 
    \end{gather}
\end{subequations}

where,
\begin{equation*}
a_{\xi}=a\|\nabla \xi\|, \quad a_{\eta}=a\|\nabla \eta\|, \quad a_{\zeta}=a\|\nabla \zeta\|.
\end{equation*}\\

The metric terms with a tilde on the top $\tilde{(\cdot)}$ present in the above matrices denote normalized quantities defined as follows.
\begin{equation} \label{norm-metrics}
\begin{array}{ll}
\tilde{\xi}_{x}=\frac{\xi_{x}}{\|\nabla \xi\|}, \quad \tilde{\xi}_{y}=\frac{\xi_{y}}{\|\nabla \xi\|}, \quad \tilde{\xi}_{z}=\frac{\xi_{z}}{\|\nabla \xi\|}, \\
\tilde{\eta}_{x}=\frac{\eta_{x}}{\|\nabla \eta\|}, \quad \tilde{\eta}_{y}=\frac{\eta_{y}}{\|\nabla \eta\|}, \quad \tilde{\eta}_{z}=\frac{\eta_{z}}{\|\nabla \eta\|}, \\
\tilde{\zeta}_{x}=\frac{\zeta_{x}}{\|\nabla \zeta\|}, \quad \tilde{\zeta}_{y}=\frac{\zeta_{y}}{\|\nabla \zeta\|}, \quad \tilde{\zeta}_{z}=\frac{\zeta_{z}}{\|\nabla \zeta\|}.
\end{array}
\end{equation}

where,
\begin{equation*}
\begin{array}{l}
\|\nabla \xi\|=\sqrt{\xi_{x}^{2}+\xi_{y}^{2}+\xi_{z}^{2}}, \quad
\|\nabla \eta\|=\sqrt{\eta_{x}^{2}+\eta_{y}^{2}+\eta_{z}^{2}}, \text{ and} \quad
\|\nabla \zeta\|=\sqrt{\zeta_{x}^{2}+\zeta_{y}^{2}+\zeta_{z}^{2}}
\end{array}
\end{equation*}

\noindent The corresponding inverse matrices or the left eigen vector matrices are,

\begin{subequations}
    \begin{gather}
        \overline{\mathbf{R}}_{\xi}^{-1}=\left[\begin{array}{ccccc}
        0 & -\tilde{\xi}_{x} \rho a & -\tilde{\xi}_{y} \rho a & -\tilde{\xi}_{z} \rho a & 1 \\
        \tilde{\xi}_{x} a^{2} & 0 & -\tilde{\xi}_{z} & \tilde{\xi}_{y} & -\tilde{\xi}_{x} \\
        \tilde{\xi}_{z} a^{2} & -\tilde{\xi}_{y} & \tilde{\xi}_{x} & 0 & -\tilde{\xi}_{z} \\
        -\tilde{\xi}_{y} a^{2} & -\tilde{\xi}_{z} & 0 & \tilde{\xi}_{x} & \tilde{\xi}_{y} \\
        0 & \tilde{\xi}_{x} \rho a & \tilde{\xi}_{y} \rho a & \tilde{\xi}_{z} \rho a & 1
        \end{array}\right], \quad  \overline{\mathbf{R}}_{\eta}^{-1}=\left[\begin{array}{ccccc}
        0 & -\tilde{\eta}_{x} \rho a & -\tilde{\eta}_{y} \rho a & -\tilde{\eta}_{z} \rho a & 1 \\
        \tilde{\eta}_{x} a^{2} & 0 & -\tilde{\eta}_{z} & \tilde{\eta}_{y} & -\tilde{\eta}_{x} \\
        \tilde{\eta}_{z} a^{2} & -\tilde{\eta}_{y} & \tilde{\eta}_{x} & 0 & -\tilde{\eta}_{z} \\
        -\tilde{\eta}_{y} a^{2} & -\tilde{\eta}_{z} & 0 & \tilde{\eta}_{x} & \tilde{\eta}_{y} \\
        0 & \tilde{\eta}_{x} \rho a & \tilde{\eta}_{y} \rho a & \tilde{\eta}_{z} \rho a & 1
        \end{array}\right] \tag{\theequation a-\theequation b} \\
        \overline{\mathbf{R}}_{\zeta}^{-1}=\left[\begin{array}{ccccc}
        0 & -\tilde{\zeta}_{x} \rho a & -\tilde{\zeta}_{y} \rho a & -\tilde{\zeta}_{z} \rho a & 1 \\
        \tilde{\zeta}_{x} a^{2} & 0 & -\tilde{\zeta}_{z} & \tilde{\zeta}_{y} & -\tilde{\zeta}_{x} \\
        \tilde{\zeta}_{z} a^{2} & -\tilde{\zeta}_{y} & \tilde{\zeta}_{x} & 0 & -\tilde{\zeta}_{z} \\
        -\tilde{\zeta}_{y} a^{2} & -\tilde{\zeta}_{z} & 0 & \tilde{\zeta}_{x} & \tilde{\zeta}_{y} \\
        0 & \tilde{\zeta}_{x} \rho a & \tilde{\zeta}_{y} \rho a & \tilde{\zeta}_{z} \rho a & 1
        \end{array}\right]. \tag{\theequation c} 
    \end{gather}
\end{subequations}


\section{Time integration} \label{sec:time-int}
For time marching the solution, the 3rd order explicit Total Variation Diminishing Range-Kutta (TVD RK-3) \cite{gottlieb1998total} time integration scheme is used. To employ this method, firstly, the residual ($Res$) is computed from the spatial discretization procedure described in the previous sections. Then the following three-stage formulation is employed where the solution corresponding to the previous RK step is used to compute the solution corresponding to the next; thus arriving at the final solution corresponding to $t+\Delta t$ at the end.

\begin{equation}
    \begin{aligned}
\hat{\mathbf{Q}}^{(1)} &=\hat{\mathbf{Q}}^{\mathbf{n}}+\Delta t \operatorname{Res}\left(\hat{\mathbf{Q}}^{\mathbf{n}}\right) J \\
\hat{\mathbf{Q}}^{(\mathbf{2})} &=\frac{3}{4} \hat{\mathbf{Q}}^{\mathbf{n}}+\frac{1}{4} \hat{\mathbf{Q}}^{(1)}+\frac{1}{4} \Delta t \operatorname{Res}\left(\hat{\mathbf{Q}}^{(1)}\right) J \\
\hat{\mathbf{Q}}^{\mathbf{n}+\mathbf{1}} &=\frac{1}{3} \hat{\mathbf{Q}}^{\mathbf{n}}+\frac{2}{3} \hat{\mathbf{Q}}^{(\mathbf{2})}+\frac{2}{3} \Delta t \mathbf{R e s}\left(\hat{\mathbf{Q}}^{(\mathbf{2})}\right) J,
\end{aligned}
\end{equation}

The Jacobian (J) used in the above formulation corresponds to the cell center of the grid, the location where time integration is performed.\\

All the simulations using this solver are performed with a Courant Friedrichs Lewy (CFL) number number of 0.2 unless a different value is specified for a particular case. As opposed to maintaining a constant CFL, a constant stable time-step ($\text{CFL}<0.5$) is maintained for supersonic jet flow calculations to facilitate Fourier analysis of pressure signals extracted from the solution. The time-step while performing Euler simulations is computed as follows.

\begin{equation}
    \Delta t=\mathrm{CFL} \times \min _{\text {cells }} \left(\frac{1}{|U|+c \|\nabla \xi\|},                                                            \frac{1}{|V|+c \|\nabla \eta\|},                                                           \frac{1}{|W|+c \|\nabla \zeta\|}\right).
\end{equation}

\noindent The time step for viscous flow computations is computed as follows.

\begin{equation}
\Delta t=\mathrm{CFL} \times \min \left[\min _{\text {cells }}\left(\frac{1}{|U|+c\|\nabla \xi\|}, \frac{1}{|V|+c\|\nabla \eta\|}, \frac{1}{|W|+c\|\nabla \zeta\|}\right)\right. ,\frac{1}{\alpha}\left.\min _{\text {cells }}\left(\frac{1}{\mu\|\nabla \xi\|^2}, \frac{1}{\mu\|\nabla \eta\|^2}, \frac{1}{\mu\|\nabla \zeta\|^2}\right)\right]
\end{equation}

\section{Metric term computation at boundaries} \label{app:one-sided}

As previously discussed in Section \ref{sec:FP}, the lack of geometric information in ghost points poses a limitation on the computation of metric terms using high-order estimates, as outlined in Equations \ref{FP-interp}, \ref{FP-interp2}, and \ref{interp-formulae}. To address this limitation, low-order approximations are employed for interpolation and derivative computation in the calculation of metric terms. An example of this approach is demonstrated in the computation of the term `$x_{\xi}$' at various boundary cell centers and interfaces.


\begin{figure}[h!]
    \centering
    \includegraphics[width=100mm]{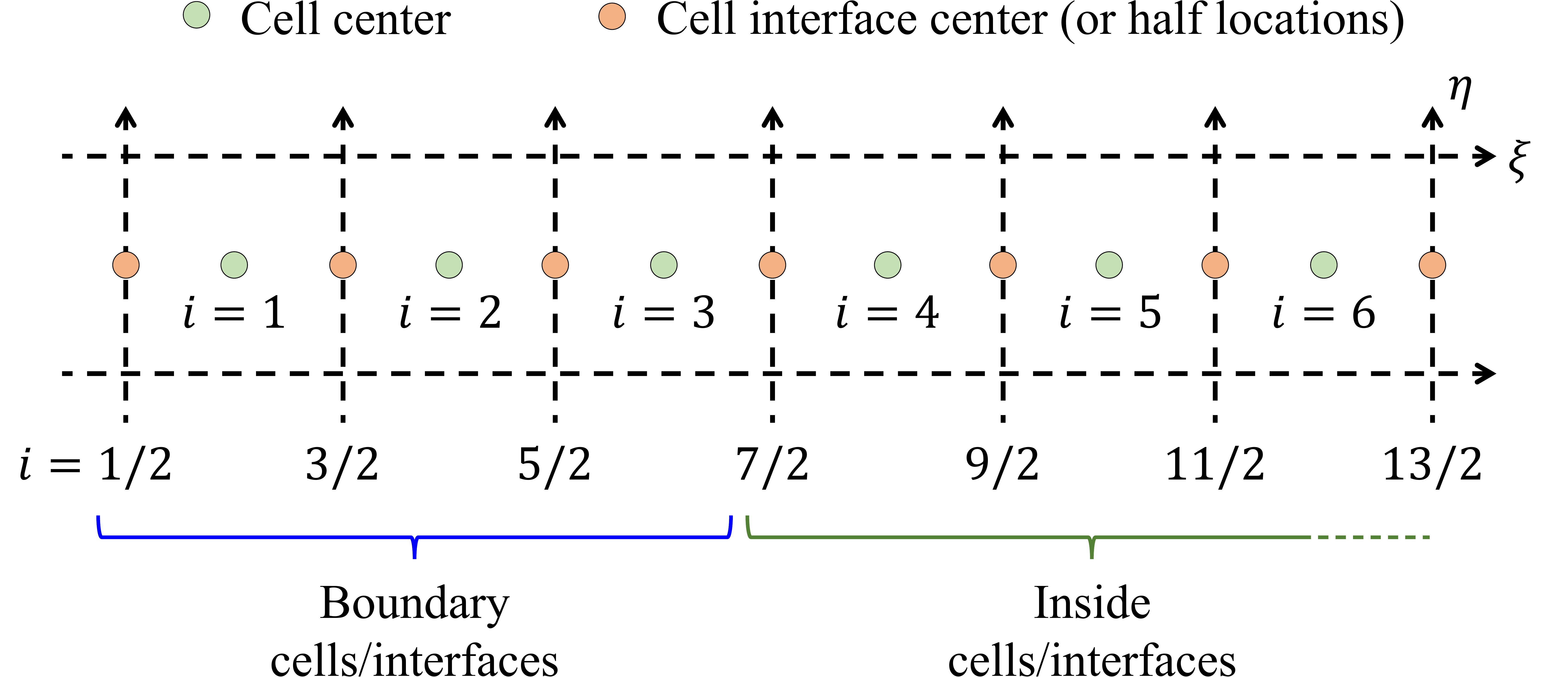}
    \caption{Representation of boundary cells and interfaces along the $\xi$ direction, with corresponding indices. Note that the first cell index begins at $1$ and the first interface index starts at `$1/2$'.}
    \label{boundary-pts}
\end{figure}

\begin{equation}
    x_{i+\frac{1}{2}} = \left\{\begin{array}{ll}
x_{i+1}-\frac{1}{2} x^{'}_{i+1} +\frac{1}{12} x^{''}_{i+1} & \text { if } \quad i=1 \\
\frac{1}{2}\left[x_{i}+\frac{1}{2} x^{'}_{i} +\frac{1}{12} x^{''}_i \right] + \frac{1}{2}\left[x_{i+1}-\frac{1}{2} x^{'}_{i+1} +\frac{1}{12} x^{''}_{i+1} \right] & \text { if } \quad i=2 \text { and } 3.
\end{array}\right.
\end{equation}

\begin{equation}
    \left(x^{'}\right)_{i} = \left\{\begin{array}{ll}
-\frac{3}{2}x_i + 2x_{i+1} -\frac{1}{2}x_{i+2} & \text { if } \quad i=1 \text { (2nd order one sided)}, \\
-\frac{1}{4}x_i - \frac{5}{6}x_{i+1} +\frac{3}{2}x_{i+2} -\frac{1}{2}x_{i+3} +\frac{1}{12}x_{i+4} & \text { if } \quad i=2 \text { (4th order partially one sided)}, \\
\frac{2}{3}\left[ x_{i+1} - x_{i-1}\right] - \frac{1}{12}\left[ x_{i+2} - x_{i-2}\right] & \text { if } \quad i=3 \text { (4th order central)}.
\end{array}\right.
\end{equation}

\begin{equation}
    \left(x^{''}\right)_i = \left\{\begin{array}{ll}
-\frac{3}{2} x^{'}_i + 2x^{'}_{i+1} -\frac{1}{2} x^{'}_{i+2} & \text { if } \quad i=1, \\
-\frac{1}{4}x^{'}_i - \frac{5}{6} x^{'}_{i+1} +\frac{3}{2} x^{'}_{i+2} -\frac{1}{2}x^{'}_{i+3} +\frac{1}{12}x^{'}_{i+4} & \text { if } \quad i=2, \\
-\frac{1}{2}\left[ x^{'}_{i+1} - x^{'}_{i-1}\right] - 2\left[ x_{i+1} - 2x_i + x_{i-1}\right]  & \text { if } \quad i=3.
\end{array}\right.
\end{equation}

and finally
\begin{equation}
    \left(x_{\xi}\right)_i = x_{i+\frac{1}{2}} - x_{i-\frac{1}{2}}
\end{equation}

After computing $x_{\xi}$ and the other required terms present in Equation \ref{GCL_metrics}, they are utilized in the computation of the metric terms through the use of conservative formulations detailed in Section \ref{conser-metrics}. However, as primitive variables are retained in the ghost cells, it is not necessary to use one-sided formulae to compute fluxes at boundary points. As a result, the high-order relations discussed in Section \ref{sec:disc} remain valid for both inviscid and viscous flux computation at boundaries.

\section{Non-conservative metric term formulation} \label{app:non-FP}
The non-conservative metric formulations used to compute metrics in the approach mentioned in section \ref{sec:FP} are:

\begin{equation}
    \begin{aligned}
&\hat{\xi}_{x}=\left(y_\eta z_\zeta-y_\zeta z_\eta\right),\quad 
 \hat{\xi}_{y}=\left(z_\eta x_\zeta-z_\zeta x_\eta\right),\quad 
 \hat{\xi}_{z}=\left(x_\eta y_\zeta-x_\zeta y_\eta\right), \\
&\hat{\eta}_{x}=\left(y_\zeta z_{\xi}-y_{\xi} z_\zeta\right), \quad
 \hat{\eta}_{y}=\left(z_\zeta x_{\xi}-z_{\xi} x_\zeta\right), \quad
 \hat{\eta}_{z}=\left(x_\zeta y_{\xi}-x_{\xi} y_\zeta\right), \\
&\hat{\zeta}_{x}=\left(y_{\xi} z_\eta-y_\eta z_{\xi}\right), \quad
 \hat{\zeta}_{y}=\left(z_\zeta x_\eta-z_\eta x_\zeta\right), \quad
 \hat{\zeta}_{z}=\left(x_{\xi} y_\eta-x_\eta y_{\xi}\right).
\end{aligned}
\end{equation}

The Jacobian is computed using Eqn. \ref{Jac-matrix}. To estimate the derivative terms involved in the above equations, the fourth order explicit central scheme is used \cite{lele1992compact,Visbal2002}. For instance, to estimate $x_{\xi}$, 

\begin{equation} \label{4thorder-grads}
    (x_{\xi})_i = \frac{\frac{2}{3}(x_{i+1}-x_{i-1}) - \frac{1}{12}(x_{i+2}-x_{i-2})}{\Delta \xi}
\end{equation}

The interpolation of metrics to cell interfaces is performed using the same formulae listed in Eqn. \ref{interp-formulae}, except the derivative terms involved in it are computed using the above mentioned fourth order central scheme.

\section{Improved MP-limiter} \label{app:MP-limit}

A modified version of the MP-limiter as described in Ref.\cite{chamx} is presented here. After evaluating the unlimited upwind interpolated values of the characteristic variables ($\mathbf{W}_{i+1/2, unlimited}^{L,R}$) in Step-3 described in Sec. \ref{Inv-disc}, the solution is limited in Step-4. For brevity the procedure is only described for the left interpolated value here.

\noindent Check the following condition to determine if a discontinuity exists:
\begin{equation}
    \left(\mathbf{W}_{i+1 / 2}^{L}-\hat{\mathbf{W}}_i\right)\left(\mathbf{W}_{i+1 / 2}^{L}-\mathbf{W}^{M P}\right) \leq 10^{-20}
\end{equation}

where,
\begin{equation}
    \begin{gathered}
\mathbf{W}^{M P}=\hat{\mathbf{W}}_i+\operatorname{minmod}\left[\hat{\mathbf{W}}_{i+1}-\hat{\mathbf{W}}_i, \beta\left(\hat{\mathbf{W}}_i-\hat{\mathbf{W}}_{i-1}\right)\right], \\
\text { and, } \operatorname{minmod}(a, b)=\frac{1}{2}(\operatorname{sign}(a)+\operatorname{sign}(b)) \min (|a|,|b|) .
\end{gathered}
\end{equation}

Compute $\mathbf{W}_{i+\frac{1}{2}}^{\min }$ and $\mathbf{W}_{i+\frac{1}{2}}^{\max }$ as follows:
\begin{equation}
    \begin{aligned}
& d_i=2\left(\hat{\mathbf{W}}_{i+1}-2 \hat{\mathbf{W}}_i+\hat{\mathbf{W}}_{i-1}\right)-0.5 \Delta x\left(\mathbf{W}_{i+1}^{\prime}-\mathbf{W}_{i-1}^{\prime}\right) \\
& d_{i+1 / 2}^M=\operatorname{minmod}\left(d_i, d_{i+1}\right) \\
& \mathbf{W}_{i+\frac{1}{2}}^{M D}=\frac{1}{2}\left(\hat{\mathbf{W}}_i+\hat{\mathbf{W}}_{i+1}\right)-\frac{1}{2} d_{i+\frac{1}{2}}^M \\
& \mathbf{W}_{i+\frac{1}{2}}^{U L}=\hat{\mathbf{W}}_i+\alpha\left(\hat{\mathbf{W}}_i-\hat{\mathbf{W}}_{i-1}\right) \\
& \mathbf{W}_{i+\frac{1}{2}}^{L C}=\frac{1}{2}\left(3 \hat{\mathbf{W}}_i-\hat{\mathbf{W}}_{i-1}\right)+\frac{4}{3} d_{i-\frac{1}{2}}^M \\
& \mathbf{W}_{i+\frac{1}{2}}^{\min }=\max \left[\min \left(\hat{\mathbf{W}}_i, \hat{\mathbf{W}}_{i+1}, \mathbf{W}_{i+\frac{1}{2}}^{M D}\right), \min \left(\hat{\mathbf{W}}_i, \mathbf{W}_{i+1 / 2}^{U L}, \mathbf{W}_{i+1 / 2}^{L C}\right)\right] \\
& \mathbf{W}_{i+\frac{1}{2}}^{\max }=\min \left[\max \left(\hat{\mathbf{W}}_i, \hat{\mathbf{W}}_{i+1}, \mathbf{W}_{i+\frac{1}{2}}^{M D}\right), \max \left(\hat{\mathbf{W}}_i, \mathbf{W}_{i+\frac{1}{2}}^{U L}, \mathbf{W}_{i+\frac{1}{2}}^{L C}\right)\right] \\
&
\end{aligned}
\end{equation}

Finally, the limited value is computed.
\begin{equation}
    \mathbf{W}_{i+\frac{1}{2}}^{\mathrm{limited}}=\mathbf{W}_{i+\frac{1}{2}}^{L}+\operatorname{minmod}\left(\mathbf{W}_{i+\frac{1}{2}}^{\min }-\mathbf{W}_{i+\frac{1}{2}}^{L}, \mathbf{W}_{i+\frac{1}{2}}^{\max }-\mathbf{W}_{i+\frac{1}{2}}^{L}\right).
\end{equation}

\bibliographystyle{elsarticle-num}
\bibliography{curvilinear}

\begin{thebibliography}{10}
\expandafter\ifx\csname url\endcsname\relax
  \def\url#1{\texttt{#1}}\fi
\expandafter\ifx\csname urlprefix\endcsname\relax\def\urlprefix{URL }\fi
\expandafter\ifx\csname href\endcsname\relax
  \def\href#1#2{#2} \def\path#1{#1}\fi

\bibitem{chamx}
A.~S. Chamarthi, Gradient based reconstruction: Inviscid and viscous flux
  discretizations, shock capturing, and its application to single and
  multicomponent flows, Computers \& Fluids 250 (2023) 105706.

\bibitem{chamarthi2021high}
A.~S. Chamarthi, S.~H. Frankel, High-order central-upwind shock capturing
  scheme using a boundary variation diminishing (bvd) algorithm, Journal of
  Computational Physics 427 (2021) 110067.

\bibitem{li2021low}
Y.~Li, L.~Fu, N.~A. Adams, A low-dissipation shock-capturing framework with
  flexible nonlinear dissipation control, Journal of Computational Physics 428
  (2021) 109960.

\bibitem{Nishikawa2010}
H.~Nishikawa, {Beyond Interface Gradient: A General Principle for Constructing
  Diffusion Schemes}, 40th Fluid Dynamics Conference and Exhibit (2010).

\bibitem{liu1994weighted}
X.-D. Liu, S.~Osher, T.~Chan, Weighted essentially non-oscillatory schemes,
  Journal of Computational Physics 115~(1) (1994) 200--212.

\bibitem{jiang1996efficient}
G.-S. Jiang, C.-W. Shu, Efficient implementation of weighted eno schemes,
  Journal of computational physics 126~(1) (1996) 202--228.

\bibitem{cai2008performance}
X.~Cai, F.~Ladeinde, Performance of weno scheme in generalized curvilinear
  coordinate systems, in: 46th AIAA Aerospace Sciences Meeting and Exhibit,
  2008, p.~36.

\bibitem{nonomura2010freestream}
T.~Nonomura, N.~Iizuka, K.~Fujii, Freestream and vortex preservation properties
  of high-order weno and wcns on curvilinear grids, Computers \& Fluids 39~(2)
  (2010) 197--214.

\bibitem{shadab2019fifth}
M.~A. Shadab, D.~Balsara, W.~Shyy, K.~Xu, Fifth order finite volume weno in
  general orthogonally-curvilinear coordinates, Computers \& Fluids 190 (2019)
  398--424.

\bibitem{cao2019gortler}
S.~Cao, I.~Klioutchnikov, H.~Olivier, G{\"o}rtler vortices in hypersonic flow
  on compression ramps, AIAA Journal 57~(9) (2019) 3874--3884.

\bibitem{cheng2005numerical}
T.~Cheng, K.~Lee, Numerical simulations of underexpanded supersonic jet and
  free shear layer using weno schemes, International Journal of Heat and Fluid
  Flow 26~(5) (2005) 755--770.

\bibitem{martin2006bandwidth}
M.~P. Mart{\'\i}n, E.~M. Taylor, M.~Wu, V.~G. Weirs, A bandwidth-optimized weno
  scheme for the effective direct numerical simulation of compressible
  turbulence, Journal of Computational Physics 220~(1) (2006) 270--289.

\bibitem{Henrick2005}
A.~K. Henrick, T.~D. Aslam, J.~M. Powers, {Mapped weighted essentially
  non-oscillatory schemes: Achieving optimal order near critical points},
  Journal of Computational Physics 207~(2) (2005) 542--567.
\newblock \href {https://doi.org/10.1016/j.jcp.2005.01.023}
  {\path{doi:10.1016/j.jcp.2005.01.023}}.

\bibitem{Fu2016}
L.~Fu, X.~Y. Hu, N.~A. Adams, {A family of high-order targeted ENO schemes for
  compressible-fluid simulations}, Journal of Computational Physics 305 (2016)
  333--359.
\newblock \href {https://doi.org/10.1016/j.jcp.2015.10.037}
  {\path{doi:10.1016/j.jcp.2015.10.037}}.

\bibitem{suresh1997accurate}
A.~Suresh, H.~Huynh, Accurate monotonicity-preserving schemes with runge-kutta
  time stepping, Journal of Computational Physics 136~(1) (1997) 83--99.

\bibitem{kawai2008localized}
S.~Kawai, S.~K. Lele, Localized artificial diffusivity scheme for discontinuity
  capturing on curvilinear meshes, Journal of Computational Physics 227~(22)
  (2008) 9498--9526.

\bibitem{kakumani2022use}
H.~C.~V. Kakumani, N.~R. Vadlamani, P.~G. Tucker, On the use of high order
  central difference schemes for differential equation based wall distance
  computations, Computers \& Fluids 248 (2022) 105666.

\bibitem{sun2016boundary}
Z.~Sun, S.~Inaba, F.~Xiao, Boundary variation diminishing (bvd) reconstruction:
  A new approach to improve godunov schemes, Journal of Computational Physics
  322 (2016) 309--325.

\bibitem{costa2007high}
B.~Costa, W.~S. Don, High order hybrid central—weno finite difference scheme
  for conservation laws, Journal of Computational and Applied Mathematics
  204~(2) (2007) 209--218.

\bibitem{karami2019high}
S.~Karami, P.~C. Stegeman, A.~Ooi, J.~Soria, High-order accurate large-eddy
  simulations of compressible viscous flow in cylindrical coordinates,
  Computers \& Fluids 191 (2019) 104241.

\bibitem{pirozzoli2011stabilized}
S.~Pirozzoli, Stabilized non-dissipative approximations of euler equations in
  generalized curvilinear coordinates, Journal of Computational Physics 230~(8)
  (2011) 2997--3014.

\bibitem{chamarthi2022importance}
A.~S. Chamarthi, S.~Bokor, S.~H. Frankel, On the importance of high-frequency
  damping in high-order conservative finite-difference schemes for viscous
  fluxes, Journal of Computational Physics 460 (2022) 111195.

\bibitem{sainadh2022spectral}
A.~S. Chamarthi, N.~Hoffmann, S.~Bokor, S.~H. Frankel, et~al., On the role of
  spectral properties of viscous flux discretization for flow simulations on
  marginally resolved grids, Computers \& Fluids 251 (2023) 105742.

\bibitem{Visbal2002}
M.~R. Visbal, D.~V. Gaitonde, {On the Use of Higher-Order Finite-Difference
  Schemes on Curvilinear and Deforming Meshes}, Journal of Computational
  Physics 181~(1) (2002) 155--185.
\newblock \href {https://doi.org/10.1006/jcph.2002.7117}
  {\path{doi:10.1006/jcph.2002.7117}}.

\bibitem{thomas1979geometric}
P.~Thomas, C.~Lombard, Geometric conservation law and its application to flow
  computations on moving grids, AIAA journal 17~(10) (1979) 1030--1037.

\bibitem{nonomura2015new}
T.~Nonomura, D.~Terakado, Y.~Abe, K.~Fujii, A new technique for freestream
  preservation of finite-difference weno on curvilinear grid, Computers \&
  Fluids 107 (2015) 242--255.

\bibitem{zhu2019free}
Y.~Zhu, X.~Hu, Free-stream preserving linear-upwind and weno schemes on
  curvilinear grids, Journal of Computational Physics 399 (2019) 108907.

\bibitem{kakumani2023gpu}
H.~C.~V. Kakumani, A.~S. Chamarthi, N.~N. Hoffmann, S.~H. Frankel,
  Gpu-accelerated numerical study of temperature effects in choked
  under-expanded supersonic jets, in: AIAA SCITECH 2023 Forum, 2023, p. 0976.

\bibitem{powell1953mechanism}
A.~Powell, On the mechanism of choked jet noise, Proceedings of the Physical
  Society. Section B 66~(12) (1953) 1039.

\bibitem{edgington2019aeroacoustic}
D.~Edgington-Mitchell, Aeroacoustic resonance and self-excitation in screeching
  and impinging supersonic jets--a review, International Journal of
  Aeroacoustics 18~(2-3) (2019) 118--188.

\bibitem{choi2012grid}
H.~Choi, P.~Moin, Grid-point requirements for large eddy simulation:
  Chapman’s estimates revisited, Physics of fluids 24~(1) (2012) 011702.

\bibitem{witherden2014pyfr}
F.~D. Witherden, A.~M. Farrington, P.~E. Vincent, Pyfr: An open source
  framework for solving advection--diffusion type problems on streaming
  architectures using the flux reconstruction approach, Computer Physics
  Communications 185~(11) (2014) 3028--3040.

\bibitem{bernardini2021streams}
M.~Bernardini, D.~Modesti, F.~Salvadore, S.~Pirozzoli, Streams: a high-fidelity
  accelerated solver for direct numerical simulation of compressible turbulent
  flows, Computer Physics Communications 263 (2021) 107906.

\bibitem{romero2020zefr}
J.~Romero, J.~Crabill, J.~Watkins, F.~D. Witherden, A.~Jameson, Zefr: A
  gpu-accelerated high-order solver for compressible viscous flows using the
  flux reconstruction method, Computer Physics Communications 250 (2020)
  107169.

\bibitem{di2021htr}
M.~Di~Renzo, S.~Pirozzoli, Htr-1.2 solver: Hypersonic task-based research
  solver version 1.2, Computer Physics Communications 261 (2021) 107733.

\bibitem{goc2021large}
K.~A. Goc, O.~Lehmkuhl, G.~I. Park, S.~T. Bose, P.~Moin, Large eddy simulation
  of aircraft at affordable cost: a milestone in computational fluid dynamics,
  Flow 1 (2021).

\bibitem{nampelly2022surface}
G.~Nampelly, A.~S. Malathi, A.~Vaid, N.~R. Vadlamani, S.~Rengarajan, K.~Kontis,
  Surface roughness effects on cavity flows, Flow, Turbulence and Combustion
  (2022) 1--25.

\bibitem{huang2009energy}
S.~Huang, S.~Xiao, W.-c. Feng, On the energy efficiency of graphics processing
  units for scientific computing, in: 2009 IEEE International Symposium on
  Parallel \& Distributed Processing, IEEE, 2009, pp. 1--8.

\bibitem{vspetko2021dgx}
M.~{\v{S}}pet'ko, O.~Vysock{\`y}, B.~Jans{\'\i}k, L.~{\v{R}}{\'\i}ha, Dgx-a100
  face to face dgx-2—performance, power and thermal behavior evaluation,
  Energies 14~(2) (2021) 376.

\bibitem{bres2022gpu}
G.~A. Br{\`e}s, S.~T. Bose, C.~B. Ivey, M.~Emory, F.~Ham, Gpu-accelerated
  large-eddy simulations of supersonic jets from twin rectangular nozzle, in:
  28th AIAA/CEAS Aeroacoustics 2022 Conference, 2022, p. 3001.

\bibitem{terrana2020gpu}
S.~Terrana, C.~Nguyen, J.~Peraire, Gpu-accelerated large eddy simulation of
  hypersonic flows, in: AIAA Scitech 2020 Forum, 2020, p. 1062.

\bibitem{laufer2022gpu}
M.~Laufer, S.~H. Frankel, D.~Greenblatt, Gpu-accelerated implicit large eddy
  simulation of a naca 0018 airfoil with active flow control, in: AIAA SCITECH
  2022 Forum, 2022, p. 0471.

\bibitem{cernetic2022high}
M.~Cernetic, V.~Springel, T.~Guillet, R.~Pakmor, High-order discontinuous
  galerkin hydrodynamics with sub-cell shock capturing on gpus, arXiv preprint
  arXiv:2208.11131 (2022).

\bibitem{Sutherland1893}
W.~Sutherland, \href{https://doi.org/10.1080/14786449308620508}{Lii. the
  viscosity of gases and molecular force}, The London, Edinburgh, and Dublin
  Philosophical Magazine and Journal of Science 36~(223) (1893) 507--531.
\newblock \href
  {http://arxiv.org/abs/https://doi.org/10.1080/14786449308620508}
  {\path{arXiv:https://doi.org/10.1080/14786449308620508}}, \href
  {https://doi.org/10.1080/14786449308620508}
  {\path{doi:10.1080/14786449308620508}}.
\newline\urlprefix\url{https://doi.org/10.1080/14786449308620508}

\bibitem{van1977towards}
B.~Van~Leer, Towards the ultimate conservative difference scheme. iv. a new
  approach to numerical convection, Journal of Computational Physics 23~(3)
  (1977) 276--299.

\bibitem{Shu1997}
C.~W. Shu, {Essentially Non-Oscillatory and Weighted Essentially
  Non-Oscillatory Schemes for Hyperbolic Conservation Laws Operated by
  Universities Space Research Association}, ICASE Report~(97-65) (1997) 1--78.

\bibitem{toro2009riemann}
E.~Toro, Riemann Solvers and Numerical Methods for Fluid Dynamics: A Practical
  Introduction, Springer Berlin Heidelberg, 2009.

\bibitem{harten1983upstream}
A.~Harten, P.~D. Lax, B.~v. Leer, On upstream differencing and godunov-type
  schemes for hyperbolic conservation laws, SIAM review 25~(1) (1983) 35--61.

\bibitem{shen2010large}
Y.~Shen, G.~Zha, Large eddy simulation using a new set of sixth order schemes
  for compressible viscous terms, Journal of Computational Physics 229~(22)
  (2010) 8296--8312.

\bibitem{Nishikawa2013}
H.~Nishikawa, {First , Second , and Third Order Finite-Volume Schemes}, Journal
  of Computational Physics (2013) 7--10.

\bibitem{lien1996multiblock}
F.~Lien, W.~Chen, M.~Leschziner, A multiblock implementation of a
  non-orthogonal, collocated finite volume algorithm for complex turbulent
  flows, International Journal for Numerical Methods in Fluids 23~(6) (1996)
  567--588.

\bibitem{ponton1997near}
M.~K. Ponton, J.~M. Seiner, M.~C. Brown, Near field pressure fluctuations in
  the exit plane of a choked axisymmetric nozzle, Tech. rep., NASA (1997).

\bibitem{jiang2014free}
Y.~Jiang, C.-W. Shu, M.~Zhang, Free-stream preserving finite difference schemes
  on curvilinear meshes, Methods and applications of analysis 21~(1) (2014)
  1--30.

\bibitem{achu2021entropically}
S.~Achu, N.~R. Vadlamani, Entropically damped artificial compressibility solver
  using higher order finite difference schemes on curvilinear and deforming
  meshes, in: AIAA Scitech 2021 Forum, 2021, p. 0634.

\bibitem{clausen2013entropically}
J.~R. Clausen, Entropically damped form of artificial compressibility for
  explicit simulation of incompressible flow, Physical Review E 87~(1) (2013)
  013309.

\bibitem{edgington2021generation}
D.~Edgington-Mitchell, J.~Weightman, S.~Lock, R.~Kirby, V.~Nair, J.~Soria,
  D.~Honnery, The generation of screech tones by shock leakage, Journal of
  Fluid Mechanics 908 (2021).

\bibitem{raman1999supersonic}
G.~Raman, Supersonic jet screech: half-century from powell to the present,
  Journal of Sound and Vibration 225~(3) (1999) 543--571.

\bibitem{borges2008improved}
R.~Borges, M.~Carmona, B.~Costa, W.~S. Don, An improved weighted essentially
  non-oscillatory scheme for hyperbolic conservation laws, Journal of
  Computational Physics 227~(6) (2008) 3191--3211.

\bibitem{ahn2021numerical}
M.~Ahn, D.-J. Lee, M.~Mihaescu, A numerical study on near-field pressure
  fluctuations of symmetrical and anti-symmetrical flapping modes of twin-jet
  using a high-resolution shock-capturing scheme, Aerospace Science and
  Technology 119 (2021) 107147.

\bibitem{ahn2018supersonic}
M.-H. Ahn, D.~J. Lee, Supersonic jet noise prediction using optimized compact
  scheme with modified monotonicity preserving limiter, in: 2018 AIAA Aerospace
  Sciences Meeting, 2018, p. 1249.

\bibitem{bodony2006analysis}
D.~J. Bodony, Analysis of sponge zones for computational fluid mechanics,
  Journal of Computational Physics 212~(2) (2006) 681--702.

\bibitem{germano1991dynamic}
M.~Germano, U.~Piomelli, P.~Moin, W.~H. Cabot, A dynamic subgrid-scale eddy
  viscosity model, Physics of Fluids A: Fluid Dynamics 3~(7) (1991) 1760--1765.

\bibitem{fureby1999monotonically}
C.~Fureby, F.~Grinstein, Monotonically integrated large eddy simulation of free
  shear flows, AIAA journal 37~(5) (1999) 544--556.

\bibitem{pack1950note}
D.~Pack, A note on prandtl's formula for the wave-length of a supersonic gas
  jet, The Quarterly Journal of Mechanics and Applied Mathematics 3~(2) (1950)
  173--181.

\bibitem{tam1995supersonic}
C.~K. Tam, Supersonic jet noise, Annual review of fluid mechanics 27~(1) (1995)
  17--43.

\bibitem{bailly2016high}
C.~Bailly, K.~Fujii, High-speed jet noise, Mechanical Engineering Reviews 3~(1)
  (2016) 15--00496.

\bibitem{gojon2019antisymmetric}
R.~Gojon, E.~Gutmark, M.~Mihaescu, Antisymmetric oscillation modes in
  rectangular screeching jets, AIAA Journal 57~(8) (2019) 3422--3441.

\bibitem{davies1962tones}
M.~Davies, D.~Oldfield, Tones from a choked axisymmetric jet. i. cell
  structure, eddy velocity and source locations, Acta Acustica united with
  Acustica 12~(4) (1962) 257--267.

\bibitem{westley1969near}
R.~Westley, J.~Woolley, The near field sound pressures of a choked jet during a
  screech cycle, in: AGARD Conference Proceedings, Vol.~42, 1969, pp. 23--1.

\bibitem{powell1992observations}
A.~Powell, Y.~Umeda, R.~Ishii, Observations of the oscillation modes of choked
  circular jets, The Journal of the Acoustical Society of America 92~(5) (1992)
  2823--2836.

\bibitem{weiss2019tutorial}
J.~Weiss, A tutorial on the proper orthogonal decomposition, in: AIAA aviation
  2019 forum, 2019, p. 3333.

\bibitem{chandravamsi2023control}
H.~Chandravamsi, S.~Bhardwaj, K.~Ramachandra, R.~Sriram, Control of bow shock
  induced three-dimensional separation using bleed through holes, Physics of
  Fluids 35~(1) (2023) 016121.
\newblock \href {https://doi.org/10.1063/5.0132358}
  {\path{doi:10.1063/5.0132358}}.

\bibitem{openacc-web1}
Openacc compilers and directives for open accelerators,
  \url{https://www.openacc.org/}, accessed: 2022-10-12.

\bibitem{openacc-web2}
Openacc learning resources, \url{https://www.openacc.org/resources}, accessed:
  2022-10-12.

\bibitem{crespo2015dualsphysics}
A.~J. Crespo, J.~M. Dom{\'\i}nguez, B.~D. Rogers, M.~G{\'o}mez-Gesteira,
  S.~Longshaw, R.~Canelas, R.~Vacondio, A.~Barreiro, O.~Garc{\'\i}a-Feal,
  Dualsphysics: Open-source parallel cfd solver based on smoothed particle
  hydrodynamics (sph), Computer Physics Communications 187 (2015) 204--216.

\bibitem{masatsuka2013like}
K.~Masatsuka, I do Like CFD, vol. 1, Vol.~1, Lulu. com, 2013.

\bibitem{gottlieb1998total}
S.~Gottlieb, C.-W. Shu, Total variation diminishing runge-kutta schemes,
  Mathematics of computation 67~(221) (1998) 73--85.

\bibitem{lele1992compact}
S.~K. Lele, Compact finite difference schemes with spectral-like resolution,
  Journal of Computational Physics 103~(1) (1992) 16--42.

\end{thebibliography}

\end{document}